\def\sqr#1#2{{\vcenter{\vbox{\hrule height.#2pt\hbox{\vrule width.#2pt 
height#1pt \kern#1pt \vrule width.#2pt}\hrule height.#2pt}}}}
\def\d{\partial}

\def\=d{\,{\buildrel\rm def\over =}\,}

\documentclass[12pt]{article}
\usepackage{amssymb}\usepackage{amsmath}\usepackage{amsthm}

\newcommand{\beq}{\begin{equation}}
\newcommand{\eeq}{\end{equation}}
\newcommand{\bg}{\begin{gather}}
\newcommand{\eg}{\end{gather}}
\newcommand{\CC}{\mathbb C}
\newcommand{\RR}{\mathbb R}

\newcommand{\NN}{\mathbb N}
\newcommand{\MM}{\mathbb M}
\newcommand{\supp}{{\rm supp\>}}
\newcommand{\C}[1]{{\mathcal #1}}
\newcommand{\ket}[1]{\lvert#1\rangle}       
\newcommand{\bra}[1]{\langle#1\rvert}       

\newtheorem{prop}{Proposition}
\newtheorem{thm}[prop]{Theorem}

\newtheorem{lemma}[prop]{Lemma}
\newtheorem{definition}{Definition}

\begin{document}
\title{Causal perturbation theory in terms of retarded products,
and a proof of the Action Ward Identity}
\author{Michael D\"utsch
\thanks{Work supported by the Deutsche 
Forschungsgemeinschaft.} \\[2mm] 
Institut f\"ur Theoretische Physik\\
Universit\"at Z\"urich\\
CH-8057 Z\"urich, Switzerland\\
{\tt \small duetsch@physik.unizh.ch}\\[2mm] 
\and Klaus Fredenhagen \\[2mm]
II. Institut f\"ur Theoretische Physik\\
Universit\"at Hamburg\\
D-22761 Hamburg, Germany\\
{\tt \small klaus.fredenhagen@desy.de}}

\date{}
\maketitle
\begin{abstract}
In the framework of perturbative algebraic quantum field theory a
local construction of interacting fields in terms of retarded products 
is performed, based on earlier work of Steinmann \cite{Ste}. 
In our formalism the entries of the retarded 
products are local functionals of the off shell classical fields,
and we prove that the interacting fields depend only on the action
and not on terms in the Lagrangian which are total derivatives, thus 
providing a proof of Stora's 'Action Ward Identity' \cite{AWI}. 
The theory depends on free parameters which flow under the 
renormalization group. This flow can be derived in our local framework 
independently of the infrared behavior, as was first established by Hollands 
and Wald \cite{HW}. We explicitly compute
non-trivial examples for the renormalization of the interaction
and the field.    

Renormalization group evolution of parameters,
11.15.Bt Gauge field theories: General properties of perturbation theory

\end{abstract}

\tableofcontents
\section{Introduction}\setcounter{equation}{0}
Among the various, essentially equivalent formulations of 
quantum field theory (QFT) the
algebraic formulation \cite{HK,Haag} seems to be the most appealing one
from the conceptual point of view, but on the other side the least
accessible one from the computational point of view. In perturbation
theory, which is still the most successful method for making contact
between theory and experiment, the approach towards a construction of
interacting fields in terms of operators on a Hilbert space was
popular in the early years, see e.g. \cite{Kall}. But later it was
largely abandoned in favor of a direct determination of Green
functions. This is partially due to some complications which arise in
the perturbative expansion of Wightman functions
compared to the time ordered functions. Moreover, if the ultimate goal
of QFT is the computation of the $S$-matrix, the approach via time
ordered functions is more direct, in view of the LSZ-formulas, and the
approach via Wightman functions seems to be a detour, which is
conceptually nice but unimportant for practitioners.

There are, however, several reasons for a revision of this prevailing 
attitude. One is the desire to understand QFT on generic curved
backgrounds. There, no general asymptotic condition \`a la LSZ exists;
moreover, even the concepts of vacuum and particles loose their
distinguished meaning, and one is forced to base the theory on the
algebra of quantum fields (see e.g. \cite{BF} and references cited 
therein). Another reason is the connection to the classical limit.
The algebra of perturbative quantum fields may be understood
in terms of deformation quantization of the underlying Poisson algebra
of free classical fields \cite{DF1,DF1a}, and one may hope that
deformation quantization applies also to non-perturbative
fields. But even in the traditional QFT on Minkowski space
the algebraic formulation has great advantages because it completely
separates the UV-problem from the IR-problem and, in principle, allows a
consistent treatment of situations (like in QED \cite{DF}) where an
$S$-matrix, strictly speaking, does not exist.

Actually, the perturbative construction of the algebras of quantum
fields is possible, e.g. using the Bogoliubov-Epstein-Glaser approach
of causal perturbation theory \cite{BS,EG}. There, the system of time
ordered products of Wick polynomials of free quantum fields is 
recursively constructed,
and interacting fields are given in terms of Bogoliubov's formula 
\cite{BS}
\begin{equation}
A_{\int\! g{\cal L}}(x)\=d \frac{1}{i}\frac{\delta}{\delta h(x)}
\Bigl(T\> e^{i\int\! g{\cal L}}\Bigr)^{-1}
\Bigl(T\> e^{i\int\! g{\cal L}+hA}\Bigr)\vert_{h =0}\ ,\label{intf}
\end{equation}
where $g$ and $h$ are test functions and $A$ is a polynomial in the
basic fields and their partial derivatives.

The Taylor series expansion of the interacting field with respect 
to the interaction defines the retarded products
\begin{equation}
    R_{n,1}({\cal L}(x_1),\ldots,{\cal L}(x_n),A(x))=
    \frac{\delta^n}{\delta g(x_1)\cdots \delta g(x_n)} 
    A_{\int\! g{\cal L}}(x)
\label{intf:R}
\end{equation}
which can, by Bogoliubov's formula, be expressed in terms of time 
ordered and anti time ordered products.
It is desirable, however, to have
a direct construction of the 
retarded products, without the detour via the time ordered products.
In particular, the approach to the classical limit simplifies enormously, 
since the retarded products are power
series in $\hbar$, whereas the time ordered products are Laurent
series (see
the simplifications in \cite{DF1a} compared to \cite{DF1}). 

Such a construction was  
performed by Steinmann \cite{Ste} (see also his recent book 
on QED \cite{Ste1}). 
We review his construction with some modifications. The most important one 
is the restriction to localized interactions (the support of the
coupling function $g$ in (\ref{intf}) is bounded) as it is characteristic 
for causal perturbation theory in the sense of Bogoliubov \cite{BS}, 
Epstein and Glaser \cite{EG}. 
Therefore we don't have to discuss the asymptotic structure. Instead we 
use the algebraic adiabatic limit introduced in \cite{BF}. This
limit relies on the observation that the algebraic structure of 
observables localized in a certain region does not depend on the 
behavior of the interaction outside of the region. This allows a 
construction of interacting fields but not of the $S$-matrix 
(Sect.~5). In this way we 
obtain a unified theory of massive and massless theories as well as of 
theories on curved space time (the latter are not discussed 
in the present paper).

We allow as interaction also terms with derivatives. Traditionally, 
in causal perturbation theory one uses, as arguments of time ordered 
and retarded products, on shell fields, i.e.~fields which are 
subject to the free field equations\footnote{An exception is Sect.~4
of the Epstein Glaser paper \cite{EG}: to interprete it consistently
the arguments of time ordered products must be off shell fields.} 
\cite{EG,Scharf,BD}. The inclusion of derivative couplings 
then leads to complications, and a consistent treatment requires a 
somewhat involved formalism \cite{BD}. Moreover, a change of the splitting 
between free and interaction terms in the Lagrangean requires a major effort.
We therefore prefer to use off shell fields, thereby following a suggestion 
of Stora \cite{AWI}. A natural question is then whether in this framework 
derivatives commute with time ordered and retarded products. This amounts 
to the problem whether the interacting fields depend only on the 
{\it action} $S$ and not on how it is written as an integral
$S=\int dx\, g(x){\cal L}(x)$, i.e. whether total derivatives in 
$g{\cal L}$ can be ignored. The corresponding Ward identity was termed 
Action Ward Identity by Stora \cite{AWI}. We will show in this paper that 
the Action Ward Identity can indeed be fulfilled.  

We even go a step further: also the {\it values} of our retarded
products are off shell fields, in contrast with the literature.
For the physical predictions this makes no difference, and we gain
technical simplifications.

The dependence of the theory on the renormalization conditions is 
usually analyzed in the adiabatic limit. In the algebraic adiabatic limit 
we can 
perform this analysis completely locally, thus avoiding all infrared 
problems (Sect.~5). Actually, such an analysis was already given by 
Hollands and Wald \cite{HW}
for theories on curved space-times where the traditional 
adiabatic limit makes no sense, in general.  
In case of the scaling transformations we illustrate the
formalism by computing (to lowest non-trivial order) examples for the
renormalization of the interaction and the field.   

Hollands and Wald \cite{HW1-2} also introduce a new concept of 
scaling transformation which applies to fields on generic curved 
space-times. This concept already entails important 
consequences for massive theories on Minkowski space.  We therefore 
adopt this point of view: among the axioms of causal perturbation 
theory we require smooth mass dependence\footnote{In even dimensions this 
cannot be satisfied if the $*$-product is defined with respect to the usual  
two-point function of the free field; a modification of the $*$-product is necessary.} 
for $m\geq 0$ and almost homogeneous scaling. These conditions ensure that  
renormalization depends only on the short distance behavior 
of the theory, in agreement with the principle of locality. 

A main question in perturbative QFT is whether symmetry with respect
to a certain group $G$ can be maintained in the process of
renormalization. In Appendix C we prove that this is possible if all
finite dimensional representations of $G$ are completely reducible.
In case of compact groups and for Lorentz-invariance we complete this
existence result by giving a construction of a symmetric
renormalization. 

\section{Axioms for retarded products}\setcounter{equation}{0}
We consider for notational simplicity the theory of a real scalar field on 
$d$ dimensional Minkowski space $\MM$, $d>2$. The classical configuration 
space $\mathcal{C}$ is the space $\mathcal{C}^{\infty}(\MM,\RR)$
and the field $\varphi$ is the evaluation functional on this space:
$(\d^a\varphi) (x)(h)=\d^ah(x),\>a\in\NN_0^d$.
Let $\mathcal{F}$ be the set of all functionals $F$ on 
$\mathcal{C}$ with values in the formal power series in $\hbar$ and 
which have the form
\begin{equation}
  F(\varphi)=\sum_{n=0}^N\int dx_1\ldots dx_n\,\varphi(x_1)
  \cdots\varphi(x_n)f_n(x_1,\ldots,x_n)\ ,\quad\quad N<\infty\ ,
  \label{F(phi)} 
\end{equation}
where the $f_n$'s are $\CC[[\hbar]]$-valued 
distributions with compact support, which are symmetric under 
permutations of the arguments and whose wave front 
sets satisfy the condition 
\begin{equation}\label{wave front}
  \mathrm{WF}(f_n)\cap \bigl(\MM^n\times
(\overline{V_+}^n\cup \overline{V_-}^n)\bigr)
   =\emptyset 
\end{equation}
and $f_0\in\CC[[\hbar]]$ (see Sect.~5.1 of \cite{DF1} and Sect.~4 of \cite{DF1a}).
The value of the functional $F(\varphi)$ for the argument $h\in {\cal C}$ is obtained by 
substituting everywhere $h$ for $\varphi$ on the right side of (\ref{F(phi)}): $F(\varphi)(h)
=F(h)$. An important example is
\begin{equation}
  f_n(x_1,...,x_n)=\left\{
  \begin{array}{ccc}
0 &  \mathrm{for} & n\not= k  \\
\int dx\, f(x)\,\prod_{j=1}^k\d^{a_j}\delta(x_j-x) & \mathrm{for} & n= k 
  \end{array}\right.
\end{equation}
(where $f\in {\cal D}(\MM)$ and $a_j\in\NN_0^d$), which gives $F(\varphi)=(-1)^{\sum_j|a_j|}
(\prod_{j=1}^k\d^{a_j}\varphi)(f)$.  $\mathcal{F}$ is a *-algebra with
the classical product $(F_1\cdot F_2)(h):=F_1(h)\cdot F_2(h)$, where $F^*$
is obtained from $F$ (\ref{F(phi)}) by complex conjugation of all $f_n$'s.

We introduce the functional
\begin{equation}
  \omega_0:\left\{
  \begin{array}{ccc}
        \mathcal{F} &  \longrightarrow  & \CC[[\hbar]]  \\
        F & \mapsto & F(0)\equiv f_{0}
  \end{array}\right. \label{vacuum}
\end{equation}
which will be interpreted as 'vacuum state'. 
$\mathcal{F}(\mathcal{O})$ denotes the space of functionals 
localized in the spacetime region $\mathcal{O}$, i.e.~which 
depend only on $\varphi(x)$ for $x\in\mathcal{O}$,
 \begin{displaymath}
        \mathcal{F}(\mathcal{O})=\{F\in\mathcal{F}\quad |\quad\supp\frac{\delta 
        F}{\delta \varphi} \subset\mathcal{O}\}\ .
 \end{displaymath}
Here, the functional derivatives of a polynomial functional $F$ 
(\ref{F(phi)}) are ${\cal F}$-valued distributions given by
\begin{gather}
\frac{\delta^k F}{\delta\varphi (x_1)\ldots\delta\varphi (x_k)}=\notag\\
\sum_{n=k}^N\frac{n!}{(n-k)!}\int dy_1\ldots dy_{n-k}\,\varphi(y_1)
  \cdots\varphi(y_{n-k})f_n(x_1,\ldots,x_k,y_1,\ldots,y_{n-k})\ .
\end{gather}
  
Let $\Delta^{(m)}_+$ be the 2-point function of the free scalar field
with mass $m$. On $\mathcal{F}$ we define an $m$-dependent 
associative product by
\begin{gather}
  (F\star_m G)(\varphi)=
  \sum_{n=0}^{\infty}\frac{\hbar^n}{n!}
  \int dx_1\ldots dx_n dy_1\ldots dy_n 
  \frac{\delta^n F}{\delta\varphi(x_1)\cdots\delta\varphi(x_n)}\notag\\
  \cdot \prod_{i=1}^n \Delta^{(m)}_+(x_i-y_i) 
  \frac{\delta^n G}{\delta\varphi(y_1)\cdots\delta\varphi(y_n)}\ ,
  \label{*-product}
\end{gather}
which induces a product $\star_m:\mathcal{F}(\mathcal{O})\times\mathcal{F}
(\mathcal{O})\rightarrow \mathcal{F}(\mathcal{O})$. The condition 
(\ref{wave front}) on the coefficients $f_n$ guarantees that the product is 
well defined, i.e. the pointwise product of distributions on the right side of
(\ref{*-product}) exists and the 'coefficients' of $(F\star_m G)$ satisfy again
(\ref{wave front}). $\star_m$
corresponds to a $\star$-product in the sense of deformation quantization 
\cite{BFFLS} (see also \cite{HH,BFFO}), and it
may be interpreted as Wick's Theorem for `off-shell fields'
(i.e.~fields which are not restricted by any field equation).
The corresponding algebras are denoted by ${\cal A}^{(m)}$ 
and ${\cal A}^{(m)}(\mathcal{O})$, respectively. For $\hbar=0$ the 
product reduces to the classical product.   
Since we understand the functionals $F,G$ and their product $(F\star_m
G)$ as formal power series in $\hbar$, each equation must hold
individually in each order of $\hbar$, in particular renormalization
(see Sect.~3) has to be done in this sense.

The algebra of Wick polynomials is obtained by dividing out the ideal 
$\mathcal{J}^{(m)}$ generated by the field equation
\begin{gather}
  \mathcal{J}^{(m)}=\{F(\varphi)=\sum_{n=1}^N
  \int dx_1\ldots dx_n \varphi(x_1)\cdots 
  \varphi(x_n)(\square_{x_1}+m^2)f_n(x_1,\ldots x_n)\> |\>\notag\\
  f_n\>\mathrm{as}\>\mathrm{above}\}\ .\label{J}
\end{gather}
The quotient algebra $\mathcal{F}^{(m)}_{0}\equiv {\cal F}/{\cal J}^{(m)}$ 
can be, for each fixed value of 
$\hbar>0$, faithfully represented on Fock space by 
identifying the classical product of fields $\pi\Bigl(\int\,dx_1...dx_n\,
\varphi(x_1)...\varphi(x_n)\,f_n(x_1,...,x_n)\Bigr)$
with the normally ordered product $\int\,dx_1...dx_n\,
:\!\varphi(x_1)...\varphi(x_n)\!:\,f_n(x_1,...,x_n)$, where $\pi$ is the 
canonical surjection $\pi:{\cal F}\rightarrow {\cal F}^{(m)}_0$ (Theorem
4.1 in \cite{DF1a}). $\omega_0$ (\ref{vacuum}) induces a state on
$\mathcal{F}^{(m)}_{0}$ which corresponds to the Fock vacuum.
$\mathcal{F}^{(m)}_{0}(\mathcal{O})$ denotes 
the image of $\mathcal{F}(\mathcal{O})$ under $\pi$.
Let ${\cal C}^{(m)}_{0}\subset{\cal C}$ be the space of
smooth solutions of the free field equation. Since
$F\in\mathcal{J}^{(m)}\Leftrightarrow F\vert_{{\cal C}^{(m)}_{0}}
=0$, the canonical surjection $\pi$ can alternatively be viewed
as the restriction of the functionals $F\in {\cal F}$ to 
${\cal C}^{(m)}_{0}$ (for details see \cite{DF2}).  

We are particularly interested in local functionals. We call a functional 
$F\in\mathcal{F}$ {\it local} if 
\begin{displaymath}
  \frac{\delta^2 F}{\delta \varphi(x)\delta\varphi(y)}=0 
   \quad\text{ for }\quad x\ne y \ . 
\end{displaymath}
Local functionals are of the form
\begin{equation}
  F=\int dx\,\sum_{i=1}^N A_i(x)h_i(x) \equiv \sum_{i=1}^{N} 
  A_{i}(h_{i})\label{F=Wf}
\end{equation}
where the $A_i$'s are polynomials of the field $\varphi$ and its 
derivatives and the $h_i$'s are test functions with compact support,
$h_i\in {\cal D}(\MM)$. The set of local functionals will 
be denoted by $\mathcal{F}_{\text{loc}}$. 
\medskip
 
\noindent {\it Remark:} There exist faithful Hilbert space representations of the 
off-shell fields. For example, a faithful representation $\pi$ of the algebra ${\cal F}$ 
(with the classical product) is obtained by interpreting ${\cal F}$ as a vector space
and the representation is defined by left-multiplication: $\pi(F)\,G:=FG$ ($F,G\in {\cal F}$). 
A possible scalar product reads 
\begin{equation}
 \langle F,G\rangle :=\omega_0(F^*\star_g G)\ ,\label{scal-prod}
\end{equation}
where $\star_g$ is the $\star$-product (\ref{*-product}) with $\Delta^{(m)}_+$
replaced by the 2-point function $\Delta^{[g]}_+$ of a generalized free field
with weight function $g\in {\cal D}(\RR_+)$, i.e. 
$\Delta^{[g]}_+(y)=\int dm^2\, g(m^2) \,\Delta^{(m)}_+(y)$. (Note that the smoothness
of $g$ excludes the case of a free field, $\Delta^{[g]}_+=\Delta^{(m_0)}_+$
for some $m_0$, in which (\ref{scal-prod}) would be degenerate.)

Compared with their Hilbert space representations, the algebras ${\cal F}$ and 
$\mathcal{F}^{(m)}_{0}$ are more flexible and more 
convenient\footnote{For example one does not need to care about domains 
of unbounded operators.}; and, as it is demostrated by this paper, they provide 
all necessary information.  
\medskip

We want to construct, for any pair of local functionals $F,G\sim\hbar^0$ the 
quantum field theoretical operator $F_{G/\hbar}$ (``interacting field'') 
which corresponds to $F$ under the interaction term\footnote{Mostly we will set
$\hbar =1$.} $G/\hbar$.   
$F_G$ should be a {\it formal power series} in $G$ 
where each term is an element of 
$\mathcal{F}(\mathcal{O})$ if $F,G\in \mathcal{F}(\mathcal{O})$. 
Here we deviate essentially from the usual formalism of perturbative
QFT: there the interacting fields are Fock space operators (which means
in our algebraic formulation that they are elements of 
$\mathcal{F}^{(m)}_{0}$). Motivated by the study of the Peierls
bracket \cite{Marolf, DF2}, we define them to be unrestricted
functionals ('off-shell fields'). This simplifies strongly the
proof of the 'Main Theorem of perturbative renormalization' (Sect.~4.2)
and e.g.~the formulation of the renormalization conditions
'Covariance' and 'Field Independence' given below.
At $\hbar=0$, the restriction $F_G\vert_{{\cal C}^{(m)}_{0}}$ is 
the (perturbative) classical retarded field as constructed in 
\cite{DF2} (see also \cite{de Witt, Marolf}).

We require the following properties, which may be motivated by their
validity in classical field theory (see \cite{DF2}):
\begin{description}
 
 \item[Initial condition:] For $G=0$ we obtain the original functional,
  \begin{displaymath}
    F_0=F\ .
  \end{displaymath}
  
 \item[Causality:] Fields are not influenced by interactions which 
  take place later:
  \begin{displaymath}
    F_{G+H}=F_{G}
  \end{displaymath}
  if there is a Cauchy surface such that $F$ is localized in its past and $H$ 
  in its future.
 
 \item[GLZ Relation:] The Poisson bracket 
  $\{F_{G/\hbar},H_{G/\hbar}\}\=d\frac{i}{\hbar}[F_{G/\hbar},H_{G/\hbar}]_{\star_m}$ 
  satisfies the GLZ relation \cite{GLZ,Ste,DF2}
  \begin{displaymath}
    \{F_{G/\hbar},H_{G/\hbar}\}=\frac{d}{d\lambda}
     (F_{(G+\lambda H)/\hbar}-H_{(G+\lambda F)/\hbar})|_{\lambda =0} \ .
  \end{displaymath}
\end{description}
Due to the GLZ relation and (\ref{*-product}) the interacting fields 
depend on the $*$-product (\ref{*-product}) and with that they depend on
the mass $m$ of the free field. 

Steinmann \cite{Ste} discovered that
by these conditions an inductive construction of the perturbative 
expansion of $F_G$ (i.e.~of the retarded products (\ref{intf:R}))
can be done up to local functionals which could be 
added in every order. But these undetermined terms correspond to the 
renormalization ambiguities which are 
there anyhow in perturbative quantum field theory. One may reduce these 
ambiguities by prescribing normalization conditions which are 
satisfied in classical field theory \cite{DF2}.
We impose the following conditions:
\begin{description}

 \item[Unitarity:] Complex conjugation induces an involution $F\mapsto F^*$ 
 of the algebra
 which after restriction to ${\cal C}_0^{(m)}$ 
 becomes the formal adjoint operation on Fock space. We require
 \begin{displaymath}
  (F_G)^*=F^*_{G^*} \ .
 \end{displaymath}
 This condition implies that a real interaction $G$ leads formally to a 
 unitary S-matrix and hermitian interacting fields (if $F^*=F$).
 
 \item[Covariance:] The Poincar\'{e} group ${\cal P}_+^\uparrow$
  has a natural automorphic action $\beta$ on $\mathcal{F}$. We require 
  \begin{displaymath}
    \beta_L(F_G)=\beta_L(F)_{\beta_L(G)}\quad
    \forall L\in {\cal P}_+^\uparrow\ .
  \end{displaymath}
  (See \cite{BFV} and \cite{HW1-2} for the formulation of covariance 
  on curved spacetime.) In addition, {\bf global inner symmetries}
  (in our case the field parity $\alpha:\varphi\mapsto -\varphi$)
  should be preserved,
  \begin{equation}
    \alpha (F_G)=\alpha(F)_{\alpha (G)}\ .\label{parity}
  \end{equation}

  \item[Field Independence:] A coherent prescription for the 
  renormalization of polynomials 
  in the basic fields and all sub-polynomials can be obtained by the 
  following condition
 \begin{equation}
    \langle h,\frac{\delta}{\delta\varphi}F_G \rangle 
    = \langle h,\frac{\delta F}{\delta\varphi}\rangle_G+
    \frac{d}{d\lambda}\vert_{\lambda=0}
    F_{G+\lambda \langle h, \frac{\delta 
    G}{\delta\varphi}\rangle}\ \ , \  h\in\mathcal{D}(\MM) \label{dF_G/dphi}
 \end{equation}
  (where $\langle h, \frac{\delta H}{\delta\varphi}\rangle\equiv
  \int dx\, h(x)\frac{\delta H(x)}{\delta\varphi (x)}$ for $H\in {\cal
  F}$).
  Below it will turn out that this condition is equivalent to the natural
  generalization to our 'off-shell formalism' of the causal Wick expansion
  given in Sect.~4 of \cite{EG}. (The latter is equivalent to 
  the condition {\bf N3} in \cite{DF}, which is also called 'relation
  to time-ordered products of sub-polynomials' in \cite {BD}.)
  
  \item[Field equation:] The renormalization ambiguities can be used to 
  fulfill the Yang-Feldman equation 'off-shell': 
  \begin{equation}
    \varphi_G(x)=\varphi(x)-\int dy\,\Delta_m^{\rm ret}(x-y)
    \Bigl(\frac{\delta G}{\delta\varphi (y)}\Bigr)_G\ .\label{yangfeldm}
  \end{equation}
  The latter condition may be enforced by requiring the validity of 
  all local identities which hold classically as a consequence of the 
  field equation. This was termed Master Ward Identity in \cite{DF2}. 
  Since the Master Ward Identity cannot always be fulfilled we do not 
  impose it  here.

\item[Smoothness in the mass $m\geq 0$:]  
the classical interacting 
fields depend smoothly on the mass of the free fields in 
the range $m\geq 0$.  
In the quantum case, in even dimensional spacetime, this is no longer 
true even for the free fields because of logarithmic singularities 
of the two-point function $\Delta^+_m$ (see Appendix A).\footnote{We are indebted 
to Stefan Hollands for pointing out to us that this fact invalidates 
our treatment of the scaling behavior in an older version of this 
manuscript.} One can remedy 
this defect by passing to an equivalent star product $\star_{m,\mu}$
which is defined by (\ref{*-product}) with $\Delta^{+(d)}_m$ 
replaced by
  \begin{equation}
H^{\mu\, (d)}_m(x)\equiv \Delta^+_m(x)-\mathrm{log}(m^2/\mu^2)\>
\gamma(x)\ ,\quad\gamma(x)\equiv m^{d-2}\>h^{(d)}(m^2x^2)\ .\label{H^mu_m}
  \end{equation}
(This procedure follows essentially \cite{Holl}.) 
$\mu >0$ is an additional mass parameter. $h^{(d)}$ is analytic and it 
is chosen such that $H^{\mu\, (d)}_m$ is smooth in $m\geq 0$.
Explicitly we choose $h^{(2l+1)}\equiv 0$
and $h^{(4+2k)}(y)\equiv \pi^{-k}\, f^{(k)}(y)$, where
$f$ is given in Appendix A by (\ref{f}).  
With this particular choice $H^{\mu\, (d)}_m$ solves the Klein-Gordon equation, 
but we point out that this is not necessary for the purposes of this 
paper,\footnote{In order that  $\star_{m,\mu}$ induces a well defined 
$\star$-product on $\mathcal{F}^{(m)}_{0}\equiv {\cal F}/{\cal J}^{(m)}$
it is needed that $H^{\mu\, (d)}_m$ solves the Klein-Gordon equation.
However, we always work with off-shell fields (i.e. with $\mathcal{F}$).}
cf.~\cite{Holl}.
To show that the product $\star_{m,\mu}$
is equivalent to $\star_{m}$, we introduce the transformation
  \begin{equation}
\Bigl(\frac{m}{\mu}\Bigr)^\Gamma =1+\sum_{k=1}^\infty \frac{1}{k!}
\Bigl(\mathrm{log}(m/\mu)\cdot\Gamma\Bigr)^k
  \end{equation}
where $\Gamma$ is the operator
  \begin{equation}
\Gamma\equiv \Gamma^{(m)}\equiv\int dx\>dy\>\gamma (x-y)\>\frac{\delta^2}
{\delta\varphi(x)\delta\varphi(y)}\ .
  \end{equation}
By using
  \begin{equation}
e^{\lambda\varphi(f)}\star_m e^{\lambda\varphi(g)}=e^{\lambda\varphi(f+g)}\>
e^{\lambda^2\, (f,\Delta^+_m g)}
  \end{equation}
(understood as formal power series in $\lambda$),
the same formula for $(\star_{m,\mu},H^{\mu}_m)$ and
  \begin{equation}
\Bigl(\frac{m}{\mu}\Bigr)^\Gamma e^{\lambda\varphi(f)}=e^{\lambda\varphi(f)}
\Bigl(\frac{m}{\mu}\Bigr)^{\lambda^2\,\int dx\>dy\>\>\gamma (x-y)\>f(x)\, f(y)}\ ,
\label{Gamma(phi)}  
\end{equation}
we find that $(\frac{m}{\mu})^\Gamma$ intertwines between $\star_{m,\mu}$
and $\star_{m}$: 
  \begin{equation}
\Bigl(\frac{m}{\mu}\Bigr)^\Gamma\Bigl(F\star_{m,\mu} G\Bigr)=
\Bigl(\frac{m}{\mu}\Bigr)^\Gamma(F)\star_{m}
\Bigl(\frac{m}{\mu}\Bigr)^\Gamma(G)\ ,
\quad F,G\in \mathcal{F}\ .\label{intertwine*}
  \end{equation}
As mentioned after (\ref{J}), the normally ordered product $:\varphi(f)^n:_m$
(where normal ordering is done with $\Delta^+_m$) agrees 
with $\pi((\varphi(f))^n)$.  
The modified normally ordered product $:\varphi(f)^n:_{m,\mu}$ (i.e. normal 
ordering is done with $H^\mu_m$) agrees with
$\pi ((\frac{m}{\mu})^{-\Gamma}\>(\varphi(f))^n)$. With that 
(\ref{Gamma(phi)}) yields that the two different kinds of 
Wick powers are related by
\begin{equation}
:e^{\lambda\,\varphi(x)}:_{m,\mu}=:e^{\lambda\,\varphi(x)}:_m
\Bigl(\frac{m}{\mu}\Bigr)^{-\lambda^2\,\gamma(0)}\ .
\end{equation}

We may now introduce interacting fields with respect to the 
modified $*$-product,
  \begin{equation}
(F_G)^{(m,\mu)}:=\Bigl(\frac{m}{\mu}\Bigr)^{-\Gamma}\Bigl(
\bigl((\frac{m}{\mu})^{\Gamma}(F)\bigr)^{(m)}_{
(\frac{m}{\mu})^{\Gamma}(G)}\Bigr)\ ,\label{intfield:modified}
\end{equation}
where $F_G^{(m)}$ denotes the interacting field with respect to 
the usual $*$-product (\ref{*-product}). Note that at $m=0$
the two kinds of interacting fields agree: $F_G^{(0,\mu)}=F_G^{(0)}$.
Since $\Gamma$ is local (i.e. $\supp\frac{\delta 
        (\Gamma F)}{\delta \varphi}\subset\supp\frac{\delta 
        F}{\delta \varphi}$), field independent (in the sense that 
$\gamma$ is field independent),  Poincar\'{e} invariant and commutes 
with the $*$-operation (since $\gamma$ is Poincar\'{e} invariant
and real), the modified interacting fields satisfy the same conditions 
as the original interacting fields (where, of course, in the GLZ 
relation the commutator with respect to the modified $*$-product has 
to be used).

Our smoothness condition now takes the following form: we require 
that the maps
  \begin{equation}
m\mapsto (F_G)^{(m,\mu)}\ ,\quad\quad F,G\in {\cal F}_{\rm loc}\ ,
\quad\mu>0\ ,\label{smoothness:intfield}
\end{equation}
are smooth (in the sense of one sided derivatives at 
$m=0$).\footnote{Due to (\ref{H^mu_m}) the interacting fields 
depend only on $m^2$, and setting $(F_G)^{(-m,\mu)}:=(F_G)^{(m,\mu)}$
the map (\ref{smoothness:intfield}) can be extended to $m<0$. It is even possible
to require that $(F_G)^{(m,\mu)}$ is smooth in $m^2$ (which is a stronger
condition than smoothness in $m$). However, note that this footnote is not valid 
for spinor fields.}
\medskip
 
\noindent {\it Remarks:} (1) The fact that the $\star$-products 
$\star_{m,\mu}$ are equivalent for different values of $\mu$ 
(which follows immediately from (\ref{intertwine*}))
may also be formulated in the following way. Introduce new fields by
$$\varphi^{\otimes n}(x_1,\ldots,x_n)_{m,\mu}=
\left(\frac{m}{\mu}\right)^{-\Gamma}\varphi(x_1)\ldots\varphi(x_n)\ .$$
Every functional $F\in{\cal F}$ may be expanded in any of these fields
$$F=\sum_n\int dx_1...dx_n\, f_n^{m,\mu}(x_1,\ldots,x_n)\, \varphi^{\otimes
n}(x_1,\ldots,x_n)_{m,\mu}$$
with suitable coefficients $f_n^{m,\mu}$. The different $\star$-products
then arise when the $\star$-product $\star_m$ is expressed in terms
of the coefficients:
\begin{gather}
F\star_mG=\sum_{n,k}\int dx_1dy_1...\, f_n^{m,\mu}(x_1,...)\,
g_k^{m,\mu}(y_1,...)\, \left(\frac{m}{\mu}\right)^{-\Gamma}\Bigl(\varphi(x_1)...\star_{m,\mu}
\varphi(y_1)...\Bigr)\notag\\
=\sum\int f^{m,\mu}\cdot (\prod H_m^\mu)\cdot g^{m,\mu}\cdot (\varphi^{\otimes})_{m,\mu}\ .\notag
\end{gather}
Hence, the choice of $\mu$ can be understood as the choice of a basis for
${\cal A}^{(m)}\equiv ({\cal F},\star_m)$. \\
\noindent (2) The introduction of the modified $*$-product and 
modified interacting fields (\ref{intfield:modified}) can be avoided by requiring 
that the function $m\mapsto F_G^{(m)}$ is {\it almost smooth for} $m\downarrow 0$
for all $F;G\in {\cal F}_{\rm loc}$. In doing so a function $\RR_+\ni m\mapsto f(m)$
is called almost smooth for $m\downarrow 0$, if for any fixed mass parameter
$\mu >0$ there exist polynomials $p_{k,\mu},\> k\in \NN_0$, such that for 
each $n\in\NN_0$ it holds
  \begin{equation}
m^{-n}\Bigl( f(m)-\sum_{k\leq n}m^k\, p_{k,\mu}({\rm log}\>\frac{m}{\mu})
\Bigr)\longrightarrow 0
\end{equation}
for $m\downarrow 0$. Note that the polynomials $p_{k,\mu}$ are
{\it uniquely} determined by this condition. Then the scaling expansion
(\ref{scal-exp}) has to be generalized correspondingly. We do not go 
this way, because the treatment of the renormalization ambiguities is much 
simpler for the modified interacting fields: the map $D^{(m)}_H$ of the Main 
Theorem (i.e.~Theorem \ref{main}), which gives a finite renormalization of the 
{\it modified} interacting fields, is free of $({\rm log}\> m)$-terms (\ref{D:scaling});
but the corresponding $D^{(m)}$ of the original interacting fields
contains such terms (\ref{D:scaling1}).
\medskip
  \item[Scaling:] Under a simultaneous scaling of the coordinates and
    the mass, $(x,m)\mapsto (\rho x,\rho^{-1}m)$, the
    interacting classical fields transform homogeneously. This
    can no longer be maintained for the quantized theory,
together with the requirement of smoothness at $m=0$, even for free
fields (in even dimensions) since $H^\mu_m$ does not scale 
homogeneously (\ref{H:scaling}).
However, we will show that the retarded products can be normalized such
that they scale almost homogeneously (i.e.~up to logarithmic terms).
 \end{description}
The last two normalization conditions were first imposed in
\cite{HW1-2} in the more general context of renormalization 
on curved spacetime. In the traditional literature
(e.g. \cite{EG,Ste,BF}) instead the weaker requirement was used that
'renormalization may not make the interacting fields more singular' 
(in the UV-region), see footnote \ref{fn:N0}. 

The listed conditions can be 
translated in a straightforward way into conditions 
on the {\bf retarded products} $R_{n,1}$  
which are by definition (\ref{intf:R}) the Taylor coefficients of the interacting 
field with respect to the interaction,
\begin{equation}
A(f)_{{\cal L}(g)}=
\sum_{n=0}^\infty\frac{1}{n!}R_{n,1}
\bigl({\cal L}(g)^{\otimes n},A(f) \bigr)\equiv  
R(e_\otimes^{{\cal L}(g)},A(f))
\label{ansatz:intfield}
\end{equation}
with $g,f\in {\cal D}(\MM)$ and 
$A, {\cal L}\in{\cal P}$, where $\mathcal{P}$ 
is the algebra of polynomials in the classical field $\varphi$ and its 
partial derivatives (with respect to pointwise multiplication).
We use (\ref{ansatz:intfield}) for both kinds of interacting fields,
$F_G^{(m)}$ and $F_G^{(m,\mu)}$, and
by writing $R$ (or $R_{n,1}$) we mean both kinds of 
retarded products, $R^{(m)}$ and $R^{(m,\mu)}$. 
The retarded product $R_{n-1,1}$ is a {\bf linear} map, 
from ${\cal F}_{\rm loc}^{\otimes n}$ into ${\cal F}$ which is 
symmetric in the first $n-1$ variables.
In the last expression of (\ref{ansatz:intfield}) the 
sequence $R\equiv (R_{n-1,1})_{n\in\NN}$ is viewed as a map
\begin{equation}
  R:T{\cal F}_{\rm loc}\longrightarrow {\cal F}\ ,\quad \mathrm{where}
\quad T{\cal F}_{\rm loc}\=d\oplus_{n=0}^\infty 
{\cal F}_{\rm loc}^{\otimes n}\label{R:sequence}
\end{equation}
and $R_{-1,1}\=d 0$ which is extended by linearity to formal power 
series. It is sometimes advantageous to interprete $R_{n-1,1}$ as 
$\mathcal{F}$-valued distribution in $n$ variables on the test 
function space $\mathcal{D}(\MM,\mathcal P)$ (cf. \cite{DF2}). 
In particular our retarded products are multi-linear in the fields
$A\in {\cal P}$.

We also use the 
symbolic notation $R_{n-1,1}(x_{1},\ldots,x_{n})$ for a 
distribution which takes values in $\mathcal{F}\otimes 
\mathcal{P}_{n}'$, where $\mathcal{P}_{n}'$ is the dual 
vector space of $\mathcal{P}^{\otimes n}$. After insertion 
of fields $A_{1},\ldots,A_{n}\in\mathcal{P}$ we obtain 
$\mathcal{F}$-valued distributions on 
$\mathcal{D}(\MM^{n})$ which are symbolically written as 
$R_{n-1,1}(A_{1}(x_{1}),\ldots,A_{n}(x_{n}))$.
 
\begin{description}
 \item[Action Ward Identity (AWI):] Since the retarded 
 products depend only on the functionals, derivatives may be 
 shifted from the test functions to the fields and vice versa. Hence,
 the associated distributions must satisfy the Action Ward Identity
\begin{equation}
  \d^x_{\mu}R_{n-1,1}(\ldots A_k(x)\ldots)= 
   R_{n-1,1}(\ldots,\d_{\mu}A_k(x),\ldots)\ .
\end{equation}

\item[Symmetry:] Motivated by (\ref{ansatz:intfield}) we require that
$R_{n-1,1}$ is symmetric in the first $(n-1)$ factors. This property
is reflected in the convention for the lower indices of $R_{n-1,1}$.

 \item[Initial condition:] $\quad R_{0,1}(F)=F$\ .

\item[Causality:]
\begin{equation}
\supp R_{n-1,1}\subset
\{(x_1,\ldots,x_{n})\in\MM^{n}\> |\> x_i\in x_{n}+\overline{V}_-,\>
\forall i=1,\ldots n-1\}\ .\label{R:supp}
\end{equation}

\item[GLZ relation:]
\begin{equation}
  R_{n-1,1}(\ldots,y,z)-R_{n-1,1}(\ldots,z,y) 
  =\hbar\> J_{n-2,2}(\ldots,y,z)\label{glz:R}
\end{equation}
where $J_{n-2,2}$ is an algebra valued distribution in $n$ 
variables, $n\ge 2$, which is defined by
\begin{equation}
\begin{split}
&J_{n-2,2}(A_1(x_1),\ldots,A_{n-2}(x_{n-2}),B(y),C(z))\=d\\
&\sum_{I\subset\{1,\ldots,n-2\}}\Bigl\{ R_{|I|,1}(A_i(x_i),i\in I,B(y)),
              R_{|I^c|,1}(A_i(x_i),i\in I^c,C(z))\Bigr\}\ .\label{J:def}
\end{split}  
\end{equation}
Here, $I^c$ is the complement of $I$ in $\{1,\ldots,n-2\}$.
Obviously, $J_{n-2,2}$ is symmetric in the first $n-2$ variables and 
antisymmetric in the last 2 variables. The properties of $J_{n-2,2}$
stated in the following Lemma are necessary conditions for the
GLZ-relation and the causality of $R_{n-1,1}$. But actually, they
are already fulfilled as a consequence 
of the definition of $J$ (\ref{J:def}), and the GLZ relation 
and Causality of the retarded products to lower orders \cite{Ste}.
\begin{lemma} (a) $J$ satisfies the Jacobi identity 
\begin{equation}
  J_{n-2,2}(\ldots,x,y,z)+\mathrm{cycl}(x,y,z)=0 \ .\label{Jacobi}
\end{equation}
(b) The support of $J_{n-2,2}$ is contained in the set 
\begin{equation}
  \{x_{i}\in x_{n}+\bar V_-,i=1,\ldots,n-1\} \cup \{x_{i}\in 
  x_{n-1}+\bar V_-,i=1,\ldots n-2,n\} \ . \label{J:supp}
\end{equation}
\end{lemma}
\begin{proof} (cf. \cite{Ste}) We start with
the Jacobi identity of the Poisson bracket and use the notation
$x_M\equiv (x_m|m\in M)$, where $M\subset \{1,...,n-1\}$. So we know
\begin{equation}
  \sum_{I\sqcup H\sqcup L=\{1,...,n-1\}}\{\{R(x_I,x),
R(x_H,y)\},R(x_L,z)\}+\mathrm{cycl}(x,y,z)=0,\label{Jac}
\end{equation}
where $\sqcup$ means the disjoint union. The sum over all 
decompositions of $K\equiv I\sqcup H
(=\mathrm{fixed})$ of the inner Poisson bracket is equal to
$J_{|K|,2}(x_K,x,y)$, $|K|\leq n-1$, 
which splits into $R(x_K,x,y)-R(x_K,y,x)$
due to the validity of the GLZ relation to lower orders. With that we
obtain
\begin{gather}
  0=\sum_{K\sqcup L=\{1,...,n-1\}}\{(R(x_K,x,y)-
R(x_K,y,x)),R(x_L,z)\}+\mathrm{cycl}(x,y,z)\notag\\
=J_{n,2}(x_1,\ldots ,x_{n-1},x,y,z)+\mathrm{cycl}(x,y,z).
\end{gather}
(b) By definition of $J$ and the support properties of 
$R$ it follows that $J_{n-2,2}(x_{1},\ldots,x_{n-2},y,z)$ vanishes if 
one of the first 
$n-2$ arguments is not in the past of $\{y,z\}$. 

It remains 
to show that it vanishes also for $(y-z)^2<0$. If 
one of the first $n-2$ arguments is different from $y$ and 
$z$, and is in the past of, say $y$, then by the Jacobi 
identity $J$ has to vanish. 

If, on the other hand, all 
arguments $x_{i}$ are sufficiently near to either $y$ or 
$z$, then they are space-like to the other point, hence all retarded 
products in the definition of $J$ 
vanish up to those where all arguments in the first factor are 
near to $y$ and all arguments in the second factor are near 
to $z$. But then the Poisson bracket of these retarded products 
vanishes, since  by assumption the retarded products are localized at 
their arguments.\footnote{The proof of 
the last fact given by Steinmann \cite{Ste} is much 
more involved since he does not assume the localization 
property of the retarded products.}
\end{proof}  
An immediate consequence of the Initial condition and the GLZ relation is
\beq
R_{n-1,1}(F_1,...,F_n)={\cal O}(\hbar^{(n-1)})\quad\mathrm{if}\quad
 F_1,...,F_n\sim\hbar^0\ .\quad
\eeq

\item[Field Independence:] The condition (\ref{dF_G/dphi}) translates 
into
\begin{equation}
    \frac{\delta}{\delta\varphi (x)}R_{n-1,1}(F_1,\dots,F_n)=
    \sum_{l=1}^nR_{n-1,1}(F_1,\ldots ,\frac{\delta F_l}{\delta\varphi (x)},
    \ldots ,F_n)\ .\label{dR/dphi}
  \end{equation}
 This condition determines the retarded product on the left side in terms
 of the retarded products on the right side, up to its value at 
 $\varphi=0$, i.e.~its vacuum expectation value. 
 Since by definition, the functionals $F_{i}$ are polynomials in 
 $\varphi$, one obtains the finite Taylor 
expansion\footnote{In (\ref{causWick}) and (\ref{causWick1}) 
  the product of fields is the classical product, 
  $\bigl(\varphi(x)\varphi(y)\bigr)(h)=h(x)h(y)$, and
  not the $\star$-product (\ref{*-product}).}
   \begin{gather}
     R_{n-1,1}(F_1,\dots,F_n)=\sum_{l_1\ldots l_n}\frac{1}{l_1!\ldots l_n!}
 \int dx_{11}\ldots dx_{1l_1}\ldots dx_{n1}\ldots dx_{nl_n}\notag\\
 \omega_0\Bigl(R_{n-1,1}\bigl(\frac{\delta^{l_1}F_1}{\delta\varphi (x_{11})
 \ldots\delta\varphi (x_{1l_1})},\ldots ,\frac{\delta^{l_n}F_n}
 {\delta\varphi (x_{n1})\ldots\delta\varphi (x_{nl_n})}\bigr)\Bigr)\notag\\
 \varphi (x_{11})\ldots\varphi (x_{1l_1})\ldots
 \varphi (x_{n1})\ldots\varphi (x_{nl_n})\ ,\label{causWick}
   \end{gather}
where the coefficients $\omega_0\bigl(R_{n-1,1}(\ldots
)\bigr)$ are restricted by the other axioms.
The retarded product on the right side is well-defined, because, due to
$F_k\in {\cal F}_\mathrm{loc}$, the support of 
$\delta^{l}F_k/\delta\varphi (x_1)\ldots\delta\varphi (x_l)$ 
is contained in the
total diagonal $x_1=x_2=\ldots =x_l$. After integrating out the
corresponding $\delta$-distributions the right side of
(\ref{causWick}) is a sum of terms of the form
\begin{equation}
  \int d^{nd} x\, \omega_0\Bigl(
R_{n-1,1}\bigl(A_1(x_1),\ldots ,A_n(x_n)\bigr)\Bigr)
\prod_{i=1}^{n}h_i(x_{i})\prod_{j_i=1}^{l_i}\d^{a_{ij_i}}
\varphi (x_i) \label{causWick1}
\end{equation}
with $h_i\in{\cal D}(\MM)$, $A_{i}\in\mathcal{P}$  and multi-indices 
$a_{ij_i}\in \NN_{0}^{d}$.
Due to translation invariance of the vacuum, the coefficients in this 
expansion depend on the relative coordinates only.
In particular, their wave front set satisfies the condition 
(\ref{wave front}) on 
the admissible coefficients in $\mathcal{F}$. 

The condition (\ref{dR/dphi}) that the retarded product 
is independent of $\varphi$, is the only axiom which relies 
on the fact that one perturbs around a theory with an 
action which is of second order in the field. Indeed, in 
the general case, the retarded products of classical field theory 
depend on the action only via its second functional 
derivative, see Proposition 1 of \cite{DF2}.

The restriction of (\ref{causWick}) to ${\cal C}^{(m)}_{0}$ is the
causal Wick expansion of Epstein and Glaser \cite{EG}. In particular
the 'coefficients' $ \omega_0(R_{n-1,1}(...))$ are exactly the same as in 
the on-shell formalism of \cite{EG}.  
\item[Smoothness in the mass $m\geq 0$:] In terms of the retarded products
  our requirement reads that 
\begin{equation}
R^{(m,\mu)}_{n-1,1}(F_1,\ldots,F_n)=
(\frac{m}{\mu})^{-\Gamma}\>\Bigl(R^{(m)}_{n-1,1}\Bigl(
(\frac{m}{\mu})^{\Gamma}(F_1),\ldots,
(\frac{m}{\mu})^{\Gamma}(F_n)\Bigr)\Bigr)\label{R^m,mu}
\end{equation}
depends smoothly on $m$ $\forall
m\geq 0$. By this we mean that $R^{(m,\mu)}_{n-1,1}(F_1,...,F_n)(h)$
is smooth as a function of $m$ for all 
$F_1,...,F_n\in {\cal F}_\mathrm{loc}$ and
all field configurations $h\in {\cal C}$, and
that the derivatives are of the form
$G(h)$ with $G\in {\cal F}$.\footnote{The latter condition 
is necessary since the distributions occurring in the 
representation (\ref{F(phi)}) of $G$ have to satisfy the wave 
front set condition (\ref{wave front}).}
In particular this smoothness implies that $R^{(0,\mu)}_{n-1,1}$
can be obtained by the limit $\lim_{m\downarrow 0}
R^{(m,\mu)}_{n-1,1}$ in the sense of distributions on ${\cal D}(\MM^n)$.
\item[Scaling:] As an introduction and for later purpose 
we first define 'almost homogeneous 
scaling' for a distribution under rescaling of the
coordinates (cf. \cite{HW1-2}):
\begin{definition} A distribution $t\in {\cal D}^{\prime} (\RR^k)$
(or ${\cal D}^{\prime} (\RR^k\setminus\{0\})$) scales almost
homogeneously with degree $D\in\RR$ and power $N\in\NN_0$ if
\begin{equation}
  (\sum_{r=1}^k z_r\d_{z_r}+D)^{N+1}t(z_1,...,z_k)=0
\label{alm-hom-scal-1}
\end{equation}
and $N$ is the minimal natural number with this property. 
For $N=0$
the scaling is homogeneous with degree $D$.
\end{definition}
The condition (\ref{alm-hom-scal-1}) is equivalent to
\begin{equation}
  0=(\rho\d_\rho)^{N+1}\Bigl(\rho^D t(\rho z_1,\ldots,
  \rho z_k)\Bigr)=\frac{\d^{N+1}}{\d(\mathrm{log}\>\rho)^{N+1}}
  \Bigl(\rho^D t(\rho z_1,\ldots,\rho z_k)\Bigr)\ .
\label{alm-hom-scal-2}
\end{equation}
Hence, $t$ scales almost homogeneously with degree $D$ and power 
$N$ if $\rho^D t(\rho z_1,\ldots,\rho z_k)$ is a 
polynomial of $\log\rho$ with degree $N$.

To formulate almost homogeneous scaling for the retarded products
under simultaneous rescalings of the coordinates and the mass $m$
we introduce some tools. The 
\textit{mass dimension of a monomial} in $\mathcal{P}$ is 
fixed by the conditions
\begin{equation}
  \mathrm{dim}(\d^a\varphi)=\frac{d-2}{2}+|a|\quad
\mathrm{and}\quad \mathrm{dim}(A_1A_2)=\mathrm{dim}(A_1)+
\mathrm{dim}(A_2)\label{UV-dim}
\end{equation}
for all monomials $A_1,A_2\in {\cal P}$. This introduces a grading for 
$\mathcal{P}$:
\begin{equation}
  \mathcal{P}=\oplus_{j}\mathcal{P}_j\ ,\label{grad(P)}
\end{equation}
where $\mathcal{P}_j$ is the linear span of all monomials with
mass dimension $j$. The mass dimension of $A=
\sum_j A_j$, with $A_j\in\mathcal{P}_j$, is the maximum
of the contributing $j$'s. We also introduce the set
of all field polynomials which are homogeneous in the mass dimension
\begin{equation}
 {\cal P}_{\text{hom}}\=d \bigcup_{j}{\cal P}_j \ .\label{P_hom}
\end{equation}
A scaling transformation $\sigma_{\rho}$ is 
introduced (in analogy to \cite{HW}) as an 
automorphism of $\mathcal{F}$ (considered as an algebra with 
the classical product) by 
\begin{equation}
  \sigma_{\rho}(\varphi(x))= 
  \rho^{\frac{2-d}{2}}\varphi(\rho^{-1}x) \ . \label{sigma}
\end{equation}
Note $\omega_0\circ\sigma_\rho=\omega_0$.
Due to $\rho^{d-2}\Delta_+^{(\rho^{-1}m)}(\rho x)=
\Delta_+^{(m)}(x)$, $\sigma_{\rho}$ is also an algebra isomorphism 
from ${\cal A}^{(\rho^{-1}m)}$ to ${\cal A}^{(m)}$.
However, denoting by  ${\cal A}^{(m,\mu)}$ the algebra $({\cal F},\star_{m,\mu})$,
$\sigma_\rho$ is an isomorphism from ${\cal A}^{(\rho^{-1}m,\rho^{-1}\mu)}$
to ${\cal A}^{(m,\mu)}$, but {\it not} from ${\cal A}^{(\rho^{-1}m,\mu)}$
to ${\cal A}^{(m,\mu)}$, since $\rho^{d-2}\>H^{\rho^{-1}\mu\, (d)}_{\rho^{-1}m}(\rho x)
=H^{\mu\, (d)}_m(x)$. For $m=0$ the coordinates are scaled only  
and $\sigma_\rho$ is an automorphism of 
${\cal A}^{(m=0)}={\cal A}^{(m=0,\mu)}$. For $A\in{\cal P}_\mathrm{hom}$, we obtain
\begin{equation}
  \rho^{\mathrm{dim}(A)}\sigma_\rho (A(\rho x))=A(x).
\label{scal:hom}
\end{equation}
So, with the identification given by $\sigma_\rho$, they 
scale homogeneously with degree given by their mass dimension. 
By inserting the definitions one finds 
\begin{equation}
\sigma_\rho\Gamma^{(\rho^{-1}m)}\sigma_\rho^{-1}=\Gamma^{(m)}\quad
\mathrm{and\>\> hence}\quad
\sigma_\rho\Bigl(\frac{m}{\mu}\Bigr)^{\Gamma^{(\rho^{-1}m)}} \sigma_\rho^{-1}=
\Bigl(\frac{m}{\mu}\Bigr)^{\Gamma^{(m)}}\ .\label{Gamma:scaling} 
\end{equation}

The property 
\begin{equation}
\sigma_\rho 
(\sigma_\rho^{-1}F\star_0\sigma_\rho^{-1}G)=F\star_0G
\end{equation}
of the $\star$-product at $m=0$ 
cannot be maintained for the retarded products, i.e.~
\begin{equation}
  R^{(0)}_{n-1,1\>\rho}:=\sigma_\rho \circ 
R^{(0)}_{n-1,1}\circ (\sigma_\rho^{-1})^{\otimes n}
\label{R^0:scaled}
\end{equation}
will differ from $R^{(0)}_{n-1,1}$, in general. 
But one can reach that $R^{(0)}_{n-1,1}$ 
{\it scales almost homogeneously
with degree zero}, 
i.e. $R^{(0)}_{n-1,1\>\rho}$ 
has polynomial behavior in $\log\rho$,
in the sense that for all $F_1,...,F_n\in {\cal F}_{\rm loc}$,
$R^{(0)}_{n-1,1\>\rho}(F_1,...,F_n)$ is
a polynomial in $\log\rho$. 
In the {\it massive case}, scaling relates retarded products for different 
masses; therefore, the condition of homogeneity gives no restriction
at a fixed mass. Our condition of almost homogeneous scaling states
that
\begin{equation}
  R^{(m)}_{n-1,1\>\rho}:=\sigma_\rho \circ 
R^{(\rho^{-1}m)}_{n-1,1}\circ (\sigma_\rho^{-1})^{\otimes n}\ ,
\label{R^m:scaled}
\end{equation}
or equivalently
\begin{equation}
  R^{(m,\mu)}_{n-1,1\>\rho}:=\sigma_\rho \circ 
R^{(\rho^{-1}m,\mu)}_{n-1,1}\circ (\sigma_\rho^{-1})^{\otimes n}=
\Bigl(\frac{m}{\rho\mu}\Bigr)^{-\Gamma^{(m)}}\circ R^{(m)}_{n-1,1\>\rho}
\circ \Bigl(\Bigl(\frac{m}{\rho\mu}\Bigr)^{\Gamma^{(m)}}\Bigr)^{\otimes n}
\label{R^m,mu:scaled}
\end{equation}
(where (\ref{Gamma:scaling}) is used),
has polynomial behavior in $\log\rho$.
We will see, that this condition together with the 
smoothness requirement, imposes
non-trivial restrictions also in the massive case.
As mentioned above, in even dimensions, the smoothness condition
requires the transition to an equivalent $*$-product which depends
on an additional mass parameter $\mu$.
\end{description}
\medskip

{\noindent} It is obvious how 
the remaining conditions on the interacting fields ({\bf Unitarity},
{\bf Covariance} and {\bf Field equation}) read in terms of the 
retarded products. With regard to the Field equation note that the retarded 
propagator is the same for the $R^{(m)}$-products and the $R^{(m,\mu)}$-products:
\begin{equation}
R^{(m,\mu)}_{1,1}(\varphi(y),\varphi(x))=\Delta_m(x-y)\, \Theta(x^0-y^0)=
\Delta^\mathrm{ret}_m(x-y)=R^{(m)}_{1,1}(\varphi(y),\varphi(x))\ .
\end{equation}
Hence, the Yang-Feldman equation (\ref{yangfeldm}) has precisely the 
same form for both kinds of interacting fields.

From Bogoliubov's definition of interacting fields (\ref{intf}) it
follows that there is a unique correspondence between retarded products
and {\it time ordered products}, see e.g. \cite{EG}. 
So the axioms given in this section can equivalently be 
formulated in terms of time ordered products. This is done in 
Appendix E. In addition we give there an explicit formula
which describes the time ordered products in terms of 
retarded products.
\section{Construction of the retarded products}\setcounter{equation}{0}
Our procedure is based on the strategies developed in 
\cite{Ste,Sto} and \cite{BF}.

\subsection{Inductive step outside of the total diagonal}
If the retarded products with less than $n$ factors are 
given, we can define the distribution $J_{n-2,2}$. But, by 
the GLZ relation, Symmetry and Causality, 
$R_{n-1,1}$ {\it is already fixed by $J_{n-2,2}$ outside of the 
total diagonal}
$\Delta_{n}=\{(x_{1},\ldots,x_{n})\in\MM^n|x_{1}=\cdots=x_{n}\}$. 
Namely, if not all points coincide, we may separate them 
into two nonempty sets which are in the past and the 
future, respectively, of a 
suitable Cauchy surface. If the last argument $x_{n}$ is in the past, then 
$R_{n-1,1}(x_{1},\ldots,x_{n})$ vanishes due to the support 
properties of retarded products. If, on the other hand, 
$x_{n}$ is in the future and $x_{k}$ for some $k\ne n$ is in 
the past, then the retarded product vanishes if the arguments $x_{k}$ and 
$x_{n}$ are permuted, hence in this case we find
\begin{equation}
  R_{n-1,1}(x_1,...,x_{n-1},x_{n})=J_{n-2,2}(x_1,...\hat
  k...,x_{n-1},x_k,x_{n})\quad\mathrm{if}\quad x_k\not\in
  x_{n}+\bar V_+\ .\label{R=J}
\end{equation}

Moreover, it is only the totally 
symmetric part $S_{n}$ of $R_{n-1,1}$ which is not completely fixed 
by lower orders,
\begin{equation}
   S_{n}(x_{1},\ldots,x_{n})\=d \frac{1}{n}\sum_{k=1}^n 
   R_{n-1,1}(x_{k+1},\ldots,x_{n},x_{1},\ldots,x_{k}) \ .\label{S_n}
\end{equation}
This follows again from the GLZ relation which yields
\begin{displaymath}
        R_{n-1,1}(x_{1},\ldots,x_{n}) = 
        S_{n}(x_{1},\ldots,x_{n}) + \frac{1}{n}\sum_{k=1}^{n-1} 
        J_{n-2,2}(x_{k+1},\ldots,x_{n-1},x_{1},\ldots,x_{k},x_{n}) \ .
\end{displaymath}

We now want to prove the {\bf existence} of $R_{n-1,1}$. In 
this Subsect.~we give the 
first step: we check that the above findings define a 
distribution on the complement of the total diagonal 
which we denote by\footnote{The construction of  $R^\circ_{n-1,1}$
(from the retarded products of lower orders) can be done for 
$R^{\circ(m)}_{n-1,1}$ as well as for $R^{\circ(m,\mu)}_{n-1,1}$, the 
results are related by (\ref{R^m,mu}).} $R^\circ_{n-1,1}$. 
For this purpose, in view of (\ref{R=J}) and the sheaf theorem 
on distributions, it is sufficient to check that in case two 
of the first $n-1$ arguments, say $x,y$, 
are different from the last argument $z$ and 
both in the past of $z$, then
\begin{displaymath}
  J_{n-2,2}(\ldots,x,y,z)=J_{n-2,2}(\ldots,y,x,z)\ .    
\end{displaymath}   
But by the Jacobi identity (\ref{Jacobi}), the difference is equal to
\begin{displaymath}
   J_{n-2,2}(\ldots,z,x,y)              
\end{displaymath}      
which vanishes since $z$ is neither in the past of $x$ nor 
$y$. So, $R^\circ_{n-1,1}$ is well defined by (\ref{R=J}) and 
fulfills the axioms Symmetry and Causality by construction.

In the next step we check that $R^\circ_{n-1,1}$ satisfies all 
other axioms. The only nontrivial points are the GLZ relation
and the Scaling. Concerning the former we have to prove that
  \begin{displaymath}
        R^\circ_{n-1,1}(\ldots,x,y,z)-R^\circ_{n-1,1}(\ldots,x,z,y) 
        =J_{n-2,2}(\ldots,x,y,z)\label{glz:R^0}
  \end{displaymath}
holds whenever the points $x,y,z$ are not identical. If $y\not= z$
then $y\not\in z+\overline{V}_+$ or $y\not\in z+\overline{V}_-$. 
In these cases the assertion follows from the construction (\ref{R=J})
of $R^\circ_{n-1,1}$. So it remains to treat the 
case $x\not= y\>\wedge\> x\not= z$, i.e.~
when $x\not\in y+\overline{V}_{\epsilon}\cup 
z+\overline{V}_{\epsilon'}$ with $\epsilon,\epsilon'\in\{+,-\}$.
In the case $\epsilon=\epsilon'=-$ all terms in the GLZ relation 
vanish (by construction of $R^\circ_{n-1,1}$ and due to the support
properties of $J$). In the case $\epsilon=-$ and $\epsilon'=+$
we analogously find $R^\circ (\ldots,x,z,y)=0$ and $J(\ldots,z,x,y)=0$. 
So, by the Jacobi identity the assertion (\ref{glz:R^0}) becomes
$R^\circ (\ldots,x,y,z)=J(\ldots,y,x,z)$ which is the construction
(\ref{R=J}) of $R^\circ_{n-1,1}$. The case $\epsilon=+$ and $\epsilon'=-$
is analogous. Finally, for  
$\epsilon=\epsilon'=+$ we apply again the Jacobi identity to 
the right side and find two terms which are just by 
(\ref{R=J}) the retarded products on the left side. 

We turn to the scaling. 
We show that $\sigma_\rho \circ
J_{n-2,2}^{(\rho^{-1}m)}\circ (\sigma_\rho^{-1})^{\otimes 
n}$ has polynomial behavior in $\log\rho$.  
This then implies the same property for 
$R^{\circ (m)}_{n-1,1}$, since the causal relations are scale 
invariant. (And from that it follows that also $R^{\circ (m,\mu)}_{n-1,1}$
scales almost homogeneously, analogously to (\ref{R^m,mu:scaled}).)
By definition, $J_{n-2,2}$ is a sum of $\star$-products of 
retarded products $R_{k-1,1}$ and $R_{n-k-1,1}$, $k=1,\ldots 
n-1$, which scale by assumption with a polynomial behavior in $\log\rho$.
The statement follows now from the fact that $\sigma_\rho$ is a 
$\star$-algebra isomorphism from ${\cal A}^{(\rho^{-1}m)}$ to ${\cal A}^{(m)}$.
\subsection{Extension to the total diagonal; the Action Ward Identity}
We now come to the main step in renormalization, namely the 
{\bf extension of the symmetric part $S^\circ_{n}$ of 
$R^\circ_{n-1,1}$ (\ref{S_n}) to a distribution on} $\MM^n$. 
For the construction of $R^\circ_{n-1,1}$ the normalization conditions 
(i.e.~the axioms Action Ward Identity, Covariance, Field Independence, 
Unitarity, Field equation, Smoothness in $m\geq 0$ and Scaling) have
not been needed, but they give guidance how to do the 
extension of $S^\circ_n$ and reduce the non-uniqueness drastically.
In particular the expansion (\ref{causWick})-(\ref{causWick1}) 
(i.e.~the axiom Field Independence) and Covariance for translations 
simplify the problem to the extension of the symmetric part 
$s^\circ_n(A_1,\ldots)(x_1-x_n,\ldots)$ of the
distribution\footnote{Using translation invariance of $\omega_0$
we denote $\omega_0(H(A_1(x_1),\ldots A_n(x_n)))$ by
$h(A_1,\ldots A_n)(x_1-x_n,\ldots)$ for
$H=R_{n-1,1},\,R^\circ_{n-1,1},\, S_n,\,S_n^\circ$ etc.~.\label{fn:r}}
$r^\circ_{n-1,1}(A_1,\ldots)(x_1-x_n,\ldots)$ 
from ${\cal D}(\RR^{d(n-1)}\setminus\{0\})$ to ${\cal D}
(\RR^{d(n-1)})$. This is the crucial problem of
perturbative renormalization, since it is this step which is
non-unique and which is the source of anomalies. Since the Smoothness
axiom applies for the $R^{(m,\mu)}$-products, but not for the $R^{(m)}$-products,
the extension is done for $S^{\circ (m,\mu)}_n$. From the resulting $S^{(m,\mu)}_n$,
the extension $S^{(m)}_n$ of $S^{\circ (m)}_n$ is obtained by (\ref{R^m,mu}).
\medskip

The basic idea to fulfill the {\bf Action Ward Identity} goes as follows:
since $\d^\mu_{x_l}s_n(\ldots,A_l,\ldots)$ is an extension of
$\d^\mu_{x_l}s_n^\circ (\ldots,A_l,\ldots)=s_n^\circ 
(\ldots,\d^\mu A_l,\ldots)$ we may define $s_n
(\ldots,\d^\mu A_l,\ldots)\=d\d^\mu_{x_l}s_n(\ldots,A_l,\ldots)$, 
provided $s_n(\ldots,A_l,\ldots)$ was already constructed. We are now
going to show that this can be done without running into 
inconsistencies. Namely, the fields in $\mathcal{P}$ are of the form
\begin{displaymath}
        A(x)=\sum_{n=0}^{N} 
        p_{n}(\d^1,\ldots,\d^n)\varphi(x_{1})
        \cdots\varphi(x_{n})|_{x_{1}=\cdots=x_{n}=x}
\end{displaymath}
with polynomials $p_{n}$ in the derivatives 
$\d_{\mu}^k=\frac{\d}{\d x_{k}^{\mu}},k=1,\ldots,n,\mu=0,\ldots,d-1,$ 
which are symmetric in the upper indices $k=1,\ldots,n$. The 
polynomials $p_{n}$ are uniquely determined by $A$.
 
Now let $\d_{\mu}=\sum_{k=1}^{n}\d_{\mu}^{k}$ denote the 
derivatives with respect to the center of mass coordinates and 
$\d_{\mu}^{ij}=\d_{\mu}^{i}-\d_{\mu}^{j},1\le i<j\le n$ the 
relative derivatives. The crucial observation is now that 
the vector space $P_{n}$ of all symmetric polynomials 
$p_{n}$ is isomorphic to the tensor product of the space $P^{\text{com}}$ 
of polynomials $p(\d)$ of the center of mass derivatives and 
the space $P_{n}^{\text{rel}}$ of symmetric polynomials 
$p_{n}(\d^{ij},1\le i<j\le n)$ of the relative derivatives. (Symmetry is meant
with respect to permutations $\sigma (\d^{ij}):=\d^{\sigma(i)\,\sigma(j)}$, 
$\sigma\in S_n$, where $\d^{ij}\equiv -\d^{ji}$.) 
The argument is straightforward for the unsymmetrized polynomials. 
Namely, the independent variables $\d^{in}$, $i=1,\ldots,n-1$ generate a 
polynomial algebra $\tilde{P}_n^{\text{rel}}$, and the linear map 
\begin{equation}
  \alpha:\left\{
  \begin{array}{ccc}
  P^{\text{com}}\otimes  \tilde{P}_n^{\text{rel}}    &  \longrightarrow  & \bigvee\{\d^1,...,\d^n\} \\
 \alpha(\d\otimes 1)=\sum_{i=1}^n \d^i  & , & \alpha(1\otimes\d^{in})=\d^i-\d^n
  \end{array}\right. 
\end{equation}
is an isomorphism onto the polynomial algebra generated by 
$\d^i$, $i=1,\ldots,n$.

This isomorphism intertwines the actions of the permutation 
group which are induced by the permutation of indices (on 
$\tilde{P}_n^{\text{rel}}$ the action is given by
\begin{displaymath}
  \sigma(\d^{in})=\d^{\sigma(i)\, n}-\d^{\sigma(n)\, n}
\end{displaymath}
with $\d^{nn}=0$) and thus restricts to an isomorphism of the invariant subspaces.
Interpreting $P_{n}^{\text{rel}}$ as a subspace of $\tilde{P}_n^{\text{rel}}$ (by the obvious
identification $\d^{ij}\equiv\d^{in}-\d^{jn}$), the invariant subspace of
$P^{\text{com}}\otimes  \tilde{P}_n^{\text{rel}}$ is just 
$P^{\text{com}}\otimes  P_n^{\text{rel}}$. 

We therefore use the space of balanced derivatives of fields \cite{BOR}   
\begin{equation}
        \mathcal{P}_{\text{bal}} \=d \{
        p_{n}(\d^{ij},1\le i<j\le n)\varphi(x_{1})
        \cdots\varphi(x_{n})|_{x_{1}=\cdots=x_{n}=x}\> |\>
        p_n\in P_{n}^{\text{rel}}\ ,\> n\in\NN\}\ .\label{bal-field}
 \end{equation}
and obtain the isomorphism of vector spaces
\begin{displaymath}
   \mathcal{P}\cong P^{\text{com}}(\d)\otimes \mathcal{P}_{\text{bal}} \ .
\end{displaymath}
In other words, every $A\in \mathcal{P}$ can {\it uniquely} be written as
\begin{displaymath}
   A=\sum_{i=1}^Np_i(\d)\, B_i\ ,\quad p_i(\d)\in P^{\text{com}}(\d)\ ,\quad
B_i\in  \mathcal{P}_{\text{bal}}\ ,\quad N<\infty\ .
\end{displaymath}
Applying this result to (\ref{F=Wf}) we obtain
\begin{prop}\label{prop:AWI}
Let $F$ be a local functional. Then there exists a {\it unique}
$h\in{\cal D}(\MM,\mathcal{P}_{\text{bal}})$ with
$F=\int dx\, h(x)$.
\end{prop}
 The Action Ward Identity can now simply be fulfilled by 
 performing the extension first only for balanced fields 
 $B\in\mathcal{P}_{\text{bal}}$ and by using the AWI and linearity 
 for the  definition of the extension for general fields 
 $A\in\mathcal{P}$. By construction this yields, in every 
 entry, a linear map from $P^{\text{com}}(\d)\otimes 
 \mathcal{P}_{\text{bal}}\cong \mathcal{P}$ to 
 distributions on $\MM$.  
\medskip

Next we recall from the literature the main statements about the extension
of a distribution $t^\circ \in{\cal D}^{\prime}(\RR^{k}\setminus
\{0\})$ to $t\in {\cal D}^{\prime} (\RR^{k})$
and give some completions. The existence and 
uniqueness of $t$ can be answered in terms of Steinmann's 
\textit{scaling degree} \cite{Ste} of $t^\circ$.
The latter is defined by
 \begin{equation}
{\rm sd}(f)\=d {\rm inf}\{\delta\in \RR\>|\>\lim_{\rho\downarrow 0}
\rho^\delta f(\rho x)=0\},\quad f\in {\cal D}'(\RR^k)
\>\>\mathrm{or}\>\> f\in {\cal D}'(\RR^k\setminus \{0\}).\label{sd}
\end{equation}
We immediately see that a distribution
which scales almost homogeneously with degree $D$ and an arbitrary
power $N<\infty$ has scaling degree $D$. In addition
one easily verifies
\begin{equation}
  {\rm sd}(\d^af)\leq{\rm sd}(f)+|a|,\quad
{\rm sd}(x^b f)\leq {\rm sd}(f)-|b|\quad\mathrm{and}
\quad {\rm sd}(\d^a\delta^{(k)})=k+|a|,\label{sd:d}
\end{equation}
where $a,b\in\NN_0^k$, $|a|\equiv\sum_{j=1}^k a_j$. Returning to the 
extension of $t^\circ$, the definition (\ref{sd}) implies immediately: 
${\rm sd}(t)\geq {\rm sd}(t^\circ)$. We are looking for extensions 
which do not increase the scaling degree.\footnote{In a large 
part of the literature (e.g. 
\cite{BS,EG,Scharf})
our axioms 'Smoothness in $m\geq 0$' and 'Scaling' are replaced 
by the weaker requirement
\begin{equation}
{\rm sd}\Bigl(r_{n-1,1}(A_1,\ldots,A_n)(x_1-x_n,\ldots)\Bigr)\leq
\sum_{j=1}^n\mathrm{dim}(A_j)\ ,\quad \forall A_j\in 
{\cal P}_\mathrm{hom}\ ,\label{axiom:sd}
\end{equation}
or an analogous normalization condition. 
For the extension problem this amounts (nearly always) to the requirement
${\rm sd}(t)={\rm sd}(t^\circ)$. We point out that (\ref{axiom:sd})
is a condition on the behavior under rescalings of the coordinates in
the UV-region only. In contrast, our Scaling axiom is with respect to
simultaneous rescalings of the coordinates and the mass, and it must
hold for all $x$ and for all $m\geq 0$.\label{fn:N0}}
\begin{thm}\label{extension} \cite{BF}: 
Let $t^\circ\in {\cal D}'(\RR^k\setminus\{0\})$.\\
(a) If ${\rm sd}(t^\circ)<k$, there exists a unique extension
$t\in {\cal D}'(\RR^k)$ with ${\rm sd}(t)={\rm sd}(t^\circ)$.\\
(b) If $k\leq {\rm sd}(t^\circ)<\infty$ there exist several extensions 
$t\in {\cal D}'(\RR^k)$ with ${\rm sd}(t)={\rm sd}(t^\circ)$.
Given a particular solution $t_0$, the most general solution reads
\begin{equation}
  t=t_0+\sum_{|a|\leq\mathrm{sd}(t^\circ)-k}
  C_a\d^a\delta^{(k)}\label{P(D)delta}
\end{equation}
with arbitrary constants $C_a\in\CC$.
\\
(c) If ${\rm sd}(t^\circ)=\infty$ there exists no extension 
$t\in {\cal D}'(\RR^k)$.
\end{thm}
Case (c) is mentioned for mathematical completeness
only. It does not appear in our construction, because
$r^\circ_{n+1,1}$ scales almost homogeneously with finite degree. 
However, there are distributions $t^\circ$ with ${\rm sd}(t^\circ)
=\infty$, e.g. $t^\circ (x)=e^{\frac{1}{|x|}}$. The proof of (a)-(b) 
given in \cite{BF} is based on \cite{EG} and it is constructive
('$W$-extension'). It is sketched in Appendix B. 

The $W$-extension has
the disadvantage that in general it does not maintain 
${\cal L}_+^\uparrow$-covariance for ${\rm sd}(t^\circ)> k$. However,
by a finite renormalization
(which does not increase the scaling degree) Lorentz
covariance can be restored (see \cite{EG,Ste,Sto, Scharf,BPP}
and also Appendix D).

The $W$-extension breaks in general also the property of 
almost homogeneous scaling. 
However, Hollands and Wald have given a
(finite) renormalization prescription to restore also this symmetry,
in detail:\footnote{This is a version
of Lemma 4.1 in the second paper of \cite{HW1-2},
which follows from the proof given in that paper.
The 'improved Epstein-Glaser renormalization' of
\cite{G-BL} maintains the almost homogeneous scaling directly.}
\begin{prop}\label{alm-hom-scal} 
Let $t^\circ \in {\cal D}'(\RR^k\setminus\{0\})$ scale
almost homogeneously with degree $D\in\RR$ and power $N\in\NN$
under coordinate rescalings (\ref{alm-hom-scal-1}).
Then $t^\circ$ has an extension $t$ to a
distribution on $\RR^{k}$ which also scales almost homogeneously with
degree $D$ and with power $N$, if $D\not\in\NN_0+k$, and with 
power $(N+1)$ or $N$, if $D\in\NN_0+k$.         
The extension is unique if $D\not\in\NN_0+k$, otherwise, the most general 
solution reads: (particular solution) $+\sum_{|a|=D-k}C_a\d^a\delta^{(k)}$,
where the $C_a$'s are arbitrary constants.
\end{prop}
Now we assume that 
$t^\circ\equiv t^{(m)\circ}\in \mathcal{D}'(\RR^{dr}\setminus\{0\})$ 
is smooth in
$m\geq 0$ and scales almost
homogeneously with degree $D$ and power $N$ under simultaneous rescalings
of the coordinates and the mass $m$:
\begin{equation}
  (\rho\d_\rho)^{N+1}\Bigl(\rho^D t^{(\rho^{-1}m)\circ}
  (\rho y)\Bigr)=0\ .\label{scal:t^mo}
\end{equation}
\begin{itemize}
\item For $m=0$ the scaling of $m$ is trivial and the Proposition can
  directly be applied. 
\item For $m>0$ the Smoothness in $m\geq 0$ of $t^\circ\equiv 
t^{(m)\circ}$ ensures the existence of the Taylor
expansion\footnote{This is the 'scaling expansion' of Hollands and
  Wald (given in the second paper of \cite{HW1-2}) in the particular
  simple case of Minkowski space.}
\begin{equation}
  t^{(m)\circ}(y)=\sum_{l=0}^{D-dr}\frac{m^l}{l!}u_l^\circ (y)+
m^{[D]-dr+1}t^{(m)\circ}_\mathrm{red}(y)\label{scal-exp}
\end{equation}
(where $[D]$ is the integer part of $D$) with
\begin{equation}
  u_l^\circ (y)\=d\frac{\d^l t^{(m)\circ}(y)}{\d m^l}\vert_{m=0}\ ,\label{u_l}
\end{equation}
where the 'reduced part' $t^{(m)\circ}_\mathrm{red}$ is smooth in
$m\geq 0$. The almost homogeneous scaling of $t^{(m)\circ}$ 
(\ref{scal:t^mo}) and the definition of $u_l^\circ$ imply
\begin{equation}
  (\rho\d_\rho)^{N+1}(\rho^{D-l}u^{\circ}_l(\rho y))=0\ .
\label{scal:u_l}
\end{equation}
Hence, by the Proposition, $u^\circ_l$ has an extension 
$u_l$ with $(\rho\d_\rho)^{N+2}(\rho^{D-l}u_l(\rho y))$ $=0$.

For the reduced part we find that
$m^{[D]-dr+1}t^{(m)\circ}_\mathrm{red}$ scales almost homogeneously with
degree $D$, because all other terms in (\ref{scal-exp}) have this
property. This gives
\begin{equation}
  \rho^{D-[D]+dr-1} t^{(m)\circ}_\mathrm{red}(\rho y)
=t^{(\rho m)\circ}_\mathrm{red}(y)+\sum_{j=1}^{N}
(\mathrm{log}\>\rho)^j l_j^{(\rho m)}(y)\label{scal:t_red}
\end{equation}
for some $l_j^{(m)}\in {\cal D}^{\prime} 
(\RR^{dr}\setminus\{0\})$ which are smooth in $m\geq 0$. Since also
$t^{(m)\circ}_\mathrm{red}$ is smooth in $m\geq 0$, we conclude
\begin{equation}
\lim_{\rho\to 0}\rho^{dr} t^{(m)\circ}_\mathrm{red}(\rho y)
=0\ ,\quad\mathrm{i.e.~}\quad 
\mathrm{sd}(t^{(m)\circ}_\mathrm{red})< dr\ .\label{sd:t_red}
\end{equation}
With that and with part (a) of Theorem \ref{extension} the 
reduced part $t^{(m)\circ}_\mathrm{red}$ can be uniquely extended. 
Due to the latter, the resulting 
$t^{(m)}_\mathrm{red}$ is also smooth in $m\geq 0$ and 
also the scaling property (\ref{scal:t_red}) is
maintained. Namely, $(\rho\d_\rho)^{N+1}\bigl(\rho^{D-[D]+dr-1} 
t^{(\rho^{-1}m)}_\mathrm{red}(\rho y)\bigr)$
has support in $\{0\}$ and its scaling degree is less than $dr$ 
for each fixed $\rho >0$.

Putting together the extensions of the individual terms we get
\begin{equation}
  t^{(m)}(y)\=d\sum_{l=0}^{D-dr}\frac{m^l}{l!}u_l (y)+
  m^{[D]-dr+1}t^{(m)}_\mathrm{red}(y)\ .\label{scal-exp:ext}
\end{equation}
By construction this is an extension of $t^{(m)\circ}$ with the wanted 
smoothness and scaling properties. 
It is unique for $D\not\in\NN_0+dr$. For $D\in\NN_0+dr$ the most 
general solution is obtained by adding
\begin{equation}
   \sum_{l=0}^{D-dr}\sum_{|a|=D-dr-l}m^lC_{l,a}\d^a\delta^{(dr)}
\label{non-unique:m>0}
\end{equation}
to a particular solution, with arbitrary constants 
$C_{l,a}$. Note that the undetermined polynomial
(\ref{non-unique:m>0}) scales even homogeneously (with degree $D$). 
If we would require almost homogeneous scaling only (and
not smoothness in $m\geq 0$), terms with $l<0$ would be admitted in 
(\ref{non-unique:m>0}). An extension with such terms increases the scaling degree:
${\rm sd}(t)>{\rm sd}(t^\circ)$, cf. footnote \ref{fn:N0}. 
\end{itemize}
If $t^{(m)\circ}$ (\ref{scal:t^mo}) is additionally Lorentz invariant, 
a slight modification of the usual proofs of Lorentz invariance 
\cite{EG,Ste,Sto,Scharf,BPP} yields that $t^{(m)}$ (\ref{scal-exp:ext}) 
can be chosen to be also Lorentz invariant (Appendix D). The conditions 
Unitarity and Symmetry can easily be included, too (see e.g. \cite{EG}). 

With this general knowledge about the extension of a 
distribution to a point we return to the extension of 
$s_n^{(m,\mu)\circ}$. For $A_1,\ldots, A_n
\in {\cal P}_\mathrm{bal}\cap{\cal P}_\mathrm{hom}$ the distribution
$s_n^{(m,\mu)\circ}(A_{1},\ldots, A_{n})$ fulfills
(\ref{scal:t^mo}) with degree $D=\sum_j\mathrm{dim}(A_{j})$
and some power $N<\infty$, and we can proceed as follows:  
\begin{itemize}
\item[Step 1:] We first extend the distributions $s_n^{(m,\mu)\circ}$ 
for homogeneous (\ref{P_hom}) balanced fields only, by applying
the above given procedure, including the finite renormalizations which
restore Lorentz covariance, almost homogeneous scaling (\ref{scal:t^mo}), 
Symmetry, Unitarity and maintain Smoothness in $m\geq 0$. Furthermore, 
the global inner symmetries (in our case the field parity (\ref{parity}))
can be preserved, cf. Appendix D.
\item[Step 2:] From that we construct $s_n^{(m,\mu)}$
for all fields by using linearity and the
Action Ward Identity.
\item[Step 3:] Finally we construct $S_n^{(m,\mu)}$ by means of the Taylor 
expansion (\ref{causWick})-(\ref{causWick1}),  
  \begin{gather}
   S_n^{(m,\mu)}(A_1(x_1),\ldots, A_n(x_n))\=d\sum_{l_1\ldots }
\frac{1}{l_1!\ldots}\int dx_{11}\ldots dx_{1l_1}\ldots \notag\\
s_n^{(m,\mu)}\Bigl(\frac{\delta^{l_1}A_1(x_1)}{\delta\varphi (x_{11})
\ldots\delta\varphi (x_{1l_1})},\ldots\Bigr)\varphi (x_{11})\ldots
\varphi (x_{1l_1})\ldots\ .\label{causWick:S}
  \end{gather}
By construction $S_n^{(m,\mu)}$ is linear in the fields and fulfills 
the axioms Covariance with respect to translations and Field Independence. 
The properties of $s_n^{(m,\mu)}$ established in the steps 1 and 2 imply that
$S_n^{(m,\mu)}$ fulfills the corresponding axioms. In particular, 
by using $\d_y\frac{\delta A(y)}{\delta\varphi (x)}=
\frac{\delta \d A(y)}{\delta\varphi (x)}$, we see
from (\ref{causWick:S}) that $S_n^{(m,\mu)}$ satisfies the AWI.
\end{itemize}

From a {\it particular} solution, the {\it general} solution is 
obtained by adding an arbitrary local polynomial of the form
(\ref{non-unique:m>0}) to $s_n^{(m,\mu)}(A_1,\ldots, A_n)$
which respects also linearity in the fields, Symmetry, 
Unitarity and the AWI. $R^{(m)}_{n-1,1}$ is obtained from 
$R^{(m,\mu)}_{n-1,1}$ by (\ref{R^m,mu}).
\medskip

The construction given so far yields the most general solution 
$R^{(m,\mu)}$ and $R^{(m)}$ of the
axioms of Sect.~2 except the Field equation (\ref{yangfeldm}). (Since the latter 
has precisely the same form for both kind of retarded products the
following procedure applies to both kinds and the results are related by
(\ref{R^m,mu}).) Due to the expansion (\ref{causWick})-(\ref{causWick1}) 
the Field equation is {\it equivalent} to
\begin{gather}
r_{n-1,1}(F_1,...,F_{n-1},\varphi(h))=\notag\\
-\int dx\, h(x)
\int dy\,\Delta^\mathrm{ret}(x-y)\sum_{k=1}^{n-1}
r_{n-2,1}(F_1,...\hat k...,F_{n-1},
\frac{\delta F_k}{\delta\varphi (y)})\ ,\label{r(phi)}
\end{gather}
for all $n\geq 2$, $F_1,\ldots,F_{n-1}\in {\cal F}_\mathrm{loc}$, 
and $h\in{\cal D}(\MM)$.
The right side gives an extension of $r_{n-1,1}^\circ 
(F_1,...,F_{n-1},\varphi(h))$, because the Field equation holds
outside the total diagonal. 
It is Lorentz covariant, symmetric in the first 
$(n-1)$ factors, unitary, smooth in $m\geq 0$, scales almost
homogeneously (even with power $\leq (n-2)$) and respects the AWI.  
From (\ref{r(phi)})
and the inductively known $\omega_0(J_{n-2,2}(F_1,\ldots ,F_{n-1},
\varphi(h)))$ we obtain $r_{n-1,1}(F_1,\ldots,\varphi(h),
\ldots,F_{n-1})$ by using the GLZ relation and the Symmetry in 
the first $(n-1)$ factors.\footnote{The result is given in part (B)
of Lemma 1 in \cite{DF}.} 
By construction this yields an extension
of $r^\circ_{n-1,1}(F_1,\ldots,\varphi(h),\ldots,F_{n-1})$
which also satisfies all axioms. 
With that $s_n(\ldots,\varphi(h))$ (\ref{S_n}) is uniquely 
determined in terms of the
inductively known $r_{n-2,1}$. 

So, in order to fulfill the Field equation we
modify the step 1 as follows: 
$s_n(A_1,\ldots,A_{n-1},\varphi)$, 
$A_1,\ldots,A_{n-1}\in {\cal P}_{\text{bal}}$, is uniquely given by
the Field equation in the just described way
and fulfills the required properties. However, the
construction of $s_n(A_1,\ldots, A_n)$ remains unchanged if
$A_1,\ldots, A_n$ are all of at least second order in $\varphi$ and
its partial derivatives. (If at least one factor is a C-number the
retarded product vanishes and hence also $s_n$, see part (C)
of Lemma 1 in \cite{DF}.) Finally steps 2 and 3 are done as 
before.
\medskip

Summing up we have proved:
\begin{thm} There exist retarded products which fulfill
   all axioms of Sect.~2.
\end{thm}

\noindent {\it Example:} 
{\bf setting-sun diagram $r_{1,1}^{(m,\mu)}(\varphi^3,\varphi^3)$ for $d=4$:} the explicit
calculation of a diagram requires usually somewhat less work if the extension is done
directly for $r^\circ$ (and not for its symmetric part $s^\circ$). By using the GLZ relation
and Causality we obtain
\begin{equation}
r^{\circ (m,\mu)}(y)\equiv
r_{1,1}^{\circ (m,\mu)}(\varphi^3,\varphi^3)(y)=-6i\>\Bigl(H^\mu_m(y)^3-
H^\mu_m(-y)^3\Bigr)\> \Theta(-y^0)\ .\label{3,3:m,mu}
\end{equation}
From (\ref{4+}) we can read off the first terms of the Taylor expansion in $m^2$ of
$H^\mu_m(y)$:
\begin{equation}
H^\mu_m(y)=D^+(y)
+m^2\>\bigl[{\rm log}(-\mu^2(y^2-iy^00))\>f(0)+F(0)\bigr] + h_m(y)\ ,
\end{equation}
where $D^+(y)\equiv -(4\pi^2(y^2-iy^00))^{-1}$ is the massless two-point function and
$h_m(y)$ is of order ${\cal O}(m^4)$ and has scaling degree ${\rm sd}(h_m)<0$.
With that we get the scaling expansion (\ref{scal-exp}) of (\ref{3,3:m,mu}):
\begin{equation}
r^{\circ (m,\mu)}(y)=u^\circ_0(y)+\frac{m^2}{2}u_2^{\circ (\mu)}(y)+
r^{\circ(m,\mu)}_\mathrm{red}(y)\ ,\label{scal-exp1}
\end{equation}
where
\begin{gather}
u^\circ_0(y)=-6i\>\Bigl(D^+(y)^3-D^+(-y)^3\Bigr)\> \Theta(-y^0)\ ,\notag\\
u_2^{\circ (\mu)}(y)=-36i\,\Bigl(D^+(y)^2\, \bigl[{\rm log}(-\mu^2(y^2-iy^00))\>f(0)+F(0)\bigr]
-(y\rightarrow -y)\Bigr)\> \Theta(-y^0)\ .\label{scal-exp2}
\end{gather}
The power of the almost homogeneous scaling (\ref{scal:t^mo})
(with degree $6$) is the power of ${\rm log}(\mu^2...)$. It is
different for the individual terms: it is $0$ for $u^\circ_0$,
$1$ for $u^\circ_2$ and $3$ for $r^{\circ(m,\mu)}_\mathrm{red}$ respectively. 
In contrast to the reduced part $r^{\circ(m,\mu)}_\mathrm{red}$, 
the renormalization of $u^\circ_0$ and $u^{\circ(\mu)}_2$ 
is non-trivial and it increases the power of the almost homogeneous scaling
by $1$. The extension of these two terms is given in Appendix B by using 
{\it differential renormalization}. An alternative method, which
relies on the K\"allen-Lehmann representation, is applied to the massless 
fish  and setting-sun diagram in Appendix C.

\medskip
We now focus on the power $N$ of the almost homogeneous scaling (\ref{scal:t^mo}).
The preceding example shows that, in the scaling expansion of
$r^{\circ (m,\mu)}$ (or $s^{\circ (m,\mu)}$), the terms for which $N$ 
may be increased in the extension, are not the terms with the maximal value of $N$.
The proof of part (ii) of the following Proposition is based on this observation.
\begin{prop}\label{power} 
\begin{enumerate}
        \item If the number $d$ of space time dimensions is {\bf odd}, the 
power $N$ of the almost homogeneous scaling of 
$R^{(m)}_{n-1,1}\equiv R^{(m,\mu)}_{n-1,1}$ is smaller than $n$,
i.e.~$R^{(m)}_{n-1,1\>\rho}$ (\ref{R^m:scaled}) is a polynomial of 
$({\rm log}\>\rho)$ with degree less than $n$.
        \item For $d=4$ the power $N$ of the almost homogeneous scaling  
(\ref{scal:t^mo}) of 
\begin{equation}
r^{(m,\mu)}_{n-1,1}(A_1,\ldots,A_n)\ ,\quad\mathit{with}\quad
A_j=\prod_{s=1}^{l_j}\d^{a_{js}}\varphi\ 
\end{equation}
and
\begin{equation}
\sum_{j=1}^n\mathrm{dim}(A_j)\leq 4(n-1)+3\ ,\label{A:restriction}
\end{equation}
is bounded by
\begin{equation}
N\leq \frac{1}{2}\>\sum_{j=1}^nl_j\quad\quad\Bigl(\leq
\frac{1}{2}\>\sum_{j=1}^n\mathrm{dim}(A_j)\Bigr)\ .\label{N:bound}
\end{equation}
\end{enumerate}
\end{prop}
Note that $\frac{1}{2}\,\sum_{j=1}^nl_j$ is the number of (internal) lines in the corresponding 
Feynman diagram. Due to the expansion (\ref{causWick})-(\ref{causWick1}) 
and the fact that $\prod_j\d^{a_j}\varphi (x_{i_j})$ scales homogeneously, part (ii) 
implies that the power $N$ of the almost homogeneous scaling of 
$R^{(m,\mu)}_{n-1,1}(A_1,\ldots,A_n)$ (with $A_j$ of the mentioned kind)
is also bounded by (\ref{N:bound}). The restriction (\ref{A:restriction}) on the $A_j$'s
is e.g.~satisfied for interacting fields $A_{{\cal L}(g)}$ if $\mathrm{dim}(A)\leq 3$
and ${\cal L}$ is renormalizable by power counting, cf. Sect.~4.1.

\begin{proof} {\it Part (i):} $R_{0,1}(A)=A$ scales homogeneously (\ref{scal:hom}).
Following our inductive construction, one verifies that an increase of $N$ may happen in 
the extension $s_n^\circ\rightarrow s_n$ only. Hence, by Proposition \ref{alm-hom-scal},
$N$ is increased at most by $1$ in each inductive step.

{\it Part (ii):} We give the proof for $A_j=\varphi^{l_j}$ only. The generalization to field 
polynomials with derivatives is straightforward, it gives only notational complications.
In our inductive construction of the $R^{(m,\mu)}$-products $N$ may now be increased
also in the construction of $j^{(m,\mu)}_{n-2,2}$, since the GLZ relation uses 
the modified star product $\star_{m,\mu}$. The vacuum expectation value of the 
GLZ relation reads 
\begin{gather}
\omega_0\Bigl(\Bigl\{R^{(m,\mu)}\bigl(\varphi^{l_1}(x_1),...,
\varphi^{l_k}(x_k)\bigr),R^{(m,\mu)}\bigl(\varphi^{l_{k+1}}(y_1),...,
\varphi^{l_n}(y_{n-k})\bigr)\Bigr\}_{\star_{m,\mu}}\Bigr)=\notag\\
\sum...r^{(m,\mu)}\bigl(\varphi^{l_1-p_1}(x_1),...,
\varphi^{l_k-p_k}(x_k)\bigr)\> r^{(m,\mu)}\bigl(\varphi^{l_{k+1}-p_{k+1}}(y_1),...,
\varphi^{l_n-p_n}(y_{n-k})\bigr)\notag\\
\cdot\>\Bigl(\prod_{j=1}^p H^\mu_m(x_{i_j}-y_{i_j})-
\prod_{j=1}^p H^\mu_m(y_{i_j}-x_{i_j})\Bigr)\ ,\label{omega(GLZ)}
\end{gather}
where $p\equiv\frac{1}{2}\,(p_1+...+p_n)$. By using the inductive assumption 
and the fact that $H^\mu_m$ scales almost homogeneously with power 
$1$ (\ref{H:scaling}) \footnote{An essential ingredient of the generalization 
to field polynomials with derivatives is that also partial derivatives 
$\d^aH^\mu_m\>(a\in\NN_0^4)$ scale almost homogeneously with power $1$.}
we find that for each term (on the right side) $N$ is bounded by
\begin{equation}
\frac{1}{2}\sum_{j=1}^k(l_j-p_j)+\frac{1}{2}\sum_{j=k+1}^n(l_j-p_j)+p=
\frac{1}{2}\sum_{j=1}^n l_j\ .\label{j:bound}
\end{equation}
We turn to the extension $s^\circ_n\rightarrow s_n$. The scaling degree of each 
term in (\ref{omega(GLZ)}) is bounded by
\begin{equation}
{\rm sd}(\ldots)\leq \sum_{j=1}^n\mathrm{dim}(A_j)\leq 4(n-1)+3\ .
\end{equation}
\begin{itemize}
\item If $p=1$ the extension is trivial and, hence, the power $N$ is not 
increased in this step.
\item If $p\geq 2$ and the scaling degree is $\geq 0$, the extension may increase 
$N$ by $1$. The terms on the right side of (\ref{omega(GLZ)}) with
the {\it maximal} value of $N$ have the property that 
$-m^2\>f(m^2y^2)\>{\rm log}(m^2/\mu^2)$ is substituted for $H^\mu_m(y)$
(\ref{4H}) in {\it all} $H^\mu_m$'s. Therefore, the scaling degree of these 
terms is lowered by $2p\geq 4$, i.e.~it is $<4(n-1)$, and we are in the case
of trivial extension. We conclude that in the extension 
$s_n^{\circ (m,\mu)}(\varphi^{l_1},...)\rightarrow s_n^{(m,\mu)}(\varphi^{l_1},...)$
the corresponding value of $N$ is not increased, i.e.~$N$
is still bounded by (\ref{j:bound}).
\end{itemize}
\end{proof}
In $d=4+2k\>(k\in\NN)$ space time dimensions analogous bounds on the power $N$
of the almost homogeneous scaling of the
$R^{(m,\mu)}$-products can be derived by the same method.
\section{Non-uniqueness}\setcounter{equation}{0}
\subsection{Counting the indeterminate parameters before the adiabatic limit}
In contrast to the literature we count the indeterminate parameters {\it
  without performing the adiabatic limit}.

The interacting fields $A_{{\cal L}(g)}(x)=
\sum_{n=0}^\infty\frac{1}{n!}R_{n,1}(({\cal L}(g))^{\otimes n};
A(x))$ are left with an indefiniteness coming
from the extension of the symmetric part of
$r^\circ_{n,1}$ to the origin (in relative coordinates). 
In general the normalization conditions restrict
this indefiniteness only, they to do not remove it completely. 

Let ${\cal L}\in\mathcal{P}_{\text{bal}}$ and 
let $N(\mathcal{L},A,n)$ be the number of indeterminate
parameters (i.e.~the constants $C_a$ in (\ref{P(D)delta}) or
(\ref{ext})) in $R_{n,1}({\cal L}(g))^{\otimes n};A(x))$ 
coming from the inductive step $(n-1,1)\rightarrow (n,1)$. 
This number depends on the choice of the normalization
conditions. In the following we presume the axioms given in 
Sect.~2 except Lorentz covariance, Unitarity and Field equation. 
We will prove
\begin{prop}
Let ${\cal L}\in\mathcal{P}_{\text{bal}}$. 
\begin{itemize}

\item[(a)] $N({\cal L},A,n)$ is bounded in $n$ $\forall A\in {\cal P}$ fixed,
iff ${\rm dim}({\cal L})\leq d$.

\item[(b)] For all $A\in{\cal P}$ there exists $n(A)$ such that
$N({\cal L},A,n)=0$ $\>\forall n>n(A)$, iff ${\rm dim}({\cal L})<d$.
\end{itemize}
\end{prop}
An interaction ${\cal L}$ with the property {\it (a)} of 
$n\mapsto N(\mathcal{L},A,n)$
is called 'renormalizable by power counting'.
In the literature (also in causal perturbation theory 
\cite{EG,Scharf,S-wiley}) the counting of indeterminate parameters
is done in terms of the $S$-matrix
{\it in the adiabatic limit} (i.e.~ $\sum_n\frac{1}{n!}\lim_{g\to 1}
T_n(({\cal L}(g))^{\otimes n})$), and the corresponding version of the Proposition
can be proved rather easily, see e.g. Sect.~28.1 of \cite{BS}. 
It does not make an essential 
difference that we count in terms of retarded products.
But, since we do not perform the adiabatic limit,
our discussion is more involved. 
\begin{proof}
Due to 
\begin{equation}
\frac{\delta A(x)}{\delta\varphi(y)}=\sum_a
\frac{\d A}{\d (\d^a\varphi)}(x)(-1)^{|a|}\d^a\delta (y-x)
\end{equation}
we understand by the sub-polynomials $U$ of $A\in {\cal P}$ all
non-vanishing polynomials
$U\equiv\frac{\d^k A}{\d(\d^{a_1}\varphi)\ldots\d(\d^{a_k}\varphi)}$,
$k\in \NN_0,\> a_j\in\NN_0^d$ and we write $U\subset A$.
Ignoring the AWI, the indefiniteness of
$R_{n,1}(({\cal L}(g))^{\otimes n};A(x))$ is precisely the 
indefiniteness of all C-number distributions
$r_{n,1}(U_1,...,U_n;U)$, $U_1,...,U_n\subset {\cal L}$ and $U\subset A$, 
due to the expansion (\ref{causWick})-(\ref{causWick1}). Note
\begin{equation}
  0\leq {\rm dim}(U_j)\leq {\rm dim}({\cal L}),\quad\quad
\mathrm{and}\quad\quad 0\leq {\rm dim}(U)\leq {\rm dim}(A).
\label{dim(subpol)}
\end{equation}
The indefiniteness of $r_{n,1}(U_1,...,U_n;U)(x_1-x,...,x_n-x)$ is a
polynomial (\ref{non-unique:m>0})
\begin{equation}
  \sum_{|a|\leq\omega (U_1,...,U_n;U)}C^n_a(U_1,...,U_n;U)
\d^a\delta (x_1-x,...,x_n-x)\label{unbest-pol}
\end{equation}
which is invariant under permutations of $(U_1,x_1),...,(U_n,x_n),
(U,x)$, where 
\begin{gather}
  \omega (U_1,...,U_n;U)=\sum_{j=1}^n {\rm dim}(U_j)+
{\rm dim}(U)-dn\notag\\
\leq {\rm dim}(U)+n({\rm dim}({\cal L})-d)\ .\label{omega(subpol)}
\end{gather}
For $m=0$ the sum (\ref{unbest-pol}) runs over $|a|=\omega$ only;
this simplifies the proof.

With (\ref{dim(subpol)})-(\ref{omega(subpol)}) the only non-obvious
statement of the Proposition is that for an interaction ${\cal L}$ with 
${\rm dim}({\cal L})=d$ the function $n\mapsto N({\cal L},A,n)$ is 
bounded $\forall A\in {\cal P}$. This statement holds even true if the 
AWI is not required, and we are now going to prove it under this 
weaker supposition. The boundedness (in
$n$) of $\omega (U_1,...,U_n;U)$ \textit{alone}, i.e.~
\begin{equation}
  \omega (U_1,...,U_n;U)= {\rm dim}(U)-\sum_{j=1}^n
(d-{\rm dim}(U_j))\leq {\rm dim}(U)\leq {\rm dim}(A)\quad\forall
n\in\NN\label{w:L=d}
\end{equation}
and $\forall U_j\subset {\cal L},\,U\subset A$,
does not imply the assertion, because

(i) the number of terms $r_{n,1}(U_1,...,U_n;U),\>U_j\subset {\cal L},
\,U\subset A$, is increasing with $n$;

(ii) the number of indices $a\in\NN_0^{dn}$ with $|a|\leq \omega_0$,
$\omega_0$ fixed, see (\ref{unbest-pol}), is also increasing with $n$.
(Hence, one might expect that there is e.g. an increasing number of
constants $C_a^n(\mathcal{L},...,\mathcal{L},A)$.)

Now let $A\in\mathcal{P}$ be fixed. (i) is no problem, since 
there are indeterminate parameters in 
the $r_{n,1}(U_1,...,U_n;U)$ for $\omega (U_1,...,U_n;U)\geq 0$
(\ref{w:L=d}) only. Hence, we solely need to consider
\begin{equation}
  \mathcal{R}_n\=d\{r_{n,1}(U_1,...,U_l,{\cal L},...,{\cal L};U)\>|
\> U_1,...,U_l\subset {\cal L},\,U\subset A\}\ ,
\end{equation}
where $l$ is given by $l\cdot {\rm dim}(\varphi)={\rm dim}(A)$.
However, the number of elements of $\mathcal{R}_n$ is constant for
all $n\geq l$.

To invalidate the objection (ii) let $U_1,...,U_l$ and $U$ be fixed. Because
\begin{equation}
  r_{n,1}(U_1,...,U_l,{\cal L},...,{\cal L};U)(y_1,...,y_n),
\quad y_j\=d x_j-x,
\end{equation}
is {\it symmetrical} in $y_{l+1},...,y_n$, the number of
constants
\begin{equation}
  C^n_a(U_1,...,U_l,{\cal L},...,{\cal L};U)\quad {\rm with}
\quad a\in\NN_0^{dn},\>|a|\leq {\rm dim}(U)-\sum_{j=1}^l
(d-{\rm dim}(U_j))
\end{equation}
is bounded in $n$. We use here a modified version of
the fact that the number of coefficients in the {\it symmetrical} 
polynomials $P(z_1,...,z_m)\> (z_j\in\RR),\>m\in\NN$,
of a fixed degree becomes independent of the number $m$ of variables 
$z_j$, if $m$ is big enough.\footnote{This becomes obvious by listing
  the symmetrical polynomials in $z_1,...,z_m$ which are homogeneous
of degree $k$ for the lowest values of $k$: e.g. for $k=4$
and for all $m\geq 4$ a basis is given by
\begin{gather}
P_1=C_1{\cal S}z_1^4\ ,\quad  P_2=C_2{\cal S}z_1^3z_2\ ,
P_3=C_3{\cal S}z_1^2z_2^2\ ,\notag\\
P_4=C_4{\cal S}z_1^2z_2z_3\ ,\quad  P_5=C_5{\cal S}z_1z_2z_3z_4\ ,
\end{gather}
where ${\cal S}f(z_1,\ldots ,z_m)\equiv\frac{1}{m!}\sum_{\pi\in S_m}
f(z_{\pi 1},\ldots ,z_{\pi m})$.}

Summing up we find that $N({\cal L},A,n)$ is bounded in $n$
for any fixed $A\in {\cal P}$. 
\end{proof}
\subsection{Main Theorem of Perturbative Renormalization}
It is one of the main insights of renormalization theory that the 
ambiguities of the renormalization process can be absorbed in a 
redefinition of the parameters of the given model. In causal
perturbation theory this was termed 
{\bf Main Theorem of Renormalization} \cite{Sto}. Different 
versions of this theorem may be found in \cite{SP,G-ML,BS,Sto,Pi,G:scal}. 
But there, in contrast to the formulation of renormalization in terms 
of the action functional, the parameters of a model are test functions. 
Therefore, the renormalization group which governs the change of 
parameters is more complicated, and it is only in the adiabatic 
limit that the more standard version of the renormalization group will 
be recovered.

Fortunately, the algebraic adiabatic limit 
\cite{BF} is sufficient for this purpose, 
as was first shown by Hollands and Wald \cite{HW}. In this way, one 
finds an intrinsically local construction of the renormalization group which 
is suited for theories on curved spacetime and for theories with a bad 
infrared behavior.

Here we give a slightly streamlined proof of the Main Theorem in 
the framework of retarded products. In Sect.~5 we discuss the consequences 
for the algebraic adiabatic limit. 
\begin{thm}\label{main}
 \begin{enumerate}
 	\item  Let $\hat{R},\> R:T{\cal
       F}_\mathrm{loc}\longrightarrow {\cal F}$ be linear 
       maps which satisfy the axioms  Symmetry, Initial 
       Condition, 
       Causality 
       and GLZ for retarded products. 
       Then there exists a unique symmetric linear map
\begin{equation}
  D:T{\cal F}_\mathrm{loc}\longrightarrow
 {\cal F}_\mathrm{loc}\label{D}
\end{equation}
with $D(1)=0$ such that for all $F,S\in {\cal F}_{\mathrm{loc}}$ the 
following intertwining relation holds (in the sense of 
formal power series in $\lambda$)
   \begin{equation}
        \hat{R}(e_{\otimes}^{\lambda S}, F) = 
        R(e_{\otimes}^{D(e_{\otimes}^{\lambda S})},
        D(e_{\otimes}^{\lambda S}\otimes F)) \ .
        \label{ren:intf}
   \end{equation}
Moreover, $D$ satisfies the conditions
  \begin{enumerate}

        \item  $D(F)=F$, $F\in\mathcal{F}_{\text{loc}}$. 

        \item  
       \begin{equation}
       \supp\frac{\delta\,D(F_1\otimes\ldots\otimes F_n)}{\delta\varphi}\subset
       \bigcap_i \supp\frac{\delta F_{i}}{\delta\varphi},\quad
       F_i\in\mathcal{F}_{\text{loc}}\ .\label{locality of D}
        \end{equation}
   \end{enumerate}
 	\item  If in addition, $R$ and $\hat{R}$ satisfy some of the conditions 
    Covariance, Unitarity, Field Independence and Field Equation, then 
    $D$ has the corresponding properties, i.e.~it is
    covariant, hermitian 
    \footnote{i.e.~$D(F^{\otimes n})^{*}=D((F^{*})^{\otimes n})$,}
    , has no explicit dependence on $\varphi$,
 \begin{equation}
\frac{\delta\,D(e_{\otimes}^F)}{\delta\varphi}=
D\Bigl(\frac{\delta F}{\delta\varphi}\otimes e_{\otimes}^F\Bigr)\ ,\label{D:field independence}
\end{equation}
 and (under the condition Field Independence) fulfills 
    $D(e_{\otimes}^{\lambda S}\otimes 
    \varphi(h))=\varphi(h)$, respectively. 

            \item If there are two families $R^{(m,\mu)},\hat{R}^{(m,\mu)}$ 
    as above, which are smooth in $m\ge 0$ and satisfy the 
    axiom Scaling, then the corresponding intertwining family $D_H^{(m)}$ is also 
    smooth in $m$, and it is independent\footnote{For this reason we write 
    $D_H^{(m)}$ instead of $D^{(m,\mu)}$.} of $\mu$ and invariant under scaling:
    \begin{equation}
    \sigma_\rho D_H^{(\rho^{-1}m)}(e_{\otimes}^{\sigma_\rho^{-1}F})=
    D_H^{(m)}(e_{\otimes}^F)\ .\label{D:scaling}
    \end{equation}
 If $R^{(m)},\hat{R}^{(m)}$ are related to  $R^{(m,\mu)},\hat{R}^{(m,\mu)}$ by
(\ref{R^m,mu}), then the corresponding intertwining $D^{(m)}$ is related to
$D_H^{(m)}$ also by (\ref{R^m,mu}) and fulfills
    \begin{equation}
    \sigma_\rho \circ D_n^{(\rho^{-1}m)}\circ (\sigma_\rho^{-1})^{\otimes n}=
    \rho^{-\Gamma^{(m)}}\circ D_n^{(m)}\circ (\rho^{\Gamma^{(m)}})^{\otimes n}\ .
    \label{D:scaling1}
    \end{equation}
 	\item  
                Conversely, given $R$ and $D$ as above , equation 
 	(\ref{ren:intf}) gives a new retarded product $\hat{R}$ 
 	with the pertinent properties.  If there are families 
 	$R^{(m,\mu)}$ and $D_H^{(m)}$ as above, then the corresponding 
 	family $\hat{R}^{(m,\mu)}$ is smooth in $m\ge 0$ and satisfies the 
 	Scaling axiom.
 \end{enumerate}
  
\end{thm}
The identity (\ref{ren:intf}) states that the (finite)
renormalization $R\rightarrow\hat R$ can be absorbed in 
the renormalizations
\begin{itemize}
\item $\lambda S\rightarrow D(e_\otimes^{\lambda S})$ of the interaction and

\item $F\longrightarrow D(e_\otimes^{\lambda S}\otimes F)$ of the field.

\end{itemize}

It is crucial that the renormalization of the interaction 
is independent of the field $F$. Looking in (\ref{ren:intf}) at the
terms of $n$-th order in $\lambda$ and using the polarization identity we
find that (\ref{ren:intf}) is equivalent to 
\begin{gather}
  \hat R(F_1\otimes ...\otimes F_n)=\notag\\
\sum_{n\in I\subset\{1,...,n\}}
\sum_{P\in\mathrm{Part}(I^c)}R\bigl(\bigotimes_{T\in
  P}D(F_{T})\otimes D(F_{I})\bigr) \label{R(D)}
\end{gather}
with $F_{J}=\bigotimes_{j\in J}F_{j}$ for an ordered index 
set $J$. It is instructive to write equation (\ref{R(D)}) in lowest orders:
\begin{eqnarray}
&(n=1)&\ \ \ \ \ \ \ \  F\equiv \hat{R}(F)=R(D(F))\equiv D(F), \notag\\
&(n=2)& \ \ \ \ \hat R(F_1\otimes F_{2})=
                R(F_1\otimes F_{2})+D(F_1\otimes F_{2}),\notag\\
&(n=3)& \ \ \ \ \ \ \hat R(F_1\otimes F_2\otimes F_{3})=
              R(F_1\otimes F_2\otimes F_{3})\notag\\
   && +R(D(F_1\otimes F_2)\otimes F_{3})+R(F_1\otimes D(F_2\otimes F_{3}))+
   R(F_2\otimes D(F_1\otimes F_{3}))\notag\\
  && +D(F_1\otimes F_2\otimes F_{3}).
\end{eqnarray}
We see that the difference between the retarded products in 
order $(1,1)$ propagates to higher orders
which gives the terms in the second last line. These terms are localized on
partial diagonals and express the change of normalization of
sub-diagrams. The term in the last line is localized on the total
diagonal $\Delta_3$ and originates from the freedom of 
normalization of retarded products
in the inductive step from $(1,1)$ to $(2,1)$.

Note that the term with $I=\{1,\ldots,n\}$ in the 
second line of (\ref{R(D)}) gives, in view of 
$R_{0,1}=\mathrm{id}$, a definition of 
$D_{n}\=d D\restriction \mathcal{F}_{\text{loc}}^{\otimes n}$ 
in terms of 
$\hat{R}_{n-1,1}$, $R_{k,1}, D_{k}$ for $k=1,\ldots,n-1$. 
It is here that our formalism seems to be superior over previous 
formulations. Namely, if the retarded products take their 
values only on shell, $R_{0,1}$ is no longer the identity 
but the canonical surjection $\pi$ with respect to the ideal generated by 
the free field equation. Then the definition of $D_{n}$ 
requires a choice of representatives. Without the Action 
Ward Identity, such a choice is rather artificial, and we 
are not aware of any place in the literature where 
this problem is treated in full generality.      

\begin{proof}[Proof of the Theorem] {\it Part (iv):} it is 
  straightforward to check that every $D$ with the properties 
  described in the Theorem defines a new retarded product 
  $\hat{R}$ via equation (\ref{ren:intf}) (or equivalently (\ref{R(D)})).

  {\it Part (i):} from equation (\ref{R(D)}) 
  one immediately sees 
  that  $\hat{R}_{0,1}=\mathrm{id}$ is equivalent to 
  $D_{1}=\mathrm{id}$, and that
  $\hat{R}_{n-1,1}$ is determined by 
  the $D_l$'s of order $l\leq n$ and by $R$. Vice versa, $D_{n}$ 
  is uniquely given in terms of $R$, $\hat{R}$ and the lower 
  order $D$'s, and obviously it is linear. If $D$ satisfies the properties mentioned in
  Part (i) (or Parts (i)-(iii) respectively), then this holds true also for its truncation
  $D^{(n)}$, which is defined by $D^{(n)}_l=D_l$ for $l\leq n$ and $D^{(n)}_l=0$
  for $l>n$. Following Part (iv), $D^{(n)}$ determines a retarded product $\hat R^{(n)}$
  with the pertinent properties, which 
  coincides with $\hat{R}$ in order $(k,1)$ for $k<n$.
  From (\ref{R(D)}) we see that
\begin{equation}
D_{n}=\hat{R}_{n-1,1}-\hat R^{(n-1)}_{n-1,1},\label{D=R-R^n}
\end{equation}
i.e. $D_n$ is the difference between two 
  possible extensions of retarded product at order $(n-1,1)$ 
  and is therefore symmetric and localized on the diagonal.
The latter implies $\mathrm{ran}\>D_n\subset  {\cal F}_\mathrm{loc}$.
 
{\it Parts (ii)-(iii):} additional properties of the retarded products 
  imply directly the corresponding properties of $D_{n}$. In particular,
\begin{itemize}
\item[(Field Equation)] we know that the Field Equation (\ref{r(phi)}) determines 
$R_{n-1,1}(\ldots,\varphi (h))$ {\it uniquely} in terms of $R_{n-2,1}$; with that
(\ref{D=R-R^n}) implies $D_n(\ldots\otimes\varphi (h))=0$;
\item[(Scaling)] from (\ref{D=R-R^n}) and the almost homogeneous scaling of 
$\hat{R}_{n-1,1}$ and $\hat R^{(n)}_{n-1,1}$ we conclude that $\omega_0
(D^{(m)}_{H\,n}(A_1(h_1)\otimes\ldots A_n(h_n))$ must be of the form (\ref{non-unique:m>0})
for $A_1,\ldots A_n\in{\cal P}_\mathrm{bal}\cap{\cal P}_\mathrm{hom}$ and 
hence it scales homogeneously with degree $\sum_i\mathrm{dim}(A_i)$;
this yields (\ref{D:scaling}) for $D^{(m)}_{H\,n}$. The corresponding coefficients
$C_{l,a}$ in (\ref{non-unique:m>0}) are independent of $\mu$ because they are 
dimensionless; this shows explicitly that $D^{(m)}_{H}$ is independent of $\mu$. 
Finally (\ref{D:scaling1}) is obtained from (\ref{D:scaling}) analogously to
(\ref{R^m,mu:scaled}).
\end{itemize}
\end{proof} 
We now want to get a more explicit expression for 
$D_H^{(m)}$ under the assumptions Smoothness in $m$ and 
Scaling. 
\begin{prop}\label{prop:main}   
  Let $(D_H^{(m)})_{m\ge 0}$ be a family of symmetric, linear 
  maps  from $T{\cal F}_\mathrm{loc}\longrightarrow
  {\cal F}_\mathrm{loc}$ which  are local 
  (in the sense of (\ref{locality of D})) and independent of $\varphi$ (\ref{D:field independence}). 
  Assume that the family is scale invariant and smooth in $m$.
  Then it admits the expansion 
  \begin{equation}
    D_{H\,n}^{(m)}(A_{1}(h_{1})\otimes\cdots \otimes A_{n}(h_{n}))=
    \sum_{l\in \NN_{0},a\in(\NN_{0}^{d})^{n}} 
    m^{l}d_{n,l,a}(A_{1}\otimes\cdots \otimes 
    A_{n})(\prod_{i=1}^{n}\d^{a_{i}}h_{i})\label{D(Wh)}
  \end{equation}
  with $A_{i}\in \mathcal{P}_{\text{bal}}$ and $h_{i}\in 
  \mathcal{D}(\MM)$, where $d_{n,l,a}$ are linear symmetric maps  
  $\mathcal{P}_{\text{bal}}^{\otimes n} \to 
  \mathcal{P}_{\text{bal}}$ 
  which are homogeneous in the 
  sense that tensor products of homogeneous fields are mapped onto 
  homogeneous fields such that the mass dimensions satisfy the 
  relation 
  \begin{equation}
        \mathrm{dim}(d_{n,l,a}(A_{1}\otimes\cdots \otimes A_{n}))  
         = \sum_{i=1}^{n} \mathrm{dim}(A_{i}) -l -|a|-d(n-1) \ .
        \label{mass dimension}
  \end{equation} 
  In particular, $d_{n,l,a}$ vanishes on tensor products of 
  fields $A_{i}$, if the right hand side is negative. Hence 
  the sum in (\ref{D(Wh)}) is finite.
\end{prop}
Note that we have on the right side of (\ref{D(Wh)}) the pointwise product of the 
test functions: $\prod_i(\d^{a_{i}}h_{i}(x))$.
\medskip

\noindent {\it Remarks}: (1) In massless models solely the terms with
$l=0$ contribute in (\ref{D(Wh)}).\\ 
(2) If our requirements Smoothness in $m\geq 0$ and Scaling are 
replaced by the upper bound (\ref{axiom:sd}) on the scaling degree
for a fixed mass $m$, then there is an analogous expansion, 
$D_{n}(A_{1}(h_{1})\otimes\cdots )=
    \sum_a d_{n,a}(A_{1}\otimes\cdots )(\prod_i\d^{a_{i}}h_{i})$,
in terms of linear symmetric maps 
$d_{n,a}$ which are no longer homogeneous, but still 
satisfy the bound $\mathrm{dim}(d_{n,a}(A_{1}\otimes\cdots ))  
         \leq \sum_i \mathrm{dim}(A_{i}) -|a|-d(n-1)$.
This change of the axioms causes only little and 
obvious modifications of the applications given in Sect.~5.

\begin{proof}
  By the field independence $D_{H\,n}^{(m)}$ admits a Taylor expansion 
  in $\varphi$ where the coefficients are vacuum expectation values of 
  functional derivatives of its entries (analogously to (\ref{causWick})). 
  Because of locality (\ref{locality of D})
  the coefficients are supported on the total diagonal and are 
  thus derivatives of the $\delta$-distribution in the relative 
  coordinates. Smoothness in $m$ implies the existence of a 
  Taylor expansion in $m$ around $m=0$. By integrating out the 
$\delta$-distribution, reordering the sums and partial integration
we can write $D_{H\,n}^{(m)}$ in the form (\ref{D(Wh)}) with 
$d_{n,l,a}(A_1\ldots)\in {\cal P}_{\text{bal}}$. Since  $D_{H\,n}^{(m)}$
is symmetric and multi-linear in the fields $A_1,\ldots ,A_n$,
the $d_{n,l,a}$'s must satisfy the corresponding properties.

The homogeneous scaling (\ref{D:scaling})
implies $d_{n,l,a}(A_1\ldots)\in {\cal P}_{\text{hom}}$ for 
$A_1,\ldots\in {\cal P}_{\text{hom}}$;
and by using (\ref{scal:hom}),
which can equivalently be written as
 \begin{equation}
\sigma_\rho^{-1}\> A(h)=\rho^{\mathrm{dim}(A)}\> A(h^{(\rho)})\quad
\mathrm{with}\quad h^{(\rho)}(x)\=d \rho^{-d}\> h(\rho^{-1}x)\ ,
 \end{equation}
we obtain
 \begin{gather}
\sigma_\rho\> D_{H\,n}^{(\rho^{-1}m)}(\sigma_\rho^{-1}\>A_{1}(h_{1})\otimes\cdots )=
\rho^{\sum_i\mathrm{dim}(A_i)}\sigma_\rho\> D_{H\,n}^{(\rho^{-1}m)}
(A_{1}(h_{1}^{(\rho)})\otimes\cdots )\notag\\
=\rho^{\sum_i\mathrm{dim}(A_i)}\sum_{l,a}\Bigl(\frac{m}{\rho}\Bigr)^l
\int dx\>(\sigma_\rho\> d_{n,l,a}(A_{1}\otimes\cdots )(x))
(\prod_i\d^{a_i}h_{i}^{(\rho)})(x)\notag\\
=\sum_{l,a}\rho^{\sum_i\mathrm{dim}(A_i)-l+d-\mathrm{dim}(d_{n,l,a}(\ldots))-dn-|a|}
m^l\int dy\>d_{n,l,a}(\ldots)(y)(\prod_i\d^{a_i}h_{i})(y)\ ,
\end{gather}
where we have set $y\equiv \rho^{-1}x$. Since this expression agrees with
$D_{H\,n}^{(m)}(A_{1}(h_{1})\otimes\cdots )$ the exponent of $\rho$ must vanish;
this yields (\ref{mass dimension}).
\end{proof}

\noindent {\it Remark:} It is instructive to formulate the Theorem for the 
$S$-matrix as the generating functional of the time 
ordered products:
${\bf S}(\lambda S)=T(e_\otimes^{i\lambda S})$.
Let $R$, $\hat R$ and $D$ 
as given in the Theorem. Then the
corresponding time ordered products $T$ and 
$\hat T$ according to (\ref{ret-prod}) and the 
associated S-matrices are related by 
\begin{equation}
  {\bf \hat S}(\lambda S)\equiv \hat T(e_\otimes^{i\lambda S})=
  T\Bigl(e_\otimes^{iD(e_\otimes^{\lambda S})}
\Bigr)\equiv {\bf S}(D(e_\otimes^{\lambda S}))\ .
\end{equation}
\medskip 

By linearity the map $D$ of Theorem \ref{main} can be extended to formal 
power series, i.e.~to a map $T({\cal F}_\mathrm{loc})[[\lambda]]\rightarrow
 {\cal F}_\mathrm{loc}[[\lambda]]$. Let $S(\lambda)\in
{\cal F}_\mathrm{loc}[[\lambda]]$ with $S(0)=0$. The renormalization 
of the interaction can be considered as a bijective analytic map
\begin{equation}
  S(\lambda)\longrightarrow Z(S(\lambda))\=d 
D(e_\otimes^{S(\lambda)})\ .\label{Z}
\end{equation}
(When the $R^{(m,\mu)}$-products are meant we 
write $Z_H^{(m)}(\cdot):=D_H^{(m)}(e_\otimes^\cdot)$.)
With that admissible renormalizations of the interaction
can be composed and give again an admissible renormalization. In detail, 
let $Z_{l,j}(\cdot )=D_{l,j}(e_\otimes^\cdot)$, $ (l,j)=(1,2),(2,3)$ be given,
with $D_{l,j}$ satisfying the properties of the Theorem. 
Moreover let ${\bf S}_1$
be an $S$-matrix fulfilling the axioms. 
Then, ${\bf S}_2(S)\=d$ ${\bf S}_1(Z_{1,2}(S))$
and ${\bf S}_3(S)\=d {\bf S}_2(Z_{2,3}(S))=
{\bf S}_1(Z_{1,2}(Z_{2,3}(S)))$ satisfy 
also the axioms (due to part (iv) of the Theorem) and, hence, $Z_{1,2}\circ
Z_{2,3}$ is a renormalization of the interaction with the 
properties given in parts (i)-(iii) of the Theorem. So we infer that 
the non-uniqueness in the construction of retarded products 
is governed by a group, which we may call the 
{\it 'Stueckelberg-Petermann renormalization group'} $\mathcal{R}$. It is the 
group of all analytic bijections $Z$ of the space of formal 
power series $S(\lambda)$ with values in $\mathcal{F}_{\text{loc}}$ and 
with $S(0)=0$ (thus $Z(0)=0$), which satisfy the conditions
  \begin{enumerate}
  	\item  $Z$ preserves the first order term, i.e.
  	\begin{displaymath}
  		\frac{d}{d\lambda}Z(S(\lambda))|_{\lambda=0} 
  		=\frac{d}{d\lambda}S(\lambda)|_{\lambda =0} \ .
  	\end{displaymath}
  	\item  $Z$ is real, i.e. $Z(S(\lambda)^{*})=Z(S(\lambda))^{*}$.
  
  	\item  $Z$ is local: it preserves the localization region,
                \begin{displaymath}
                \supp\frac{\delta\, Z(S(\lambda))}{\delta\varphi}=
                \supp \frac{\delta\, S(\lambda)}{\delta\varphi}\ ,
  	\end{displaymath}
               and it is additive on sums of terms with disjoint localizations, 
  	\begin{displaymath}
  		Z(S_{1}(\lambda)+S_{2}(\lambda))= Z(S_{1}(\lambda))+ 
  		Z(S_{2}(\lambda)) 
  	\end{displaymath}
  	if $\supp\frac{\delta\, S_1(\lambda)}{\delta\varphi}\>\cap\>
                \supp\frac{\delta\, S_2(\lambda)}{\delta\varphi}=\emptyset$.
  
  	\item  $Z$ is Poincar\'{e} invariant.
  	 
  	\item $Z$ does not explicitely depend on $\varphi$.

  	\item  $Z$ acts trivially on $\varphi$ in the sense that
  	\begin{displaymath}
  	Z(S(\lambda)+\lambda\varphi(h))= Z(S(\lambda))+\lambda\varphi(h) \ .
                \end{displaymath}
	\item In the preceding conditions it is not assumed that the retarded 
                products satisfy the axioms Smoothness and Scaling. If the latter are 
                included, $Z$ is replaced by a family $Z_H=(Z_H^{(m)})_{m\geq 0}$,
                where each component $Z_H^{(m)}$ must fulfill (i)-(vi) and
                additionally it must hold
                \begin{displaymath}
                Z_H^{(\rho m)}=\sigma_{\rho}\circ Z_H^{(m)}\circ \sigma_{\rho^{-1}}  
  	\end{displaymath}
                and that $Z_H^{(m)}$ is smooth in $m\geq 0$.  
  \end{enumerate}
In terms of $Z$ and its first derivative $Z^\prime$ the transformation 
formula (\ref{ren:intf}) of the retarded products reads
\begin{equation}
\hat{F}_S=(Z^\prime (S)F)_{Z(S)}\label{ren:intf:Z}
\end{equation}
where
\begin{equation}
Z^\prime (S)F:=\frac{d}{d\tau}Z(S+\tau F)\vert_{\tau=0}\ .\label{Zprime}
\end{equation}
\section{The algebraic adiabatic limit and 
the renormalization group}\setcounter{equation}{0}
Of particular interest are certain factor groups of the 
Stueckelberg-Petermann renormalization group $\mathcal{R}$
introduced in the preceding section.      
Up to now our renormalization group transformations act on 
explicitely spacetime dependent interaction Lagrangians 
$g=\sum g_{i}\mathcal{L}_{i}\in 
\mathcal{D}(\MM,\mathcal{P}_{\text{bal}})$. We want to 
extract from this information the action of the renormalization group 
on constant Lagrangians $\mathcal{L}\in \mathcal{P}_{\text{bal}}$.
This requires a kind of adiabatic limit. The usual
adiabatic limit $g\rightarrow k$ (where $k$ 
is constant) needs a good infrared behavior (see
e.g. \cite{EG,EG1,BlSe,Scharf}). 
We therefore work in the algebraic adiabatic limit.

To explain the {\bf algebraic adiabatic limit} let ${\cal O}\subset \MM$ 
be a causally closed, open region. 
Let $S\in\mathcal{F}_{\text{loc}}$. Due to Proposition 
\ref{prop:AWI} there is a unique function $g\in 
\mathcal{D}(\MM,\mathcal{P}_{\text{bal}})$ with $\int g=S$;
we may therefore write $A(h)_g$ instead of $A(h)_S$ for the 
interacting fields and $Z(g)$, $Z^\prime (g)$ for
$Z(\int g)$ (\ref{Z}) and $Z^\prime (\int g)$ (\ref{Zprime}). 
We introduce the algebra ${\cal A}_{g}
({\cal O})$ of interacting fields belonging to ${\cal O}$
as the sub-algebra of ${\cal A}^{(m)}$ which is generated by 
the interacting fields $A(h)_{g}$ with $A\in {\cal 
P}$ and $h\in {\cal D}({\cal O})$.
Note that ${\cal A}_{g}({\cal O})$ depends on
the chosen retarded products.
For $\mathcal{L}\in \mathcal{P}_{\text{bal}}$ we define
\begin{eqnarray}
  {\cal G}_{\mathcal{L}}({\cal O})
  &\=d& \{g\in \mathcal{D}(\MM,\mathcal{P}_{\text{bal}}) \> |\> \notag\\
&&g(x)=\mathcal{L} \text{ for all }  x \text{ in a neighborhood 
of the closure of } {\cal O}\}\ .\notag
\end{eqnarray}
The algebraic adiabatic limit
relies on the following observation \cite{BF}\footnote{An alternative
proof of (\ref{adlim}) which is based on our axioms for 
retarded products (Sect.~2) is given in \cite{DF1a}. It deals with on-shell
valued retarded products; however it applies also to our ${\cal F}$-valued
retarded products because it uses the Symmetry, Causality and GLZ
relation only.}:
for any $g_1,g_2\in {\cal G}_{\mathcal{L}}({\cal O})$ there 
exists a set $\mathrm{Aut}_{g_1,g_2}$ of automorphisms $\alpha$ 
of ${\cal A}^{(m)}[[\lambda]]$ with
\begin{equation}
  \alpha \bigl(A(h)_{g_{1}}\bigr)=A(h)_{g_2}\quad
\forall \alpha\in \mathrm{Aut}_{g_1,g_2},\>\>
\forall h\in {\cal D}({\cal O}),\>\>\forall A\in {\cal P}\ .\label{adlim}
\end{equation}
This has the important consequence that the {\it algebraic structure 
of} ${\cal A}_{g}({\cal O})$ {\it is independent 
of the choice of} $g\in {\cal G}_{\mathcal{L}}({\cal O})$. 

   Following \cite{BF,DF1} we may formalize the 
   algebraic adiabatic limit in the following way: Consider the bundle 
   of algebras ${\cal A}_{g}(\cal O)$ over the space of 
   compactly supported Lagrangians 
   $g\in\mathcal{G}_{\mathcal{L}}(\mathcal{O})$ where  
   $\mathcal{L}\in\mathcal{P}_{\mathrm{bal}}$ is a constant Lagrangian.

   A section $B=(B_{g})$ is called covariantly constant 
   if it holds
   $$\alpha (B_{g_1})=B_{g_2}\quad
   \forall\alpha\in \mathrm{Aut}_{g_1,g_2}\ .$$
   In particular the interacting fields are covariantly constant sections.
   The local algebra $\mathcal{A}_{\mathcal{L}}(\mathcal{O})$ is now 
   defined as the algebra of covariantly constant sections of the bundle 
   introduced above. 
 
   To get a net of local algebras in the sense of the Haag-Kastler axioms 
   \cite{Haag} one has in addition to fix the embeddings 
   into algebras of larger regions.
   Let $\mathcal{O}_1\subset\mathcal{O}_2$. Then we define the 
   embeddings of algebras
    $$ i_{{\cal O}_2{\cal O}_1}: 
    {\cal A}_{\cal L}({\cal O}_1) \to {\cal A}_{\cal L}({\cal O}_2)$$
   by restricting a section $B$ to Lagrangians 
    $g\in\mathcal{G}_{\mathcal{L}}(\mathcal{O}_2)$.
 
   It is easy to see that these embeddings satisfy the consistency condition
   \begin{displaymath}
     i_{{\cal O}_3{\cal O}_2}\circ i_{{\cal O}_2{\cal O}_1} =
     i_{{\cal O}_3{\cal O}_1} 
   \end{displaymath}
   for $\mathcal{O}_1\subset\mathcal{O}_2\subset\mathcal{O}_3$.
   Moreover, the net is covariant under Poincar\'{e} transformations 
   provided the Lagrangian $\mathcal{L}$ is Lorentz invariant 
   \cite{DF1}. It also 
   satisfies the condition of local commutativity as a consequence of the
   conditions GLZ relation and Causality.

   We may also look for local fields associated to the net. By definition,    
   a local field associated to the net is a family of distributions
   $(A_{\cal O})$ with values in ${\cal A}_{\cal L}(\cal O)$ 
   such that
   $$i_{{\cal O}_2{\cal O}_1}(A_{{\cal O}_1}(h))= A_{{\cal O}_2}(h)$$
   if $\supp\  h \subset \mathcal{O}_1$ and which transform
   covariantly under Poincar\'e transformations \cite{BFV}.
   Examples for local fields are given in terms of the classical fields 
   $A\in\mathcal{P}$ by the sections                            
   $$(A_{\cal O}(h))_{g}:= (A(h))_{g}\ , \ 
     g\in\mathcal{G}_{\mathcal{L}}(\mathcal{O})\ .$$
   It is an open question whether there are other local fields. This 
   amounts to a determination of the Borchers class for perturbatively 
   defined interacting field theories.
   \medskip

We are now going to investigate what happens with the 
Stueckelberg-Petermann renormalization group $\mathcal{R}$
in the algebraic adiabatic limit. For this purpose we insert a
$g\in \mathcal{G}_{\mathcal{L}}(\mathcal{O})$ into (\ref{D(Wh)})
and find
\begin{equation}
Z_H(\lambda g)=D_H(e_\otimes^{\lambda\int g})\in 
\mathcal{G}_{z(\lambda\mathcal{L})}(\mathcal{O})\label{eq:littlez}
\end{equation}
with
\begin{equation}
     z(A)=\sum_{n,l}\frac{1}{n!} m^l d_{n,l,0}(A^{\otimes n})\ ,
\quad A\in \lambda\mathcal{P}_{\mathrm{bal}}[[\lambda]]\ .
\label{eq:littlezex}
   \end{equation}
(The terms $a\not= 0$ with derivatives of the test functions in
 (\ref{D(Wh)}) do not contribute to $z$ because $g\vert_{\cal O}=$
constant.) Hence,
   the renormalization group transformations $Z_H\in\mathcal{R}$ induce 
   transformations 
\begin{equation}
z:\lambda\mathcal{P}_{\mathrm{bal}}[[\lambda]]
\longrightarrow \lambda\mathcal{P}_{\mathrm{bal}}[[\lambda]]\> :
\> z(\lambda {\cal L})=\lambda {\cal L}+{\cal O}(\lambda^2).
\label{z:map}
\end{equation}
Since $Z_H$ is invertible this holds also for $z$.          
The map $\gamma:Z_H\mapsto z$ is a homomorphism of  
$\mathcal{R}$ to the {\it renormalization group in 
   the adiabatic limit} $\mathcal{R}_\mathrm{adlim}:=\gamma ({\cal R})$.   
As one might expect, the kernel of this 
   homomorphism acts trivially on the local nets. Actually, this holds
   already in case the given Lagrangian is left invariant, in detail:
   \begin{thm}
      Let ${\cal A}_{{\cal L}}$ and 
      $\hat{{\cal A}}_{{\cal L}}$ be 
      two local nets
      which are defined by renormalization prescriptions 
      $R^{(m,\mu)}$ and $\hat R^{(m,\mu)}$ which are related by a renormalization 
      group transformation $Z_H$ (\ref{Z}) such that 
      $\gamma(Z_H)(\lambda {\cal L})=\lambda {\cal L}$. Then the nets are equivalent, i.e. there
exist isomorphisms $\beta_{\cal O}:{\cal A}_{{\cal L}}({\cal O})\rightarrow
\hat{{\cal A}}_{{\cal L}}({\cal O})$ with $\beta_{{\cal O}_2}\circ  i_{{\cal O}_2{\cal O}_1}=
 \hat{i}_{{\cal O}_2{\cal O}_1}\circ\beta_{{\cal O}_1}$.           
   \end{thm}
   \begin{proof}  
   Since $\hat{F}_{g}=(Z_H'(g)F)_{Z_H(g)}$ (\ref{ren:intf:Z})
   where $Z_H'(g)$ is an invertible linear transformation on the 
   space of local functionals, the algebras
   $\hat{{\cal A}}_{g}({\cal O})$ and 
   ${\cal A}_{Z_H(g)}({\cal O})$ coincide.
   In addition note that $g\in {\cal G}_{\mathcal{L}}({\cal O})$
   is equivalent to $Z_H(g)\in {\cal G}_{\mathcal{L}}({\cal O})$.
   We are now going to show that a section $B$ in 
   ${\cal A}_{\mathcal{L}}({\cal O})$ 
   can be mapped to a section $\beta_{\cal O}(B)$ in 
   $\hat{{\cal A}}_{\mathcal{L}}({\cal O})$ by
   $$\beta_{\cal O}(B)_{g}=B_{Z_H(g)}\ .$$
   Obviously, this map is an isomorphism of the algebras 
   of sections. It remains to 
   prove, that $B$ is covariantly constant if and only if 
   $\beta_{\cal O}(B)$ is. But this follows from the fact that 
   $Z_H'(g)F=D_H(e_\otimes^{\int g}\otimes F)$ is independent 
   of the choice of $g\in {\cal G}_{\mathcal{L}}({\cal O})$ 
   if $F$ is localized within ${\cal O}$.   
   Therefore the conditions on the intertwining isomorphisms 
   $\alpha$ are identical: $\widehat{\rm Aut}_{g_1,g_2}({\cal O})=
   {\rm Aut}_{Z_H(g_1),Z_H(g_2)}({\cal O})$.
 \end{proof} 

Due to this Theorem, $\gamma(Z_H)=z$ is the {\bf renormalization 
of the interaction in the algebraic adiabatic limit}. From
(\ref{mass dimension}) and (\ref{eq:littlezex}) we immediately find 
\begin{equation}
\mathrm{dim}({\cal L})\leq d\quad\Rightarrow\quad
\mathrm{dim}(z(\lambda {\cal L}))\leq d\ .\label{z:dim}
\end{equation}
\medskip

\noindent {\it Example:} 
{\bf renormalization of ${\cal L}=\lambda\varphi^4$ in $d=4$ dimensions}.  
In case $R^{(m,\mu)}$ and $\hat R^{(m,\mu)}$ respect the field parity
(\ref{parity}), this holds also for the corresponding map $D_H$ of 
Theorem \ref{main} and hence for $z$ (\ref{eq:littlezex}).
Taking this, Lorentz covariance, Unitarity, (\ref{z:map}) 
and (\ref{z:dim}) into account we obtain\footnote{According  
to (\ref{bal-field}) there is (up to multiplication with a constant)
precisely one Lorentz invariant balanced field with two factors
$\varphi$ and two derivatives, namely $\d^{12\mu}\d^{12}_\mu
\varphi(x_{1})\varphi(x_{2})|_{x_{1}=x_{2}=x}=2(\varphi\square\varphi - 
(\d^\mu\varphi)\d_\mu\varphi)(x)$.}
\begin{equation}
z(\lambda\varphi^4)=\lambda\bigl[(1+a) \varphi^4 + 
b ((\partial \varphi)^2-\varphi\square\varphi) + 
m^2 c \varphi^2 +m^4 e{\bf 1}\bigr]\ ,\label{phi^4} 
\end{equation}
where $a,b,c,e\in\lambda\RR[[\lambda]]$. The term $m^4 e{\bf 1}$
is irrelevant because $R^{(m,\mu)}_{n,1}(\ldots {\bf 1}\ldots)=0$ for $n\geq 1$
(see e.g. \cite{DF}, Lemma 1(C)). $a$ describes 
coupling constant renormalization, $b$ and 
$c$ wave function and mass renormalization. As usual, the 
latter renormalizations can be absorbed in a redefinition of 
the free theory, so that there is only one free parameter left.   
\medskip

\noindent {\it Remark:} The invariance of the Lagrangian is sufficient but not 
   necessary for the invariance of the net (in the 
   example (\ref{phi^4}) the parameter $e$ is irrelevant for the 
   structure of the net). We plan to  determine 
   the corresponding subgroup of $\mathcal{R}_\mathrm{adlim}$ \cite{BDF}.   
\medskip

The {\bf field renormalization in the algebraic adiabatic limit}
is a map
\begin{equation}
z^{(1)}:\lambda\mathcal{P}_{\mathrm{bal}}[[\lambda]]\times
\mathcal{P}_{\mathrm{bal}}
\longrightarrow \mathcal{P}[[\lambda]]\>:\>
(\lambda {\cal L},A)\mapsto z^{(1)}(\lambda {\cal L})A
=A+{\cal O}(\lambda)
\end{equation}
which is determined by the requirement that the transformation 
formula (\ref{ren:intf}) or (\ref{ren:intf:Z}) takes the simpler form
\begin{equation}
(z^{(1)}(\lambda {\cal L})A)(h)_{z(\lambda{\cal L})}=
\hat{A}(h)_{\lambda{\cal L}}\ ,\quad\forall h\in {\cal D},\>
\forall A,{\cal L}\in \mathcal{P}_{\mathrm{bal}}\ ,\label{z^1}
\end{equation}
in the algebraic adiabatic limit.
This condition has a unique solution: in the formula (\ref{D(Wh)}) 
for $D_{H\,n}((\lambda\int g)^{\otimes n}\otimes A(h))$ we specialize to  
$g\in {\cal G}_{\mathcal{L}}({\cal O})$ and $h\in {\cal D}({\cal O})$
and perform partial integrations. This yields
\begin{equation}
z^{(1)}(\lambda {\cal L})A=\sum_{n,l,a\in\NN_{0}^{d}}\frac{1}{n!} 
m^{l}(-1)^{|a|}\d^ad_{n+1,l,(0,\ldots,0,a)}
((\lambda {\cal L})^{\otimes n}\otimes A)\ .\label{z^1explicit}
\end{equation}
Hence, the algebraic adiabatic limit simplifies the field 
renormalization $A(h)\mapsto D_H(e_\otimes^{\lambda\int g}
\otimes A(h))$ to the {\bf linear} map
$\mathcal{P}_{\mathrm{bal}}\ni A\mapsto z_1(\lambda {\cal L})A
\in \mathcal{P}[[\lambda]]$, i.e. the test function remains 
unchanged. By using the definition
\begin{equation}
z^{(1)}(\lambda {\cal L})\d^aA:=\d^az^{(1)}(\lambda {\cal L})A\ ,\quad
A\in \mathcal{P}_{\mathrm{bal}}\ ,\label{z^1d}
\end{equation}
and linearity we extend $z^{(1)}$ to a map
$\mathcal{P}_{\mathrm{bal}}[[\lambda]]\times
\mathcal{P}\rightarrow \mathcal{P}[[\lambda]]$ and with that
the relation (\ref{z^1}) holds even for all $A\in {\cal P}$. From
(\ref{mass dimension}) and (\ref{z^1explicit}) we conclude
\begin{equation}
\mathrm{dim}({\cal L})\leq d\quad\Rightarrow\quad
\mathrm{dim}(z^{(1)}(\lambda {\cal L})A)= \mathrm{dim}(A)\label{z^1:dim}
\end{equation}
(where $d$ is the number of space time dimensions).
In a {\it massless} model with ${\cal L}\in {\cal P}_d$ 
(\ref{grad(P)}) we find
\begin{equation}
  z(\lambda {\cal L})\in {\cal P}_d\quad\mathrm{and}
\quad A\in {\cal P}_j\Rightarrow z^{(1)}(\lambda {\cal L})A
\in {\cal P}_j\ ,\label{dim:m=0}
\end{equation}
due to Remark (1) after Proposition \ref{prop:main}. We point out that
$z$ and $z^{(1)}$ are uniquely determined by $R^{(m,\mu)}$ and $\hat{R}^{(m,\mu)}$ and
that they are universal, i.e. independent of ${\cal O}$.
\medskip

\noindent {\it Example:} {\bf renormalization of $\varphi$ and 
$\varphi^2$ in a massless model with ${\cal L}\in {\cal P}_d$}. In case 
the retarded products satisfy the Field equation, the renormalization 
of $\varphi$ is the identity, due to part (ii) of Theorem \ref{main}.
Here we do not use this assumption, instead we work out
some consequences of (\ref{dim:m=0}). Since
${\cal P}_1=\{a\varphi | a\in\RR\}$
the renormalization of $\varphi$ has the simple form
\begin{equation}
  \varphi\longrightarrow z^{(1)}_\varphi \varphi\ \ ,\quad 
z^{(1)}_\varphi \in\RR[[\lambda]]\ .\label{Z_phi}
\end{equation}
For $A=\varphi^2$ the renormalized field is of the form
\begin{equation}
  z^{(1)}(\lambda {\cal L})\varphi^2=z^{(1)}_0\varphi^2 + 
z^{(1)}_{1\mu}\d^\mu\varphi\ , \quad\quad 
z^{(1)}_0,\>z^{(1)}_{1\mu}\in\RR[[\lambda]]\ .\label{Z_phi^2}
\end{equation}
In case ${\cal L}$, $R^{(m,\mu)}$ and $\hat{R}^{(m,\mu)}$ 
are Lorentz covariant, the right side
of (\ref{Z_phi^2}) must also have this symmetry, i.e. $z^{(1)}_{1\mu}=0$.
(Alternatively, the latter follows also if ${\cal L}$ is even in $\varphi$ and
$R^{(m,\mu)}$ and $\hat{R}^{(m,\mu)}$ preserve the field parity (\ref{parity}).)
With that the renormalization of $\varphi^2$ is 'diagonal', too. Usually 
$z^{(1)}_0$ (\ref{Z_phi^2}) is non-trivial  and it cannot be
related to $(z^{(1)}_\varphi)^2$ (\ref{Z_phi}).
\medskip

{\bf Scaling transformations:} in Sect.~3 we have shown 
that one can fulfill almost homogeneous
scaling of $R^{(m,\mu)}_{n,1}$ for all $m\geq 0$.
The results of this
Section yield far reaching additional information about
the connection of  $R^{(m,\mu)}_{\rho}$ 
(\ref{R^m,mu:scaled}) and $R^{(m,\mu)}$ (cf. \cite{HW} and
\cite{G:scal}). The basic observation is the following: if $R^{(m,\mu)}$
fulfills the axioms given in Sect.~2 (Lorentz covariance,
global inner symmetries and the Field
equation may be excluded or included), then the same axioms hold true
for $R^{(m,\mu)}_{\rho}$, too, as can be verified straightforwardly.
Therefore, there exists a sequence $D^{(m)}_{H\>\rho}$ (\ref{D}) with the 
properties mentioned in the Main Theorem. So, the
{\it scaling transformations on a given renormalization 
   prescription induce a one parameter subgroup of the renormalization 
   group} ${\cal R}_\mathrm{adlim}$, which may be 
called {\it 'Gell-Mann-Low Renormalization Group'}. 

With the scaling as renormalization transformation we are now going to compute 
to lowest non-trivial order the renormalization of the fields $\varphi^2,\> \varphi^3,\> 
\varphi^4$ and of the interaction $\lambda {\cal L}$ for ${\cal L}=\varphi^4$
(in $d=4$ dimensions) and $m=0$. We assume that 
$R\equiv R^{(0)}\equiv R^{(0,\mu)}$ fulfills all axioms of Sect.~2.

{\bf Renormalization of $\varphi^2$:} by (\ref{R(D)})
and the expansion (\ref{causWick})-(\ref{causWick1}) we obtain
\begin{gather}
   D_{\rho}(\varphi^4(x_1)\otimes\varphi^2(x))=
R_\rho (\varphi^4(x_1),\varphi^2(x))-
R(\varphi^4(x_1),\varphi^2(x))\notag\\
=6\bigl[\rho^4 r(\varphi^2,\varphi^2)(\rho (x_1-x))
-r(\varphi^2,\varphi^2)(x_1-x)\bigr]
\varphi^2(x_1)\notag\\
=\frac{6}{(2\pi)^2}\>\mathrm{log}\,\rho\>\>\delta(x_1-x)\>\varphi^2(x_1)
\ ,\label{D(4,2)}
\end{gather}
where we use a symbolic notation for $D_\rho$ and 
the result (\ref{r:scaling}) of 
Appendix C. The tree diagram of $R(\varphi^4,\varphi^2)$
does not contribute, because
it scales homogeneously. Going over to the algebraic adiabatic limit the 
formula (\ref{z^1explicit}) yields
\begin{equation}
  z^{(1)}_\rho (\lambda\varphi^4)\varphi^2=
\Bigl(1+\lambda\frac{6}{(2\pi)^2}\>\mathrm{log}\,\rho\>+\>{\cal O}(\lambda^2)
\Bigr)\varphi^2\ ,\label{d^1_2}
\end{equation}
independently of the normalization of the fish diagram
$r(\varphi^2,\varphi^2)$.
 
{\bf Renormalization of $\varphi^3$:} analogously to (\ref{D(4,2)}) we get
\begin{gather}
   D_{\rho}(\varphi^4(x_1)\otimes\varphi^3(x))=\notag\\
18\bigl[\rho^4 r(\varphi^2,\varphi^2)(\rho (x_1-x))-
r(\varphi^2,\varphi^2)(x_1-x)\bigr]
\varphi (x)\varphi^2(x_1)\notag\\
+4[\rho^6 r(\varphi^3,\varphi^3)(\rho (x_1-x))-
r(\varphi^3,\varphi^3)(x_1-x)]\varphi(x_1)=\notag\\
+\frac{18}{(2\pi)^2}\>\mathrm{log}\,\rho\>\delta(x_1-x)\varphi^3(x_1)\notag\\
+\frac{3}{2(2\pi)^4}\>(\mathrm{log}\,\rho)\>\square\delta (x_1-x)
\varphi(x_1)\ ,\label{Z_phi^3}
\end{gather}
where we have inserted the results (\ref{r:scaling}) and
(\ref{r_0^1:scal}) of Appendix C. In
the algebraic adiabatic limit this gives
\begin{eqnarray}
z^{(1)}_\rho (\lambda\varphi^4)\varphi^3
& = & \Bigl(1+\lambda\frac{18}{(2\pi)^2}\>\mathrm{log}\,\rho
+{\cal O}(\lambda^2)\Bigr)\>\varphi^3\notag\\
& + & \Bigl(\lambda\frac{3}{2(2\pi)^4}\>\mathrm{log}\,\rho
+{\cal O}(\lambda^2)\Bigr)\>\square\varphi    \label{Zphi^3}
\end{eqnarray}
by means of (\ref{z^1explicit}).
Other terms than $\varphi^3$ and $\square\varphi$ do not appear
to higher orders either, due to (\ref{dim:m=0}), the maintenance
of the field parity and Lorentz invariance. So, the field renormalization of 
$\varphi^3$ is non-diagonal. However, since $z^{(1)}_\rho (\lambda\varphi^4)
\square\varphi =\square\varphi$ (due to the validity of the Field equation and 
(\ref{z^1d})), the field $(\varphi^3+\frac{1}{12\,(2\pi)^2}\square\varphi)$ is an eigenvector of 
$z^{(1)}_\rho (\lambda\varphi^4)$ with eigenvalue $\bigl(1+
\lambda\frac{18}{(2\pi)^2}\>\mathrm{log}\,\rho +{\cal O}(\lambda^2)\bigr)$.

{\bf Renormalization of ${\cal L}=\varphi^4$:}
we continue the Example (\ref{phi^4})
for the particular case of the scaling transformations and $m=0$. 
Since the corresponding tree diagrams scales homogeneously we obtain
\begin{gather}
 D_{\rho}(\varphi^4(x_1)\otimes\varphi^4(x))=\notag\\
36\Bigl\{\rho^4 r(\varphi^2,\varphi^2)(\rho (x_1-x))-
r(\varphi^2,\varphi^2)(x_1-x)
\Bigr\}\varphi^2(x_1)\varphi^2(x)\notag\\
+16\Bigl\{\rho^6 r(\varphi^3,\varphi^3)(\rho (x_1-x))-
r(\varphi^3,\varphi^3)(x_1-x)\Bigr\}\varphi(x_1)\varphi(x)\notag\\
+\Bigl\{\rho^8 r(\varphi^4,\varphi^4)(\rho (x_1-x))-
r(\varphi^4,\varphi^4)(x_1-x)\Bigr\}{\bf 1}\notag\\
=\frac{36}{(2\pi)^2}\>(\mathrm{log}\,\rho)\>\delta (x_1-x)
\varphi^2(x_1)\varphi^2(x)\notag\\
+\frac{6}{(2\pi)^4}\>(\mathrm{log}\,\rho)\>(\square\delta) (x_1-x)\>
\varphi(x_1)\varphi(x)+\ldots \square\square\delta (x_1-x){\bf 1}   \label{R(rho)-R}
\end{gather}
by using (\ref{r:scaling}) and (\ref{r_0^1:scal}); the form of the last term follows
from Lorentz invariance and that it must scale homogeneously with degree $8$.
In the algebraic adiabatic limit we get
the following renormalization of the interaction:
\begin{eqnarray}
z_\rho (\lambda\varphi^4) &=& \Bigl(\lambda+\lambda^2\frac{18}{(2\pi)^2}\>
\mathrm{log}\,\rho +{\cal O}(\lambda^3)\Bigr)\varphi^4\notag\\
&+& \Bigl(-\lambda^2\frac{3}{2(2\pi)^4}\>\mathrm{log}\,\rho
+{\cal O}(\lambda^3)\Bigr)\>\bigl((\d^\mu\varphi)\d_\mu\varphi\ 
-\varphi\square\varphi\bigr),\label{kren-2}
\end{eqnarray}
where we apply (\ref{eq:littlezex}) and take into account
that $z$ takes values in $\mathcal{P}_\mathrm{bal}$.
Due to (\ref{dim:m=0}) there is no mass renormalization, i.e.~the constant 
$c$ in (\ref{phi^4}) vanishes to all orders.

{\bf Field Renormalization of $\varphi^4$:}
from $D_{\rho}(\varphi^4\otimes\varphi^4)$ (\ref{R(rho)-R}) we
can also read off the field renormalization of $\varphi^4$ to first order in $\lambda$:
\begin{equation}
  z^{(1)}_\rho (\lambda\varphi^4)\varphi^4
  =\varphi^4 +\lambda\Bigl(\frac{36}{(2\pi)^2}\mathrm{log}\>\rho\>\varphi^4
  +\frac{6}{(2\pi)^4}\mathrm{log}\>\rho\>\varphi\square\varphi\Bigr)
+{\cal O}(\lambda^2) \ .
\end{equation}
\section{Outlook}\setcounter{equation}{0}
The construction of a renormalized perturbative quantum field theory, in the sense of 
algebraic quantum field theory \cite{Haag}, 
was carried through without ever meeting infrared 
problems. In particular, the renormalization group (in the sense of Stueckelberg 
and Petermann) could be constructed in purely local terms. This in variance with 
standard techniques of perturbation theory which typically rely on global properties.

Given the algebra of interacting fields, one may then, in a second step, look for 
states of interest, for instance vacuum or particle states. This amounts
to perform the adiabatic limit in the conventional sense and was done for massive 
theories by Epstein and Glaser \cite{EG,EG1} and, on the basis of retarded 
products, by Steinmann \cite{Ste}. (For QED see \cite{BlSe} for the construction of 
the vacuum state and \cite{Ste1} for the analysis of scattering.) One then may relate 
the global renormalization parameters, as masses and coupling constants at 
e.g. zero momentum, to the local parameters involved in our construction.
One may also look for other situations, for instance at finite temperature or with 
non-trivial boundary conditions. Then, other global parameters are of interest, but the 
local parameters remain the same.

On a generic curved space-time it seems that a 
{\it completely local procedure} is by far the best way to construct 
perturbative quantum fields; in particular the large ambiguity 
of renormalization in theories without translation invariance has recently been 
removed (up to few parameters) by requiring the generally covariant locality 
principle \cite{HW1-2,HW,BFV}.

In connection with the renormalization group the following
topics will be studied in a subsequent paper \cite{BDF}.
\begin{itemize}
\item The map $D$ of the Main Theorem (Theorem \ref{main}) scales 
homogeneously for the {\it modified} interacting fields only. For this reason
most applications of this Theorem given in Sects.~4 and 5 are
restricted to the modified interacting fields. These results can be translated
into statements about the original interacting fields by the transformation formula
(\ref{R^m,mu}) (which holds also for $(D^{(m)},D^{(m,\mu)})$).
\item The generator of the Gell-Mann-Low Renormalization Group
(i.e.~the subgroup of  ${\cal R}_\mathrm{adlim}$
induced by the scaling transformations on a given renormalization prescription $R$)
is related to the $\beta$ function. The Gell-Mann-Low subgroups belonging to
different renormalization prescriptions $R$ and $\hat{R}$ are 
   conjugate to each other. The generator starts with a term of 
   second order which is universal.
\item The absorption of the $b$- and $c$- term of (\ref{phi^4})
in a redefinition of the free theory (wave function and mass renormalization)
requires that the {\it physical predictions are independent of the splitting 
of the action in a free and an interacting part} (where the free part is always quadratic 
in $\d^a\varphi$ ($a\in\NN_0^d$)). It turns out that the latter is an additional
(re)normalization condition, which is part of the 'Principle of Perturbative Agreement'
required by Hollands and Wald \cite{HW4}.  
\item The scaling transformations are the bridge to Wilson's renormalization 
group. This has to be investigated as well as the connection to the 
   Buchholz-Verch scaling limit.
\item Hollands and Wald made a corresponding analysis for curved space-times 
\cite{HW}. But the present formalism is not yet fully adapted to general 
   Lorentzian space-times.
\end{itemize}

\newpage

\begin{center}
{\Large\bf Appendices}
\end{center}
\begin{appendix}
\section{Mass dependence of the two-point function}
\setcounter{equation}{0}
To investigate the mass dependence of the two-point function 
$\Delta^{+(d)}_m(x_1-x_2)\equiv\omega_0(\varphi(x_1)\star_m\varphi(x_2))$ 
in $d$-dimensions we compute the Fourier transformation\footnote{An
analogous (unpublished) computation of the commutator function by K.-H.~Rehren
and M.~D.~was very helpful for writing essential parts of this Appendix.}
\begin{equation}
  \Delta^{+(d)}_m(y)=\frac{1}{(2\pi)^{d-1}}\int d^dp\,\Theta(p^0)
\delta(p^2-m^2)e^{-ipy}
\label{Delta^+}
\end{equation}
in the sense of distributions. We perform the $p^0$-integration and use
\begin{equation}
\int d^{d-1}\vec{p}\ldots =|S_{d-3}|\int_0^\infty dp\,p^{d-2}
\int_0^\pi d\theta\,(\mathrm{sin}\,\theta)^{d-3}\ldots\ ,
\end{equation}
where
\begin{equation}
|S_k|=\frac{2\pi^{\frac{k+1}{2}}}{\Gamma(\frac{k+1}{2})}
\end{equation}
is the surface of the unit ball in $\RR^{k+1}$. With $y\equiv (t,\vec{y})$,
$r\equiv|\vec{y}|$ this gives 
\begin{equation}
  \Delta^{+(d)}_m(y)=\frac{|S_{d-3}|}{(2\pi)^{d-1}}\int_0^\infty dp\,p^{d-2}
\int_0^\pi d\theta\,(\mathrm{sin}\,\theta)^{d-3}\>
e^{ipr\mathrm{cos}\,\theta}\>\frac{e^{-i\omega t}}{2\omega}
\vert_{\omega=\sqrt{p^2+m^2}}\ .
\label{Delta^+a}
\end{equation}
It is well known that $\Delta^{+(d)}_m$ is the limit of a function 
which is analytic in the forward tube $\RR^d-iV^{(d)}_+$. (This 
is e.g.~a consequence of the Wightman axioms.)
Taking additionally Lorentz covariance into account we conclude that 
$\Delta^{+(d)}_m$ is of the form
\begin{equation}
\Delta^{+(d)}_m(y)=\lim_{\epsilon\to 0}f(y^2-iy^0\epsilon)\ ,
\label{Delta^+a:lim}
\end{equation}
where $f(z)$ is analytic for $z\in\CC\setminus U$ for some $U\subset\RR$.
With that it suffices to compute $\Delta^{+(d)}_m$ for
$y=(t,\vec{0})$, $t>0$. Namely, for $d=3$ we obtain
\begin{gather}
\Delta^{+(3)}_m(t,\vec{0})=\frac{1}{4\pi}\int_0^\infty dp\, p\,
\frac{e^{-i\omega t}}{\omega}
\vert_{\omega=\sqrt{p^2+m^2}}=\frac{i}{4\pi}\frac{\d}{\d t}\int_0^\infty dp\, 
p\,\frac{e^{-i\omega t}}{\omega^2}\notag\\
=\frac{i}{4\pi}\frac{\d}{\d t}\int_{mt}^\infty du\,\frac{e^{-iu}}{u}
=\frac{-i}{4\pi}\,\frac{e^{-imt}}{t}\ .
\end{gather}
Due to (\ref{Delta^+a:lim}) $\Delta^{+(3)}_m(y)$ is obtained for arbitrary
$y$ by replacing $it$ (in the latter formula) by $\sqrt{-(y^2-iy^00)}$.  
This gives
\begin{equation}
\Delta^{+(3)}_m(y)=\frac{1}{4\pi\>\sqrt{-(y^2-iy^00)}}\>
e^{-m\sqrt{-(y^2-iy^00)}}\ .\label{+3}
\end{equation}

Analogously, for $d=4$ one obtains 
\begin{equation}
\Delta^{+(4)}_m(y)=\frac{-1}{4\pi^2(y^2-iy^00)}
+{\rm log}(-m^2(y^2-iy^00))\>m^2\>f(m^2y^2)
+m^2F(m^2y^2)\ ,\label{4+}
\end{equation}
where $f$ and $F$ are analytic functions, see e.g.~Sect.~15.1 of \cite{BS}.
$f$ can be expressed in terms of the Bessel function $J_1$ of order $1$,
namely
\begin{equation}
f(z)\equiv \frac{1}{8\pi^2\sqrt{z}}\>J_1(\sqrt{z})
=\sum_{k=0}^\infty C_k\>z^k\ ,\quad C_k\in\RR\ ;\label{f}
\end{equation}
and $F$ is given by a power series
\begin{equation}
F(z)\equiv -\frac{1}{4\pi}\sum_{k=0}^\infty\{\psi(k+1)+\psi(k+2)\}
\frac{(-z/4)^k}{k!(k+1)!}\ ,
\end{equation}
where the Psi-function is related to the Gamma-function by
$\psi(x)\equiv\Gamma^\prime (x)\, /\, \Gamma (x)$.

We see that $\Delta^{+(3)}_m$ is smooth in $m\geq 0$, but
$\Delta^{+(4)}_m$ is not smooth at $m=0$ (it is only continuously 
differentiable)! However,
\begin{gather}
H^{\mu\, (4)}_m(y)\equiv\Delta^{+(4)}_m(y)-
m^2\>f(m^2y^2)\>{\rm log}(m^2/\mu^2)\label{4H}
\end{gather}
(where $\mu>0$ is a {\bf fixed} mass-parameter) is smooth in $m\geq 0$.
In addition, 
\begin{itemize}
\item $H^{\mu\, (4)}_m(y)-\Delta^{+(4)}_m(y)$ is a smooth function of $y$,
i.e.~the wave front sets of $H^{\mu\, (4)}_m(y)$ and $\Delta^{+(4)}_m(y)$
agree and, hence, $(H^{\mu\, (4)}_m(y))^k,\>k\in\NN$ exists;
\item the antisymmetric part of $H^{\mu\, (4)}_m$ is the same as for
$\Delta^{+(4)}_m$ (namely $=i\Delta^{(4)}_m/2$, where $\Delta^{(d)}_m$ 
is the commutator function);
\item $H^{\mu\, (4)}_m$ is  Poincar\'{e} invariant;
\item $H^{\mu\, (4)}_m$ satisfies 
the Klein-Gordon equation since $(\square_y +m^2)\>f(m^2\,y^2)=0$
(by using Bessel's differential equation);
\item $H^{\mu\, (4)}_m$ does not scale homogeneously:
\begin{equation}
\rho^2\>H^{\mu(4)}_{\rho^{-1}m}(\rho y)-H^{\mu(4)}_m(y)
= {\rm log}(\rho)\>2m^2\>f(m^2y^2)\ ;
\label{H:scaling}
\end{equation}
\item and $H^{\mu\, (4)}_{m=0}=\Delta^{+(4)}_{m=0}$.
\end{itemize}

To investigate the mass dependence of the two-point function in dimensions
$d\geq 5$ we derive a recursion relation which expresses
$\Delta^{+(d+2)}_m$ in terms of $\Delta^{+(d)}_m$. From (\ref{Delta^+a})
we find
\begin{gather}
  (\d_r^2-\d_t^2-m^2)\Delta^{+(d)}_m(y)=|S_{d-3}|\int_0^\infty dp\,p^{d}
\int_0^\pi d\theta\,(\mathrm{sin}\,\theta)^{d-1}
e^{-ipr\mathrm{cos}\,\theta}\frac{e^{i\omega t}}{2\omega}\notag\\
=(2\pi)^2\>\frac{|S_{d-3}|}{|S_{d-1}|}\> \Delta^{+(d+2)}_m(y)\ .
\label{Delta^+b}
\end{gather}
By using $\square_y^{(d)}=(\d_t^2-\d_r^2-\frac{d-2}{r}\d_r+$ derivatives with respect to 
the angles of $\vec{y}$) (the latter vanish in $\square^{(d)}\Delta^{+(d)}_m$)
and $(\square^{(d)}+m^2)\Delta^{+(d)}_m=0$ we obtain
\begin{equation}
  \Delta^{+(d+2)}_m(y)=\frac{-1}{2\pi r}\>\partial_r\>\Delta^{+(d)}_m(y)\ .
\end{equation}
Because of $\rho^{d-2}\Delta^{+(d)}_{\rho^{-1}m}(\rho y)
=\Delta^{+(d)}_m(y)$
and  Poincar\'{e} invariance, $\Delta^{+(d)}_m$ is of the form
\begin{equation}
\Delta^{+(d)}_m(y)=m^{d-2}\> F^{(d)}(m^2\>(y^2-iy^00))\ .
\end{equation}
With that we obtain
\begin{equation}
  \Delta^{+(d+2)}_m(y)=\frac{1}{2\pi(y^2-iy^00)}\>(m\partial_m+2-d)\>
\Delta^{+(d)}_m(y)\ .\label{recursion}
\end{equation}
The explicit formulas for $\Delta^{+(3)}_m,\>H^{\mu(4)}_m,\>\Delta^{+(4)}_m$ 
and the recursion relation (\ref{recursion}) imply that
\begin{itemize}
\item  (in odd dimensions) $\Delta^{+(2l+1)}_m$ is smooth in $m\geq 0$;

\item (in even dimensions) $\Delta^{+(2l)}_m$ contains a term which behaves as
$m^{2(l-1)}\>{\rm log}(m^2/\mu^2)$ for $m\to 0$;

\item 
\begin{equation}
H^{\mu\, (4+2k)}_m(y)\equiv\Delta^{+(4+2k)}_m-
\pi^{-k}\>m^{2(k+1)}\>f^{(k)}(m^2y^2)\>
{\rm log}(m^2/\mu^2)\label{2dH}
\end{equation}
(where $f^{(k)}$ is the $k$-th derivative of $f$ (\ref{f}))
is smooth in $m\geq 0$ and has the same properties as $H^{\mu\, (4)}_m$:
its antisymmetric part is $=i\Delta^{(d)}_m/2$
(where $d\equiv 4+2k$), it is  Poincar\'{e} invariant,
${\rm WF}(H^{\mu\, (d)}_m)={\rm WF}(\Delta^{+(d)}_m)$, it solves 
the Klein-Gordon equation (since $(\square_y^{(4+2k)} +m^2)
\>f^{(k)}(m^2\,y^2)=0$), it scales almost homogeneously with 
degree $(d-2)$ and power 1, and $H^{\mu\, (d)}_{m=0}=\Delta^{+(d)}_{m=0}$.
Due to the statements about the antisymmetric part, Poincar\'{e} invariance
and the wave front set, $H^{\mu\,(d)}_m$ can be used for the definition
(\ref{*-product}) of the $*$-product.
\end{itemize}
\section{Extension of a distribution to a point}
\setcounter{equation}{0}
We review in this Appendix the proofs of Theorem 
\ref{extension} and Proposition \ref{alm-hom-scal} 
(given in \cite{BF} and \cite{HW1-2} respectively) and add 
some completions. Similar or related techniques can be found in the older works
\cite{Gu,Ste} and \cite{EG}.

In case (a) of Theorem \ref{extension} the extension 
is obtained by the following limit:
let $\chi$ be a smooth function on $\RR^k$ such that $0\leq\chi\leq
1,\quad \chi (x)=0$ for $|x|<1$ and $\chi (x)=1$ for $|x|>2$. One can
show that the following limit
\begin{equation}
  (t,h)\=d\lim_{\rho\to\infty}(t^\circ(x),\chi(\rho x)h(x))
\label{fall(a)}
\end{equation}
(note $\chi(\rho x)h(x)\in {\cal D}(\RR^k\setminus\{0\})$)
exists, and that the so defined $t$ fulfills ${\rm sd}(t)={\rm sd}(t^\circ)$.

The construction of the extensions in the most interesting case
$k\leq {\rm sd}(t^\circ)<\infty$ (part (b) of Theorem \ref{extension})
proceeds as follows. Let
\begin{equation}
  \omega\=d {\rm sd}(t^\circ)-k\ ,
\quad\mathcal{D}_\omega (\RR^k)\=d
\{h\in \mathcal{D} (\RR^k)\>|\>\d^ah(0)=0\>\forall|a|\leq\omega\}.
\label{D_w}
\end{equation}
We will see that $t^\circ$ has a unique extension $t_\omega$ to
$\mathcal{D}_\omega\equiv\mathcal{D}_\omega (\RR^k)$ and that
each projector $W$ from $\mathcal{D}\equiv\mathcal{D}(\RR^k)$ onto 
$\mathcal{D}_\omega$ yields an extension $t\in\mathcal{D}^{\prime}
(\RR^k)$ (with ${\rm sd}(t)={\rm sd}(t^\circ)$) by $(t,h)\=d
(t_\omega ,Wh)$ (which is called '$W$-extension').

There are many possibilities to construct such a projector $W$ or,
equivalently, to choose a corresponding complementary space 
$\mathcal{E}=\mathrm{ran}(1-W)$ of 
$\mathcal{D}_\omega$ in $\mathcal{D}$. (By 'complementary space'
we mean: $\mathcal{D}=\mathcal{D}_\omega\oplus\mathcal{E}$).
The following Lemma gives a parametrisation of this possibilities in 
terms of a set of functions:
\begin{lemma}\label{W-projector} \cite{BF}: (a) For any set of functions
\begin{equation}
  \{w_a\in \mathcal{D}\>|\> a\in\NN_0^k,|a|\leq\omega,
\d^bw_a(0)=\delta^b_a\>\forall b\in \NN_0^k\},\label{w}
\end{equation}
the linear map
\begin{equation}
  W:\mathcal{D}\longrightarrow\mathcal{D}\>:\>Wh (x)=h(x)-
\sum_{|a|\leq\omega}(\d^ah)(0)w_a(x)\label{W}
\end{equation}
is a projector onto $\mathcal{D}_\omega$.\\
(b) Conversely, given a projector $W$ from $\mathcal{D}$ onto 
$\mathcal{D}_\omega$ (or equivalently a complementary space
$\mathcal{E}$ of $\mathcal{D}_\omega$ in $\mathcal{D}$), then there exist 
functions $(w_a)_a$ with the properties (\ref{w}), such that
$W$ can be expressed in terms of the $(w_a)_a$ by (\ref{W}).
\end{lemma}
An example for the functions $(w_a)_{|a|\leq\omega}$ is $w_a(x)
=\frac{x^a}{a!}w(x)$ where $w\in {\cal D}(\MM)$ and $w\vert_{\cal
  U}\equiv 1$ for some neighborhood ${\cal U}$ of $x=0$.

\begin{proof}\footnote{The idea of proof is given in \cite{BF}.} (a) is
obvious. To prove (b) we first show that there exists a basis
$(w_a)_{|a|\leq \omega}$ of the vector space 
$\mathcal{E}=\mathrm{ran}(1-W)$ with $\d^bw_a(0)
=\delta^b_a$. The decomposition
$\mathcal{D}=\mathcal{D}_\omega\oplus\mathcal{E}$ induces
a decomposition of the dual space $\mathcal{D}'=\mathcal{D}'
_\omega\oplus \mathcal{D}_\omega^\perp$ by the prescriptions
$(f_2,h_1)=0\>\wedge\>(f_1,h_2)=0,\>\forall f_2\in\mathcal{D}'
_\omega ,\> h_1\in \mathcal{E},\> f_1\in \mathcal{D}_\omega^\perp,
\> h_2\in\mathcal{D}_\omega$. A basis of 
$\mathcal{D}_\omega^\perp$ is given by $(\d^a\delta)_{|a|\leq \omega}$.
We define $((-1)^{|a|}w_a)_{|a|\leq \omega}$ to be the
dual basis (in $\mathcal{E}$), and it obviously has the properties
(\ref{w}).

So, for any $h\in\mathcal{D}$, $(1-W)h$ can be written as
$(1-W)h (x)=\sum_a c^aw_a(x)$ with $c_a\in\CC$,
and we find $\d^bh (0)=\d^b (1-W)h (0)=c^b,\>|b|\leq\omega$. Hence,
$Wh (x)=h(x)-\sum_a(\d^ah)(0)w_a(x)$. 
\end{proof}
We split any $h\in\mathcal{D}$ into
$h=h_1+h_2,\>h_1=\sum_{|a|\leq\omega}(\d^ah)(0)
w_a\in\mathcal{E},h_2\in\mathcal{D}_\omega$.
$h_2$ has the form
\begin{equation}
  Wh(x)\equiv h_2(x)=\sum_{|a|=[\omega]+1}x^ag_a(x),\quad\mathrm{with}
\quad g_a\in\mathcal{D}.\label{f=xp}
\end{equation}
This decomposition of $h_2$ is non-unique in general, however, 
we will see that this does not matter.
From (\ref{sd:d}) we recall $\mathrm{sd}(x^bf)\leq\mathrm{sd}(f)-|b|,\>
\forall f\in {\cal D}'(\RR^k)$ or ${\cal D}'(\RR^k\setminus\{0\})$.
For $|a|=[\omega]+1$ we find $\mathrm{sd}(x^at^\circ)\leq\omega+k-
([\omega]+1)<k$. Therefore, from part (a) of Theorem \ref{extension} 
we know that
$x^at^\circ\in {\cal D}'(\RR^k\setminus\{0\})$ has a unique extension 
$\overline{x^at^\circ}\in\mathcal{D}'(\RR^k)$. Now we define 
$t\in\mathcal{D}'(\RR^k)$ by
\begin{equation}
  (t,h)\=d \sum_{|a|=[\omega]+1}(\overline{x^at^\circ},g_a)+
\sum_{|a|\leq\omega}C_a(\d^ah)(0),\quad h\in\mathcal{D}(\RR^k),
\label{ext}
\end{equation}
where the $C_a\in\CC$ are arbitrary constants. By means of
(\ref{fall(a)}) we find for the first term
\begin{equation}
  (t,Wh)=\sum_{|a|=[\omega]+1}(\overline{x^at^\circ},g_a)=
\lim_{\rho\to\infty}(t^\circ(x),\chi(\rho x)Wh(x)).\label{t,Wh}
\end{equation}
Hence, $\sum_{|a|=[\omega]+1}(\overline{x^at^\circ},g_a)$ is independent
of the choice of the decomposition (\ref{f=xp}). Obviously, $t$
is an extension of $t^\circ$, and in $\cite{BF}$ it is proved
${\rm sd}(t)={\rm sd}(t^\circ)$. $t\vert_{\mathcal{D}_\omega}
\equiv t_\omega$ is uniquely fixed (\ref{t,Wh}), but
$t\vert_\mathcal{E}$ is completely arbitrary, namely
$(t,w_a)=C_a$ can be arbitrarily chosen. This is the freedom
of normalization in perturbative renormalization (\ref{P(D)delta}), 
which gives rise to the renormalization group (see Sect.~4).

In case (c) of Theorem \ref{extension} 
there exists a linear functional $\bar t$
on $\mathcal{D}(\RR^k)$ which fulfills $\bar t(h)=(t^\circ,h)\quad
\forall h\in {\cal D}(\RR^k\setminus\{0\})$, according to 
the Hahn-Banach theorem. But $\bar t$ is not continuous, i.e.~it is
not a distribution.

It is useful to know that given an extension $t$ of $t^\circ$, there
exists a projector $W$ from $\mathcal{D}$ onto $\mathcal{D}_\omega$
such that $t=t\circ W$ (i.e.~ all constants $C_a$ in
(\ref{ext}) vanish), in detail:
\begin{lemma} Let $\omega\=d {\rm sd} (t^\circ)-k$ and let an extension
$t$ of $t^\circ$ be given with ${\rm sd}(t)={\rm sd}(t^\circ)$.\\
(a) Then there exists a complementary space $\mathcal{E}$ of
$\mathcal{D}_\omega$ in $\mathcal{D}$ with
$t\vert_\mathcal{E}=0$.\\
(b) There exist functions $w_a\in \mathcal{D},\>
a\in\NN_0^k, |a|\leq\omega$ with $\d^bw_a(0)=\delta^b_a$
and $t=t\circ W$, where $W$ is given in terms of the
$(w_a)_a$ by (\ref{W}).
\end{lemma}
\begin{proof} By part (b) of Lemma \ref{W-projector}, 
the statement (b) is a
consequence of (a). To prove (a) let $\mathcal{E}_1$ be a complementary
space of $\mathcal{D}_\omega$ in $\mathcal{D}$ and $W_1$ the
corresponding projector on $\mathcal{D}_\omega$. 
We choose a $g\in\mathcal{D}_\omega$
with $(t,g)=1$. Now we set
\begin{equation}
  \mathcal{E}\=d\{k -(t,k)g\>|\>k\in \mathcal{E}_1\}.
\end{equation}
Obviously $\mathcal{E}$ is a vector space and it holds $t\vert_
\mathcal{E}=0$. To see $\mathcal{D}=\mathcal{D}_\omega +\mathcal{E}$
we decompose any $h\in \mathcal{D}$ into $h=h_1+h_2,\>
h_1\in\mathcal{E}_1, h_2\in \mathcal{D}_\omega$. Then,
$h=(h_1-(t,h_1)g)+(h_2+(t,h_1)g),\>
(h_1-(t,h_1)g)\in\mathcal{E},\> (h_2+(t,h_1)g)
\in \mathcal{D}_\omega$. It remains to show $\mathcal{E}\cap
\mathcal{D}_\omega=\{0\}$. Let $l\in \mathcal{E}\cap
\mathcal{D}_\omega$. So, $l=k -(t,k)g$ for some
$k\in \mathcal{E}_1$, and on the other hand
$l=W_1l=W_1k -(t,k)W_1g=-(t,k)g$. We find
$k =0$ and hence $l =0$. 
\end{proof}
 
Obviously the $W$-extension (\ref{t,Wh}) is a {\it non-local} renormalization prescription:
it depends on $t^\circ\vert_{{\cal D}(U)}$ where $U:=\cup_{|a|\leq\omega}\supp w_a$. 
In contrast the condition of almost homogeneous scaling ensures that the 
extension depends on the short distance behavior of $t^\circ$ only.
We now prove that the latter condition can be maintained, following to a 
large extent  \cite{HW1-2}.
\begin{proof}[Proof of Proposition \ref{alm-hom-scal}]
Let $t_1$ be any 
extension of $t^\circ$ with ${\rm sd}(t_1)={\rm sd}(t^\circ)=D$.
Since $t^\circ$ scales almost homogeneously (with power $N$) the support
of $(x\d_x+D)^{N+1}t_1(x)$ must be contained in $\{0\}$. In addition
it holds ${\rm sd}((x\d_x+D)^{N+1}t_1)={\rm sd}(t_1)=D$, and hence
\begin{equation}
  (\sum_r x_r\d_{x_r}+D)^{N+1}t_1(x)=\sum_{|a|\leq D-k}C_a\d^a
\delta^{(k)}(x)\ .
\end{equation}
We will frequently use
\begin{equation}
  (\sum_r x_r\d_{x_r}+D)\d^a\delta^{(k)}(x)=
(D-k-|a|)\d^a\delta^{(k)}(x)\ .
\end{equation}
\begin{itemize}
\item For $D\not\in\NN_0+k$ we may set
  \begin{equation}
    t\=d t_1-\sum_{|a|\leq D-k}\frac{C_a}{(D-k-|a|)^{N+1}}
    \d^a\delta^{(k)}\ ,\label{ext-alm-hom-scal}
  \end{equation}
   This is an extension which maintains even the power of the almost
   homogeneous scaling.
\item If $D\in\NN_0+k$ the subtraction of the $\d^a\delta$-terms in
   (\ref{ext-alm-hom-scal}) works not for $|a|=D-k$. We can only
   perform the finite renormalization
  \begin{equation}
    t\=d t_1-\sum_{|a|< D-k}\frac{C_a}{(D-k-|a|)^{N+1}}
    \d^a\delta^{(k)}\ .
  \end{equation}
With that
\begin{equation}
  (\sum_r x_r\d_{x_r}+D)^{N+1}t=\sum_{|a|=D-k}C_a\d^a\delta^{(k)}\ .
\end{equation}
However, applying the operator $(x\d_x+D)$ once more we get zero,
i.e. $t$ scales almost homogeneously with power $\leq N+1$.   
\end{itemize}
The statements about the uniqueness of $t$ are obvious, because
$\d^a\delta^{(k)}$ scales homogeneously with degree $(k+|a|)$. 
\end{proof}

How to find an extension $t$ of $t^\circ$ (with ${\rm sd}(t)={\rm sd}
(t^\circ)$) in practice? For ${\rm sd}(t^\circ)<k$ this is trivial:
$t$ is given by the same formula, the domain may be extended by
continuity (\ref{fall(a)}). But for $k\leq {\rm sd}(t^\circ)<\infty$
the map $W$ (\ref{W}) gets complicated in explicit calculations.
(An exception are purely massive theories in which one may choose
$w_a=\frac{x^a}{a!}\not\in {\cal D}$ in (\ref{w})-(\ref{W});
this gives the 'central solution' of Epstein and Glaser \cite{EG}.) A  
construction of $R_{1,1}$ (or equivalently $T_2$) is given in Appendix
B. It uses the K\"allen-Lehmann representation of the commutator
of two Wick polynomials, and hence, it is unclear how to
generalize this method to higher orders. It seems that {\bf differential 
renormalization} \cite{FJL,LMV,P} is a practicable way to trace back 
the case $k\leq {\rm sd}(t^\circ)<\infty$ to the trivial case 
${\rm sd}(t^\circ)<k$ in arbitrary high orders. The idea is to write 
$t^\circ$ as a derivative
of a distribution $f^\circ\in {\cal D}'(\RR^k\setminus\{ 0 \})$ with
${\rm sd} (f^\circ)<k$; more precisely
\begin{equation}
  t^\circ =Df^\circ\quad\quad\mathrm{where}\quad\quad
D=\sum_{|a|=l}C_a\d^a \quad (C_a\in\CC)\label{t^0=Df^0}
\end{equation}
such that ${\rm sd}(f^\circ)={\rm sd}(t^\circ)-l<k$. Let $f\in 
{\cal D}'(\RR^k)$ be the unique extension of $f^\circ$ with 
${\rm sd}(f)={\rm sd}(f^\circ)$. Then,
\begin{equation}
  t\=d Df\label{t=Df}
\end{equation}
solves the extension problem with ${\rm sd}(t)={\rm sd}(t^\circ)$. 
The non-uniqueness of $t$ shows up in the non-uniqueness of $f^\circ$:
one may add to $f^\circ$ a distribution $g^\circ\in {\cal D}'
(\RR^k\setminus\{ 0 \})$ with ${\rm sd}(g^\circ)={\rm sd}(f^\circ)$
and $Dg^\circ =0$ on $\RR^k\setminus\{ 0 \}$. The (unique)
extension $g$ of $g^\circ$ (with ${\rm sd}(g)={\rm sd}(g^\circ)$)
fulfills $\supp Dg\subset\{ 0\}$ and, hence, the addition
of $g^\circ$ can change $t$ (\ref{t=Df}) only by a local term.
For example let ${\rm sd}(t^\circ)=k$ and $D=\square$. Then, $g$ is of the
form: $g=\alpha D^{\rm ret}+\>$(solution of the
homogeneous differential equation) ($\alpha\in\CC$),
and this yields $t_\mathrm{new}\equiv D(f+g)=t_\mathrm{old}+\alpha\>\delta$.
\medskip

\noindent {\it Example:} {\bf differential renormalization of the massless fish and setting-sun
diagram,}
\begin{gather}
r^\circ_0(y)=j_0(y)\>\Theta(-y^0)\ ,\quad j_0(y)\equiv\Bigl(
\frac{1}{(y^2-iy^00)^2}-\frac{1}{(y^2+iy^00)^2}\Bigr)\ ,\notag\\
r^\circ_1(y)=j_1(y)\>\Theta(-y^0)\ ,\quad j_1(y)\equiv\Bigl(
\frac{1}{(y^2-iy^00)^3}-\frac{1}{(y^2+iy^00)^3}\Bigr)\ ,
\end{gather}
{\bf and of}
\begin{equation}
r^\circ_2(y)=j_2(y)\>\Theta(-y^0)\ ,\quad j_2(y)\equiv\Bigl(
\frac{{\rm log}(-\mu^2(y^2-iy^00))}{(y^2-iy^00)^2}-(y\rightarrow -y)\Bigr)
\end{equation}
{\bf for $d=4$}, cf.~Appendix C. These $r^\circ$-distributions appear in the Example 
(\ref{3,3:m,mu})-(\ref{scal-exp2}).\footnote{We omit constant pre-factors.} In agreement
with Lemma 1(b) the $j_l$'s have support in $\bar V_+\cup \bar V_-$.
We are looking for distributions $J_l$ with
\begin{equation}
{\rm sd}(J_l)<4\ ,\quad \supp J_l\subset (\bar V_+\cup \bar V_-)
\quad\mathrm{and}\quad j_l=D_l\>J_l
\end{equation}
where $D_l$ is a power of the wave operator. Due to the lowered scaling degree
of $J_l$, the product $J_l(y)\,\Theta(-y^0)$ exits in ${\cal D}'(\RR^4)$
and one easily verifies that
\begin{equation}
r_l(y)\=d D_l\>\bigl(J_l(y)\>\Theta(-y^0)\bigr)
\end{equation}
is a Lorentz invariant extension of $r^\circ_l$ with the same scaling degree.
With some trial and error one finds
\begin{gather}
j_0(y)=\square_y\,\Bigl(\frac{-{\rm log}(-\mu^2(y^2-iy^00))}{4\,(y^2-iy^00)}
-(y\rightarrow -y)\Bigr)\ ,\notag\\
j_1(y)=\square_y\square_y\,\Bigl(\frac{-{\rm log}(-\mu^2(y^2-iy^00))}{32\,(y^2-iy^00)}
-(y\rightarrow -y)\Bigr)\ ,\notag\\
j_2(y)=\square_y\,\Bigl(\frac{-({\rm log}(-\mu^2(y^2-iy^00)))^2
-2\, {\rm log}(-\mu^2(y^2-iy^00))}{8\,(y^2-iy^00)}
-(y\rightarrow -y)\Bigr)\ .\notag\\
\end{gather}
In case of the fish and the setting sun diagram a scale $\mu >0$ is introduced;
this cannot be avoided by using other methods of renormalization either, cf.~Appendix C.
If we would replace $(-\mu^2)$ by $\mu^2$ in $J_0$ and $J_1$ the relation
$j_l=D_l\>J_l$ would still hold, but $J_0$ and $J_1$ would have support
in $\{ y|y^2\leq 0\}$. This alternative possibility to fulfill $j_l=D_l\>J_l$
reflects the peculiarity that $j_0$ and $j_1$ vanish on ${\cal D}(\{y|y^2>0\})$.
All $J_l$'s scale almost homogeneously with degree $2$ and the corresponding 
power is the power of
${\rm log}(\mu^2...)$. We explicitly see that in all three examples the extension
increases this power by $1$. For the breaking of homogeneous scaling of 
the fish and setting sun diagram we obtain 
\begin{gather}
\rho^4\>r_0(\rho y)-r_0(y)=i\pi\>{\rm log}\>\rho\>\>\square_y
\Bigl(\Theta (-y^0)\>\delta(y^2)\Bigr)=i2\pi^2\>{\rm log}\>\rho\>\>\delta(y)\ ,\\
\rho^6\>r_1(\rho y)-r_1(y)=\frac{i\pi}{8}\>{\rm log}\>\rho\>\>\square_y\square_y
\Bigl(\Theta (-y^0)\>\delta(y^2)\Bigr)=
\frac{i\pi^2}{4}\>{\rm log}\>\rho\>\>\square\delta(y)\ ,
\end{gather}
where we use $\Theta (-y^0)\>\delta(y^2)\sim D^{\rm ret}(-y)$
and $\square D^{\rm ret}=\delta$. In Appendix C these results are
obtained by means of another method of renormalization,
which is more straightforward.
\medskip

\noindent {\it Remark:} The construction in the proof of Theorem
\ref{main} yields also an
extension $t$ of the given $t^\circ$ if one works in (\ref{D_w}),
(\ref{f=xp}) and (\ref{ext}) with an $\omega$ which is strictly greater
than $({\rm sd}(t^\circ)-k)$; but then it holds genericly ${\rm sd}(t)=
\omega +k>{\rm sd}(t^\circ)$. (This is called an 'over-subtracted' 
extension.)
\section{Extension of two-point functions}
\setcounter{equation}{0}
In this Appendix the number of space time dimensions is $d=4$.  
The $x$-space method which we give here to renormalize the fish diagram
\begin{equation}
  r_0(y)\=d r_{1,1}(\varphi^2,\varphi^2)(y)\ ,
\quad\quad y\equiv x_1-x\ ,
\end{equation}
and the setting-sun diagram
\begin{equation}
  r_1(y)\=d r_{1,1}(\varphi^3,\varphi^3)(y)\ ,
\end{equation}
can be used for arbitrary first order terms $r_{1,1}(A_1,A_2),
\>A_1,A_2\in {\cal P}$. (See footnote \ref{fn:r} for the notation.)
We treat here the massless case, which needs additional care to 
avoid IR-divergences. We first compute the fish diagram by following
our inductive construction of Sect.~3. A
straightforward calculation yields
\begin{gather}
  j_0(y)\=d \omega_0\Bigl([\varphi^2(x_1),\varphi^2(x)]_{\star}
\Bigr)\notag\\
=2\bigl(D^+(y)^2-D^+(-y)^2\bigr)=\frac{i}{2(2\pi)^2}\int_0^\infty
dm^2\,\Delta_m(y)\ ,\label{KL}
\end{gather}
which is the K\"allen-Lehmann representation.
We recall $\Delta_m(y)=
\Delta^\mathrm{ret}_m(y)-\Delta^\mathrm{ret}_m(-y)$ with $(\square+m^2)
\Delta^\mathrm{ret}_m=\delta$ and $\supp
\Delta^\mathrm{ret}_m\subset\bar V_+$. According to the axioms
Causality, GLZ relation and Scaling the 
wanted distribution $r_0$ is determined by
\begin{equation}
  \supp r_0\subset\bar V_-\ ,\quad (r_0,h)=-i(j_0,h)\quad\forall h\in
{\cal D}(\bar V_-\setminus \{0\})\label{defprop-r}
\end{equation}
and 
\begin{equation}
  (\rho\d_\rho +4)^2r_0(\rho y)=0\ .\label{fish:scal}
\end{equation}
If we replace $\Delta_m(y)$ in (\ref{KL}) by $-\Delta^\mathrm{ret}_m(-y)$
the $m^2$-integral becomes UV-divergent.
However, for $\mu>0$ one easily verifies that
\begin{equation}
  r_{0\,\mu} (y)\=d\frac{1}{2(2\pi)^2}(\square_y-\mu^2)
\int_0^\infty dm^2\,\frac{\Delta^\mathrm{ret}_m(-y)}{m^2+\mu^2}
\label{r:int}
\end{equation}
exists as a distribution and solves (\ref{defprop-r}). 
To avoid an IR-divergence, we need to
introduce a scale $\mu >0$, which breaks homogeneous scaling
(because of $\rho^4 r_{0\,\mu} (\rho y)=r_{0\,\rho\mu}(y)$). To verify 
the scaling requirement (\ref{fish:scal}) we compute
\begin{gather}
  \rho^4 r_{0\,\mu} (\rho y)-r_{0\,\mu} (y)=r_{0\,\rho\mu}(y)-r_{0\,\mu} (y)\notag\\
=\frac{1}{2(2\pi)^2}\Bigl(((\square_y-\mu^2)+(1-\rho^2)\mu^2)\int dm^2\,
\frac{\Delta^\mathrm{ret}_m(-y)}{m^2+\rho^2\mu^2}\notag\\
-(\square_y-\mu^2)\int
dm^2\,\frac{\Delta^\mathrm{ret}_m(-y)}{m^2+\mu^2}\Bigr)\notag\\
=\frac{1}{2(2\pi)^2}\Bigl[\int dm^2\,
(\square_y-\mu^2)\Delta^\mathrm{ret}_m
(-y)\Bigl(\frac{1}{m^2+\rho^2\mu^2}-\frac{1}{m^2+\mu^2}\Bigr)\notag\\
+(1-\rho^2)\mu^2\int
dm^2\,\frac{\Delta^\mathrm{ret}_m(-y)}{m^2+\rho^2\mu^2}\Bigr]=
\frac{1}{(2\pi)^2}\>\mathrm{log}\,\rho\>\>\delta (y)\ .\label{r:scaling}
\end{gather}
Hence, (\ref{fish:scal}) is indeed satisfied for all $\mu >0$. So we 
get explicitly the breaking of homogeneous scaling without really
computing the integral (\ref{r:int}). As a byproduct the calculation
(\ref{r:scaling}) shows explicitly that the choice of $\mu>0$
is precisely the choice of the indeterminate parameter $C$ in the general
solution $r_0(y)+C\delta(y)$.\footnote{Note that
\begin{gather}
    r_{0\,\mu\nu} (y)\equiv\frac{-1}{2(2\pi)^2}(-\square_y+\mu^2)(-\square_y+\nu^2)
\int_0^\infty dm^2\,\frac{\Delta^\mathrm{ret}_m(-y)}{(m^2+\mu^2)
(m^2+\nu^2)}\notag\\
=r_{0\,\mu} (y)+\frac{1}{2(2\pi)^2}\int_0^\infty dm^2\,
\frac{1}{(m^2+\mu^2)(m^2+\nu^2)}\cdot (\square -\mu^2)\delta (y)
\end{gather}
solves also (\ref{defprop-r}), but it violates (\ref{fish:scal}): 
the term $\sim\square\delta (y)$ scales homogeneously with degree $6$
(instead of $4$).}
\medskip

We proceed analogously for the setting-sun diagram:
\begin{gather}
  j_1(y)\=d \omega_0\Bigl([\varphi^3(x_1),
\varphi^3(x)]_{\star}\Bigr)\notag\\
=6\bigl(D^+(y)^3-D^+(-y)^3\bigr)=\frac{3i}{16(2\pi)^4}\int_0^\infty
dm^2\,m^2\>\Delta_m(y)\ .
\end{gather}
Obviously,
\begin{equation}
  r_{1\,\mu_1,\mu_2} (y)=\frac{-3}{16(2\pi)^4}(\square_y-\mu_1^2)
(\square_y-\mu_2^2)\int_0^\infty dm^2\,
\frac{m^2\,\Delta^\mathrm{ret}_m(-y)}{(m^2+\mu_1^2)(m^2+\mu_2^2)}
\label{r^1}
\end{equation}
(where $\mu_1,\mu_2 >0$) solves (\ref{defprop-r}). But we are looking
for solutions $r_1$ of (\ref{defprop-r}) which additionally scale 
almost homogeneously with degree $6$ and power $\leq 1$, i.e.~
$(\rho\d_\rho +6)^2 r_1(\rho y)=0$. For $r_{1\,\mu,\mu}$
the breaking of homogeneous scaling is equal to
\begin{gather}
\rho^6 r_{1\,\mu,\mu}(\rho y)-r_{1\,\mu,\mu}(y)=
  (r_{1\,\rho\mu,\rho\mu}(y)-r_{1\,\mu,\rho\mu}(y))
+(r_{1\,\mu,\rho\mu}(y)-r_{1\,\mu,\mu}(y))\notag\\
=\frac{-3}{16(2\pi)^4}\Bigl[-\square_y\delta (y)\>(\mathrm{log}\,\rho^2)+
\delta (y)\>\mu^2\>(\rho^2-1)\Bigr]\ ,\label{ssun:scal}
\end{gather}
where the method (\ref{r:scaling}) is used twice. We see that
$r_{1\,\mu,\mu}$ violates our scaling condition and,
by a generalization of the calculation (\ref{ssun:scal}),
one finds that this holds true even for all $r_{1\,\mu_1,\mu_2}$,
$(\mu_1,\mu_2)\in\RR^+\times\RR^+$. However, from the result
(\ref{ssun:scal}) we read off that
\begin{equation}
  r_1(y)\=d r_{1\,\mu,\mu}(y)+\frac{3}{16(2\pi)^4}
\>\mu^2\>\delta (y)+C_2\square\delta (y)\label{r_0^1}
\end{equation}
(where $C_2\in\RR$ is arbitrary) fulfills our requirements. (\ref{r_0^1})
is the most general solution which is additionally Lorentz invariant
and unitary. For the breaking of homogeneous scaling we obtain
\begin{equation}
  \rho^6 r_1(\rho y)-r_1(y)=\frac{3}{8(2\pi)^4}\>
(\mathrm{log}\,\rho)\>\>\square\delta (y)\ .\label{r_0^1:scal}
\end{equation}
A general fact shows up in the results (\ref{r:scaling}) and
(\ref{r_0^1:scal}) (which is also valid for $m>0$): 
the breaking of homogeneous scaling
is independent of the normalization (i.e.~of the choice of $\mu$
in (\ref{r:int}) and $C_2$ in (\ref{r_0^1})), because the undetermined
polynomial $\sum_{|a|+l=\omega}C_{a,l}m^l\d^a
\delta (y)$ scales homogeneously.
\section{Maintenance of symmetries in the extension 
of distributions}
\setcounter{equation}{0}
In contrast to a large part of the literature we work in this
Appendix with our normalization conditions Smoothness in $m\geq 0$
and Scaling, instead of the upper bound 
(\ref{axiom:sd}) on the scaling degree. However, by obvious
modifications, the procedure given here can just as well be based
on the latter normalization condition.
We investigate the question whether symmetries can be
maintained in the process of renormalization. Or in mathematical
terms: given a $t^\circ\equiv t^{(m)\circ}\in 
{\cal D}'(\RR^k\setminus\{0\})$ which is smooth in $m\geq 0$
and scales almost homogeneously with degree $D$ and power $N$
(\ref{scal:t^mo}), does there
exist an extension $t\equiv t^{(m)}\in {\cal D}'(\RR^k)$ 
with the same symmetries and smoothness (in $m$) as
$t^\circ$ and which scales almost homogeneously with $D$
and power $\leq (N+1)$?    

{\bf Existence of a symmetric extension:}
let $V$ be a representation of a group $G$ on ${\cal D}(\RR^k)$ under
which ${\cal D}(\RR^k\setminus\{0\})$ and $t^\circ$ are invariant,
\begin{equation}
  (t^\circ,V(g)h)=(t^\circ,h)\quad\quad\forall h\in {\cal
    D}(\RR^k\setminus\{0\})\ ,\quad g\in G\ .\label{t^*:inv}
\end{equation}
We denote by $V^T$ the transposed representation of $G$ on 
${\cal D}'(\RR^k)$
\begin{equation}
  (V^T(g)s,h)=(s,V(g^{-1})h)\ ,\quad\quad s\in {\cal D}'(\RR^k)\ ,
\end{equation}
and we additionally assume that smoothness in $m\geq 0$ and the scaling
behavior (\ref{scal:t^mo}) are maintained under $V^T(g)$,
$\forall g\in G$.

Let $t\in {\cal D}'(\RR^k)$ be an arbitrary extension of $t^\circ$ with
the required smoothness and scaling properties. For $h\in {\cal D}
(\RR^k\setminus\{0\})$ we know
\begin{equation}
  (V^T(g)t,h)=(t,V(g^{-1})h)=(t^\circ,V(g^{-1})h)=(V^T(g)t^\circ,h)\ .
\end{equation}
So, $V^T(g)t$ is an extension of $V^T(g)t^\circ=t^\circ$. With
(\ref{non-unique:m>0}) we conclude\footnote{The case $m=0$ is included
by using $m^l\vert_{m=0}=\delta_{l,0}$.}
\begin{equation}
  l(g)\=d V^T(g)t-t\in {\cal D}^\perp_\omega (\RR^k)\=d
\Bigl\{\sum_{|a|+l=\omega}m^lC_{l,a}\d^a\delta^{(k)}\> |\> 
C_{l,a}\in\RR\>\mathrm{or}\>\CC\Bigr\}\label{def:l}
\end{equation}
where $\omega\=d D-k$. For an $\tilde l\in
{\cal D}^\perp_\omega (\RR^k)$ we find $V^T(g)\tilde l=\bigl(
V^T(g)(t+\tilde l)-V^T(g)t\bigr)\in
{\cal D}^\perp_\omega (\RR^k)$, since both terms are an extension of
$t^\circ$. Hence, 
\begin{equation}
  G\ni g\longrightarrow \pi(g)\=d
  V^T(g)\vert_{{\cal D}^\perp_\omega (\RR^k)}\label{subrep:D}
\end{equation}
is a sub-representation of $V^T$.

We are searching an $l_0\in {\cal D}^\perp_\omega (\RR^k)$ such that
$t+l_0$ is invariant,
\begin{equation}
V^T(g)(t+l_0)=t+l(g)+V^T(g)l_0=t+l_0\ ,\label{l_0}
 \end{equation}
i.e.~ $l_0$ must fulfill 
\begin{equation}
  l(g) = l_{0} - \pi(g) l_{0}\ ,\quad\quad\forall g\in G\ .\label{eq:l_0}
\end{equation}
From (\ref{def:l}) it follows that $l(g)$ has the property
\begin{equation}
  l(gh) = V^{T}(gh)t -t = V^{T}(g)\left( 
        V^{T}(h)t-t\right) + V^{T}(g)t-t 
        = \pi(g) l(h) +l(g)\ .\label{cocycle} 
\end{equation}
A solution of such an equation is called a 'cocycle'. If $l(g)$ is
of the form (\ref{eq:l_0}) for some $l_0$, it is called a
'coboundary', and such an $l(g)$ solves automatically the cocycle 
equation (\ref{cocycle}). The space of the cocycles modulo the
coboundaries is called the cohomology of the group with respect to the
representation $\pi$. Summing up, {\it the invariance of $t^\circ$ with
  respect to the representation $V^T$ of $G$ can be maintained in the
  extension if the cohomology of $G$ with respect to $\pi$ (\ref{subrep:D})
  is trivial.}

We are now going to show that {\it this supposition holds true if all
finite dimensional representations of $G$ are completely reducible.}
For this purpose we consider the restriction of $V^T(g)$ to the space
$(\CC\cdot t)\oplus {\cal D}^\perp_\omega (\RR^k)$. 
From (\ref{def:l}) we see that this is a finite dimensional
representation of $G$, which may be identified with the matrix 
representation
\begin{equation}
g\longrightarrow \overline{\pi} (g)=\left(
        \begin{array}{cc}
                1 & 0  \\
                l(g) & \pi(g)
        \end{array}
        \right)\ .\label{over-D}
\end{equation}
Due to the complete reducibility of $\overline{\pi}$, there exists a
1-dimensional invariant subspace $U$ which is complementary to the
representation space of $\pi$ (\ref{subrep:D}). Such a subspace is of
the form $U=\CC\cdot\left(
\begin{array}{c}
        1  \\
        l_{0}
\end{array}
\right)$. So there exists an $l_0\in {\cal D}^\perp_\omega
(\RR^k)$ with
\begin{equation}
\overline{\pi} (g)\left(
\begin{array}{c}
        1  \\
        l_{0}
\end{array}
\right)=\left(
        \begin{array}{c}
                1  \\
                l(g)+\pi(g)l_0
        \end{array}
        \right)
\in\CC\cdot\left(
\begin{array}{c}
        1  \\
        l_{0}
\end{array}
\right)\ .
\end{equation}
Hence, $\overline{\pi} \vert_U= {\bf 1}$, which means that $l_0$ solves 
(\ref{eq:l_0}). $\quad\square$

For the Lorentz group ${\cal L}_+^\uparrow$ all finite dimensional 
representations are
completely reducible and, hence, Lorentz invariance can be maintained
in perturbative renormalization. 

However, for the scaling transformations one has to consider the 
representations of $\RR_+$ as a multiplicative group. They are not
always completely reducible. An example for a reducible but not
completely reducible representation is
\begin{equation}
        \RR_{+}\ni\rho\mapsto \left(
        \begin{array}{cc}
                1 & 0  \\
                \mathrm{ln}\>\rho & 1
        \end{array}
        \right)\ .
\end{equation}
The existence of such representations can be understood as the reason
for the breaking of homogeneous scaling.
\medskip

\noindent {\it Example:} The action of massless QED is invariant with
respect to the following $U(1)$ transformations
\begin{equation}
  U(1)_V\> :\>\psi(x)\rightarrow e^{i\alpha(x)}\psi(x)\ ,\quad\quad
U(1)_A\> :\>\psi(x)\rightarrow e^{i\beta(x)\gamma_5}\psi(x)\ ,\label{trafo} 
\end{equation}
$\alpha(x),\beta(x)\in [0,2\pi)$, and $\bar\psi$ is transformed 
correspondingly. According to Noether's Theorem the corresponding 
currents
\begin{equation}
  j^\mu_{V\, g{\cal L}}=-(\bar\psi\gamma^\mu\psi)_{g{\cal L}}\quad
\mathrm{and}\quad j^\mu_{A\, g{\cal L}}=-(\bar\psi\gamma^\mu
\gamma^5\psi)_{g{\cal L}}
\end{equation}
are conserved in classical field theory. In QFT the just derived
result implies that invariance of the retarded products with respect to
$U(1)_V\times U(1)_A$ transformations (\ref{trafo}) can be realized,
because this group is compact. But it is well known that conservation
of $j^\mu_{A\, {\cal L}}$ is not compatible with conservation of
$j^\mu_{V\, {\cal L}}$, this is the axial anomaly. So, in general
Noether's Theorem cannot be fulfilled in QFT, see however \cite{BDL}.
\medskip

{\bf Construction of a Lorentz-invariant extension:}
there remains the question how to find the solution $l_0$ of the
equation (\ref{eq:l_0}) (if it exists). For compact groups this can be
done in the following way: we set
\begin{equation}
        l_{0}\=d \int_{G}\mathrm{d} g\,l(g)\ ,\label{l_0:comp}
\end{equation}
where $\mathrm{d} g$ is the uniquely determined measure
on $G$ which has norm 1 and is invariant under left- and
right-translations (Haar-measure). To verify that (\ref{l_0:comp})
solves (\ref{eq:l_0}) we use the cocycle equation:
\begin{equation}
        l_{0}-\pi(g)l_{0} = \int_{G}\mathrm{d} h \left(l(h)- 
        \pi(g)l(h)\right) = \int_{G}\mathrm{d} h \left( 
        l(h)-l(gh)+l(g)\right) \ .
\end{equation}
Due to the translation invariance of the Haar-measure, the integrals
over the first two terms cancel, and since the measure is normalized
we indeed obtain (\ref{eq:l_0}).

For the Lorentz-group this method fails. Epstein and Glaser \cite{EG}
use that the renormalized distributions are boundary values of
functions which are analytic in a certain region of the complexified
Minkowski space and which are additionally invariant under the complex
Lorentz group. It suffices then to take into account the invariance 
with respect to the compact subgroup $\mathrm{SO}(4)$.
\medskip

We give here an alternative method which has some resemblance with 
\cite{Ste} and \cite{BPP}: the strategy is to construct the projector
$P$ on the space ${\cal D}_\mathrm{inv}$ of all invariant vectors in
the representation $\overline{\pi}$ (\ref{over-D}). Then, with $t$ being
the above used arbitrary extension of $t^\circ$ (with the mentioned
smoothness and scaling properties), the definition
\begin{equation}
  t_\mathrm{inv} \=d Pt\ ,\label{t-inv}
\end{equation}
yields a ${\cal L}_+^\uparrow$-invariant distribution. 
And it is also an extension of $t^\circ$ with the required 
properties, because $(t_\mathrm{inv}-t)\in {\cal D}^\perp_\omega 
(\RR^k)$. The latter is shown below. 

To find $P$ note that for a
Casimir operator $C$ (in any representation) the invariant 
vectors ${\cal D}_\mathrm{inv}$ are a subspace
of its kernel $C^{-1}(0)$, because $C$ is built from the elements of
the Lie algebra. Below we will find a Casimir operator $C_0$
of the Lorentz group with ${\cal D}_\mathrm{inv}=C_0^{-1}(0)$
in each finite dimensional representation. With that our
method relies on the fact that the operator $c^{-1}(c{\bf 1}-C_0)$ 
annihilates the eigenvectors (of $C_0$) to the eigenvalue $c\not= 0$,
and on $C_0^{-1}(0)$ it is $={\bf 1}$. Therefore, in each finite
dimensional representation of ${\cal L}_+^\uparrow$ 
and in particular for $\overline{\pi}$ (\ref{over-D}), the operator
\begin{equation}
        P\=d\prod_{c\ne 0}\frac{c{\bf 1}-\overline{\pi}(C_0)}{c}\label{P-inv}
\end{equation}
($c$ runs through all eigenvalues $\not= 0$ of $C_0$) is a projector
on ${\cal D}_\mathrm{inv}$.

To show $(t_\mathrm{inv}-t)\in {\cal D}^\perp_\omega (\RR^k)$ we write
\begin{equation}
  t_\mathrm{inv}=Pt=P(t+l_0)-Pl_0=t+l_0-Pl_0\ ,
\end{equation}
where we use that $t+l_0$ is invariant (\ref{l_0}). It follows from 
(\ref{over-D}) and (\ref{P-inv}) that 
the upper right coefficient of the matrix $P$
is $=0$. Hence, $Pl_0\in {\cal D}^\perp_\omega (\RR^k)$,
and this gives the assertion. 

To determine a Casimir operator $C_0$ of the Lorentz group with 
$C_0^{-1}(0)={\cal D}_\mathrm{inv}$ in each finite dimensional 
representation, we first note that there are 
two quadratic Casimirs,
\begin{equation}
        C_0=\vec{L}^2-\vec{M}^2\quad\mathrm{and}\quad 
C_1=\vec{L}\cdot\vec{M} \ ,\label{casimirs}
\end{equation}
where $\vec{L}$ denotes the infinitesimal rotations and $\vec{M}$ 
the infinitesimal Lorentz-boosts. On ${\cal D}(\RR^4)$ it holds
\begin{equation}
        \vec{L}=\frac{1}{i}\vec{x}\times \vec{\d} \ , \quad\quad
        \vec{M}= \frac{1}{i}(x^{0}\vec{\d}-\vec{x}\d_{0}) \ .\label{L,M} 
\end{equation}
The irreducible finite dimensional representations of the Lorentz
group ${\cal L}^{\uparrow}_{+}$ are those irreducible finite dimensional 
representations of $\mathrm{SL}(2,\CC)$ which represent the matrix 
$-1$ by $1$. The irreducible finite dimensional representations of
$\mathrm{SL}(2,\CC)$ are indexed by two spin quantum numbers 
$j_{1},j_{2}\in\frac{1}{2}\NN_{0}$ and have the form 
\begin{equation}
        \pi_{j_{1}j_{2}}(A) \bigl(\xi^{\otimes 2j_{1}} \otimes 
        \eta^{\otimes 2j_{2}}\bigr) = (A\xi)^{\otimes 2j_{1}} \otimes 
        ((A^{*})^{-1}\eta)^{\otimes 2j_{2}}\label{D_j_1j_2}
\end{equation}      
with $\xi,\eta\in\CC^2$. For $j_{1}+j_{2}\in\NN_{0}$ 
this yields a representation of the Lorentz group.
We denote by $\vec{L}_{i},\vec{M}_{i}$ the representations
of $\vec{L}$ and $\vec{M}$ on the left ($i=1$) and the right
factor ($i=2$) respectively. In the fundamental 
representation of $\mathrm{SL}(2,\CC)$ (i.e.~ $(j_1,j_2)=
(\frac{1}{2},0)$) we have
\begin{equation}
        \vec{L}=\frac{1}{2}\vec{\sigma} \ , \quad\quad 
        \vec{M}=\frac{i}{2}\vec{\sigma} \ ,
\end{equation}
and in the conjugated representation (i.e.~ $(j_1,j_2)=
(0,\frac{1}{2})$) it holds
\begin{equation}
        \vec{L}=\frac{1}{2}\vec{\sigma} \ , \quad\quad 
        \vec{M}=-\frac{i}{2}\vec{\sigma} \ .
\end{equation}
So we have
\begin{equation}
        \vec{M}_{1}=i\vec{L}_{1}\ ,\quad\quad \vec{M}_{2}=-i\vec{L}_{2}\ ,
\end{equation}
and the validity of these relations goes over to all representations 
$\pi_{j_1,j_2}$, $j_{1},j_{2}\in\frac{1}{2}\NN_{0}$ (\ref{D_j_1j_2}). 
With that we obtain for the Casimir operators
\begin{equation}
        C_0= (\vec{L}_{1}+\vec{L}_{2})^2 - 
        (\vec{M}_{1}+\vec{M}_{2})^2 = 2\vec{L}_{1}^2 + 
        2\vec{L}_{2}^2 = 2(j_{1}(j_{1}+1)+ j_{2}(j_{2}+1))\label{C_0} 
\end{equation}
and
\begin{equation}
        C_1= (\vec{L}_{1}+\vec{L}_{2})\cdot 
        (\vec{M}_{1}+\vec{M}_{2}) = i\vec{L}_{1}^2 - 
        i\vec{L}_{2}^2 = i(j_{1}(j_{1}+1) -j_{2}(j_{2}+1)) \ .
\end{equation}
We find indeed that $C_0$ vanishes on the trivial 
representation $j_{1}=j_{2}=0$ only. However, the kernel of 
$C_1$ is much bigger.

As an example let us consider a Lorentz invariant 
distribution $t_{\omega}$ on $\C{D}_{\omega}(\RR^4)$, $\omega=2$. Let
\begin{displaymath}
	W= 1-\sum_{|a|\leq\omega}(-1)^{|a|}\ket{w_a}\bra{\d^a\delta}
\end{displaymath}
be a projector on $\C{D}_{\omega}(\RR^4)$. The representation of the
Lorentz group on $\C{D}_{\omega}(\RR^4)^{\bot}=
\{\sum_{|a|\leq 2}C_a\d^a\delta|C_a\}\subset    
\C{D}'(\RR^4)$ has the irreducible sub-representations
\begin{displaymath}
	(j_{1},j_{2})= (0,0),(\frac{1}{2},\frac{1}{2}),(1,1)
\end{displaymath}
with the eigenvalues $c=0,3,8$ of the Casimir operator 
$C_0$ (\ref{C_0}). The latter reads
\begin{displaymath}
	\begin{split}
		C_0 &= -\frac{1}{2}(x_{\mu}\d_{\nu}-x_{\nu}\d_{\mu})
		                 (x^{\mu}\d^{\nu}-x^{\nu}\d^{\mu})\\
		  &=(x^{\mu}\d_{\mu})^2+2x^{\mu}\d_{\mu}-x^2 \square               
	\end{split}
\end{displaymath}
by using (\ref{casimirs})-(\ref{L,M}). Following (\ref{t-inv}) and (\ref{P-inv})    
a Lorentz invariant extension of $t_{\omega}$ is obtained by
\begin{displaymath}
	t_\mathrm{inv}:=P(t_{\omega}\circ W) \quad\mathrm{with}\quad	
                P:=(1-\frac{1}{3}C_0)(1-\frac{1}{8}C_0) \ .
\end{displaymath}
\section{Time ordered products}
\setcounter{equation}{0}
One can define 
time ordered products ('$T$-products') $(T_n)_{n\in\NN}$ by a 
direct translation of the axioms for retarded products (given in
Sect.~2) with the following modifications:
\begin{itemize}
\item $T_n$ is required to be symmetrical in {\it all} factors.

\item Causality is expressed by causal factorization:
\begin{gather}
T(A_1(x_1),...,A_n(x_n))=\notag\\
T(A_1(x_1),...,A_k(x_k))\star T(A_{k+1}(x_{k+1}),...,A_n(x_n))\label{caus}
\end{gather}
if $\{x_1,...,x_k\}\cap (\{x_{k+1},...,x_n\}+\bar V_-) =\emptyset$. 

\item Among the defining properties of $T$-products there is none
corresponding to the GLZ relation (\ref{glz:R}). 
The information corresponding to (\ref{glz:R})
is (\ref{intf})-(\ref{intf:R}), i.e.~ the definition of retarded
products as coefficients of the interacting fields 
which are obtained from the $T$-products by Bogoliubov's formula. (In 
Proposition 2 of \cite{DF} we start from the $T$-products and derive the GLZ
relation (\ref{glz:R}). This derivation uses only (\ref{intf})-(\ref{intf:R}), and linearity
and Symmetry are needed in order that (\ref{intf:R}) determines $R_{n,1}$
also for non-diagonal entries.)
\end{itemize}
The generating functional of the $T$-products is the (local)
$S$-matrix ${\bf S}(F)\=d$ $T(e_\otimes^{iF})$, $F\in {\cal F}_\mathrm{loc}$.

These defining properties of $T$-products are {\bf equivalent} to our
defining properties of retarded products in the sense of
the {\it unique} correspondence
\begin{equation}
  (T_n)_{n\in\{1,2,\ldots ,N+1\}}\longleftrightarrow 
(R_{n,1})_{n\in\{0,1,\ldots ,N\}}\label{T-R}
\end{equation}
given by (\ref{intf})-(\ref{intf:R}). Using the anti-chronological
products $(\bar T_n)_{n\in\NN}$, which are defined\footnote{Note 
that $\overline{T}_m$ is uniquely determined in terms 
of the $T$-products $T_l,\> 1\leq l\leq m$.} by 
$\bar T(e_\otimes^{-iF})\=d {\bf S}(F)^{-1}$, the correspondence
(\ref{T-R}) can be written more explicitly:
\begin{equation}
R_{n,1}(F_1\otimes...\otimes F_n;F)\=d
i^n\sum_{I\subset \{1,...,n\}}(-1)^{|I|}
\bar T_{|I|}(\otimes_{l\in I}F_l)
T_{|I^c|+1}((\otimes_{j\in
  I^c}F_j)\otimes F)\ .\label{ret-prod}
\end{equation}
This formula can also be used to construct inductively the
$T$-products from the retarded products: it yields $T_{n+1}$ in terms of
$R_{n,1}$ and $T_l,\bar T_k$ with $k,l\leq n$. Alternatively, to
obtain a direct formula for $T_n$ in terms of the $\{ R_{l,1}|0\leq
l\leq n-1\}$, we write Bogoliubov's formula (\ref{intf})
in the form
\begin{equation}
  F_{\lambda F}=-i {\bf S}(\lambda F)^{-1}\frac{d}{d\lambda}{\bf S}(\lambda F)\ .
\end{equation}
This differential equation is solved by the Dyson series
\begin{equation}
  {\bf S}(\lambda F)=\mathbf{1}+\sum_{k=1}^\infty i^k\int_0^\lambda
  d\lambda_k\int_0^{\lambda_k} d\lambda_{k-1}...\int_0^{\lambda_2} 
  d\lambda_1\> F_{\lambda_1 F}...F_{\lambda_{k-1} F}F_{\lambda_k F}\ .
\end{equation}
For $\lambda =1$ the term of $n$-th order in $F$ reads
\begin{gather}
  T_n(F^{\otimes n})=\sum_{k=1}^n i^{k-n}\sum_{l_1+...+l_k=n-k}
\frac{n!}{l_1!l_2!...l_k!}\notag\\
\cdot\frac{1}{(l_1+1)(l_1+l_2+2)...(l_1+l_2+...+l_k+k)}
R_{l_1,1}(F^{\otimes l_1},F)...R_{l_k,1}(F^{\otimes l_k},F)\ ,\label{T=R...R}
\end{gather}
where we have used $R_{l,1}((\lambda F)^{\otimes l},F)=\lambda^l
R_{l,1}(F^{\otimes l},F)$ and computed the $\lambda$-integrals. (\ref{T=R...R})
agrees with formula\footnote{We thank Christian Brouder for showing 
us this formula.}  (55) of \cite{EGS}, which is derived there in a different 
way. 
\end{appendix}

\vskip0.5cm
{\bf Acknowledgments:} We profitted from discussions with 
Dan Grigore, Henning Rehren, Othmar Steinmann and Raymond Stora. 
In particular the correspondence
with Raymond Stora was very helpful to find the proof of the AWI.
We particularly thank Romeo Brunetti for useful hints and
careful reading of the manuscript, and Stefan Hollands for helpful
comments and pointing out an error in an earlier version. During working at 
this paper M.~D. was mainly at the university of G\"ottingen;
he thanks for warm hospitality.

\end{document}